\documentclass[final,5p,times,twocolumn,authoryear]{elsarticle}

\usepackage{amssymb}
\usepackage{amsmath}

\usepackage[nolist]{acronym}
\usepackage{url}
\usepackage{subcaption}
\usepackage{placeins}
\usepackage{xcolor}
\usepackage{listings}
\definecolor{codegreen}{rgb}{0,0.6,0}
\definecolor{codegray}{rgb}{0.5,0.5,0.5}
\definecolor{codepurple}{rgb}{0.58,0,0.82}
\definecolor{backcolour}{rgb}{0.95,0.95,0.92}
\lstdefinestyle{myStyle}{
    commentstyle=\color{codepurple},
    keywordstyle=\color{codepurple},
    morekeywords={},
    numberstyle=\tiny\color{codegray},
    stringstyle=\color{codegreen},
    basicstyle=\ttfamily\footnotesize,
    emph={None},
    emphstyle=\color{codepurple},
    breakatwhitespace=false,         
    breaklines=true,                 
    keepspaces=true,                               
    showspaces=false,                
    showstringspaces=false,
    showtabs=false                  
}
\journal{Icarus}

\begin{document}

\begin{frontmatter}



\title{Debiasing astro-Photometric Observations with Corrections Using Statistics (DePhOCUS)} 

\author[affil1,affil2]{Tobias Hoffmann\corref{cor1}} 
\ead{tobias.hoffmann3@uol.de}
\cortext[cor1]{Corresponding author}

\author[affil2]{Marco Micheli} 
\author[affil3]{Juan Luis Cano} 
\author[affil2]{Maxime Devogèle} 
\author[affil4]{Davide Farnocchia}
\author[affil5]{Petr Pravec}
\author[affil6]{Peter Vereš}
\author[affil1]{Björn Poppe}

\affiliation[affil1]{organization={Division for Medical Radiation Physics and Space Environment, Carl von Ossietzky Universität Oldenburg},
            city={Oldenburg},
            postcode={26111},
            country={Germany}}
\affiliation[affil2]{organization={NEO Coordination Centre, ESA/ESRIN},
            addressline={Largo Galileo Galilei 1}, 
            city={Frascati (RM)},
            postcode={00044},
            country={Italy}}

\affiliation[affil3]{organization={Planetary Defence Office, ESA/ESOC},
            addressline={Robert-Bosch-Straße 5}, 
            city={Darmstadt},
            postcode={64293},
            country={Germany}}

\affiliation[affil4]{organization={Jet Propulsion Laboratory, California Insititute of Technology},
            addressline={4800 Oak Grove Dr.}, 
            city={Pasadena},
            postcode={91109},
            state={CA},
            country={United States}}

\affiliation[affil5]{organization={Astronomical Institute, Academy of Sciences of the Czech Republic},
            addressline={Fričova 1}, 
            city={Ondřejov},
            postcode={25165},
            country={Czech Republic}}

\affiliation[affil6]{organization={Minor Planet Center, Harvard-Smithsonian Center for Astrophysics},
            addressline={60 Garden Street}, 
            city={Cambridge},
            postcode={02138},
            state={MA},
            country={United States}}

\begin{abstract}
Photometric measurements allow the determination of an asteroid's absolute magnitude, which often represents the sole means to infer its size. Photometric observations can be obtained in a variety of filters that can be unique to a specific observatory. Those observations are then calibrated into specific bands with respect to reference star catalogs. In order to combine all the different measurements for evaluation, photometric observations need to be converted to a common band, typically V-band. Current band-correction schemes in use by \acs{IAU}'s \ac{MPC}, \acs{JPL}'s \ac{CNEOS} and \acs{ESA}'s \ac{NEOCC} use average correction values for the apparent magnitude derived from photometry of asteroids as the corrections are dependent on the typically unknown spectrum of the object to be corrected. By statistically analyzing the photometric residuals of asteroids, we develop a new photometric correction scheme that does not only consider the band, but also accounts for reference catalog and observatory. We analyzed nearly 500\,000 observations submitted to the \ac{MPC} from 468 asteroids with published and independently determined high confidence $H$ and $G$ values. We describe a new statistical photometry correction scheme for asteroid observations with debiased corrections. Testing this scheme on a reference group of asteroids, we see a $36\%$ reduction in the photometric residuals. Moreover, the new scheme leads to a more accurate and debiased determination of the $H$-$G$ magnitude system and, in turn, to more reliable inferred sizes. We discuss the significant shift in the corrections with this ``\acs{DePhOCUS}'' debiasing system, its limitations, and the impact for photometric and physical properties of all asteroids, especially \aclp{NEO}.
\end{abstract}




\end{frontmatter}

\begin{acronym}
 	\acro{aBCO}{Asteroid-weighted Band-Catalog-Observatory analysis}
	\acro{ADES}{Astrometry Data Exchange Standard}
    \acro{ATLAS}{Asteroid Terrestrial-impact Last Alert System}
    \acro{CNEOS}{Center for Near Earth Object Studies}
    \acro{CSS}{Catalina Sky Survey}
	\acro{DePhOCUS}{\textbf{De}biased astro-\textbf{Ph}otometic \textbf{O}bservations with \textbf{C}orrections \textbf{U}sing \textbf{S}tatistics}
	\acro{ESA}{European Space Agency}
	\acro{IAU}{International Astronomical Union}
	\acro{JPL}{Jet Propulsion Laboratory}
    \acro{LCDB}{Asteroid Lightcurve Database}
 	\acro{MPC}{Minor Planet Center}
 	\acro{NEO}[NEO]{Near-Earth Object}
	\acroplural{NEO}[NEOs]{Near-Earth Objects}
	\acro{NEOCC}{NEO Coordination Centre}
    \acro{Pan-STARRS}{Panoramic Survey Telescope And Rapid Response System}
	\acro{SDSS}{Sloan Digital Sky Survey}
    \acro{SNR}{Signal-to-Noise Ratio}
    \acro{UCAC}{USNO CCD Astrograph Catalog}
    \acro{ZTF}{Zwicky Transient Facility}
\end{acronym}


\section{Introduction}
To track and characterize asteroids, observers rely on ephemerides and brightness predictions. Asteroid orbit determination is a well-studied astrodynamical problem \citep{Milani2010}. It consists of the computation of the orbital elements of an object and propagating the orbit in time based on the gravitational force of the sun. It can also include the perturbations from other celestial bodies and non-gravitational effects such as the solar radiation pressure and the Yarkovsky effect \citep{Vokrouhlicky2000,DelVigna2018}. With sufficiently precise determined orbital parameters, the position of an object on the sky (ephemeris) can be reliably predicted.

Brightness predictions, on the other hand, are usually harder to predict. The predictions are based on previous photometric measurements and the orbital dynamics. Since the position and distance of the object with respect to Sun and Earth affects the reflected sunlight on its surface - more precisely, the light flux decreases quadratically with distance - accurate positional knowledge is a prerequisite for accurate brightness estimates. 

Photometry also serves as basis for the derivation of physical properties. More specifically, the asteroid's size is determined using the absolute magnitude $H$ -- the normalized magnitude at 1 au distance from the asteroid to Sun and Earth, respectively, for $0^\circ$ phase angle (angle between incident and reflected light) -- together with the object's albedo \citep{Bowell1989, Harris1997}. Albedo is not an easy parameter to derive. Most of albedo determinations are obtained using thermal observations \citep{Harris2002}, which are also strongly dependent on the $H$ magnitude determinations \citep{Masiero_2021}. On the other hand, other methods such as polarimetric observations is independent on the $H$ magnitude \citep{Cellino_2015}.

The physical description of the asteroid's brightness depends on two parameters in the $H$-$G$ model by \citet{Bowell1989}: $H$ is the absolute magnitude and $G$ is the slope parameter, describing the steepness of the $H$-$G$ phase function with increasing solar phase angle (cf. Fig~\ref{fig:methods.phasefunction} and Eq.~\ref{eq:phasecurve} in the section \ref{sec:methods.pre}-\ref{sec:methods.analysis}). Even though this method is well established, the model, its fitting procedure and results have considerable uncertainties and limitations:
\begin{enumerate}
    \item Most objects lack measurements at different phase angles and are poorly analyzed, which makes it especially challenging to determine the slope parameter $G$ confidently \citep{Buchheim2010, Pravec2012}. Therefore a canonical value $G=0.15$ is assumed for a vast majority of objects \citep{Veres2015}.
    \item Due to the irregular shape of asteroids or binary systems, rotation results in a varying reflective surface and magnitude, needed to be added to the model \citep{Pravec2006,Carry2024}.
    \item Observations use different (photometric) color-bands. Due to the different surface material of asteroids and their reflectance spectra, the color deviates from the Sun's spectra and varies among asteroids, resulting in intrinsic magnitude differences in different bands.
    \item The underlying magnitude estimates used by \ac{MPC} for the $H$ and $G$ parameter determination vary a lot in accuracy and contain systematic errors as they were made with different observational techniques and by many different observers over many years \citep{Muinonen2010,Veres2015}. We call these photometric estimates, which are provided with the astrometric observations of minor planets from the \ac{MPC}, ``astro-photometry'' \citep{Williams2013}.
\end{enumerate}

All of this shows that determining the phase curve of an asteroid, achieving highly accurate (and calibrated) photometry \citep{Buchheim2010}, whilst including the asteroid's rotational light curve and covering a wide range of phase angles, poses major difficulties for astronomers and surveys that primarily are focused on astrometry.

Moreover, the $H$-$G$ model itself has some intrinsic limitations. The so-called opposition effect and the light back-scattering in general, especially for high- and low-albedo asteroids \citep{Harris1989,Harris1989b,Shevchenko1997,Belskaya2000}, is not well considered by the model \citep{Muinonen2002,Muinonen2010}. The \ac{IAU} officially adopted the updated three-parameter $H, G_1, G_2$ model by \citet{Muinonen2010} in 2012. With an updated fitting algorithm and a two-parameter version ($H, G_{12}^*$) by \citet{Penttila2016}, the $H$-$G$ model by \citet{Bowell1989} is theoretically replaced by now. But still, as most phase curves are too inaccurate to determine even one of the $G$ values, the $H$-$G$ model is still the most frequently used model when dealing with astro-photometry only datasets.

Despite all these uncertainties, the large astro-photometric datasets available at the \ac{MPC} offer a lot of opportunities to understand the (physical) properties for a majority of asteroids as otherwise dedicated photometry is only available for a limited number of asteroids. The absolute magnitude and the slope parameter correlate with physical characteristics of the asteroid like the diameter, reflectivity and properties of its surface \citep{Oszkiewicz2011}. Most prominent example is the already mentioned determination of the object's size based on the $H$ magnitude combined with an estimate on the visual geometric albedo \citep{Harris2002}. The size-frequency distribution of asteroids is one of the most important relations in planetary and solar system science and informs our understanding of the development of the Solar System and the population of \acp{NEO} \citep{Harris2021}. Also, this affects the impact flux of small solar system bodies with Earth and the determination of size and impact effects of potential (imminent) impactors \citep{Nesvorny2024,Nesvorny2024-2}. Therefore, the topic is highly relevant for planetary defense activities.

Systematic biases in astro-photometry were first found by \citet{Juric2002} with significant errors in the \ac{MPC}'s absolute magnitudes, which cause high uncertainties in the size derivation, showing the need to deal with this topic.
In order to achieve reliable results, the systematic biases and errors in astro-photometry need to be understood, so observations can ideally be corrected to debias them and improve their accuracy. This is the aim of the present paper. 

Large surveys like the \ac{Pan-STARRS} \citep{Kaiser2010}, \ac{ATLAS} \citep{Tonry2018} and the \ac{ZTF} \citep{Bellm2019} were used to derive phase curves from asteroids with their own astro-photometry, trying to infer the slope parameter in order to classify taxonomic types and to correlate with the albedo \citep{Veres2015,Mahlke2021,Carry2024}. 

\citet{Colazo2021} used measurements of asteroids also from Gaia in the Gaia DR2 data-release to determine the $H$-$G$ phase function parameters with the instrument's $g$ magnitude for nearly 14\,000 objects. As the study also included ground-based observations in order to have near-opposition phase angle data, different photometric systems needed to be used and converted. 
When combining multiple color-band data, a conversion to the V-band with color corrections is typically used. Inaccurate color corrections might introduce systematic errors.

In other studies, like in \citet{Oszkiewicz2011}, multiple observatories were used (using V- and R-band observations), which were calibrated with accurate broad-band photometry from the \ac{SDSS}. It turned out that the photometric calibration for almost all 11 analyzed observatories varies over time. Also, residuals deviate largely for objects at both the bright and faint end of the observed magnitude. This shows that several parameters affect photometry and need to be considered in the development of a correction algorithm.

Statistical analyses of astrometric residuals based on different parameters, such as time, apparent magnitude and rate of motion have been carried out for multiple stations \citep{Veres2017,Stronati2023}. The results from these analyses were also used as basis of a new weighting scheme for orbit determination by the \ac{MPC} \citep{Veres2017}. Besides observatory-specific uncertainties, also (astrometric) reference catalogs have been investigated for biases \citep{Farnocchia2015,Eggl2020}. Corrections for position and proper motion errors in 26 astrometric catalogs led to significant improvements in the astrometric residuals and the ephemeris prediction. We propose to perform a similar assessment on photometric residuals.

The \ac{MPC} and others demonstrated the need to establish observatory-dependent corrections to the reported apparent magnitudes from observers \citep{Juric2002,Veres2015}, but these methods have not yet been implemented to correct photometry from all of the more than 2500 observatories registered with the \ac{MPC}. We are now going to quantify the required observatory-dependent biases in astro-photometry and implement a correction algorithm of these biases in order to reduce statistical uncertainties and systematical errors in the $H$, $G$ parameter fitting.

\section{Methods}
Observational data submitted to the \ac{MPC} by observers throughout the world is used for the analysis. We use all observations with valid photometric measurements, i.e. apparent magnitudes in a specified color band. These values are first pre-processed, stored and then analyzed in comparison to an accurate photometric reference source containing 468 asteroids. With that we are able to compute the photometric residuals of all observations from different observatory stations. The residuals undergo a statistical analysis to determine possible significant deviations and uncertainties. The results for the systematic biases will then be implemented in a correction algorithm.

\subsection{Data sources and pre-processing pipeline}\label{sec:methods.pre}
\begin{figure}[htb]
    \centering
    \includegraphics[width=\linewidth, trim=0mm 0mm 0mm 2mm,clip]{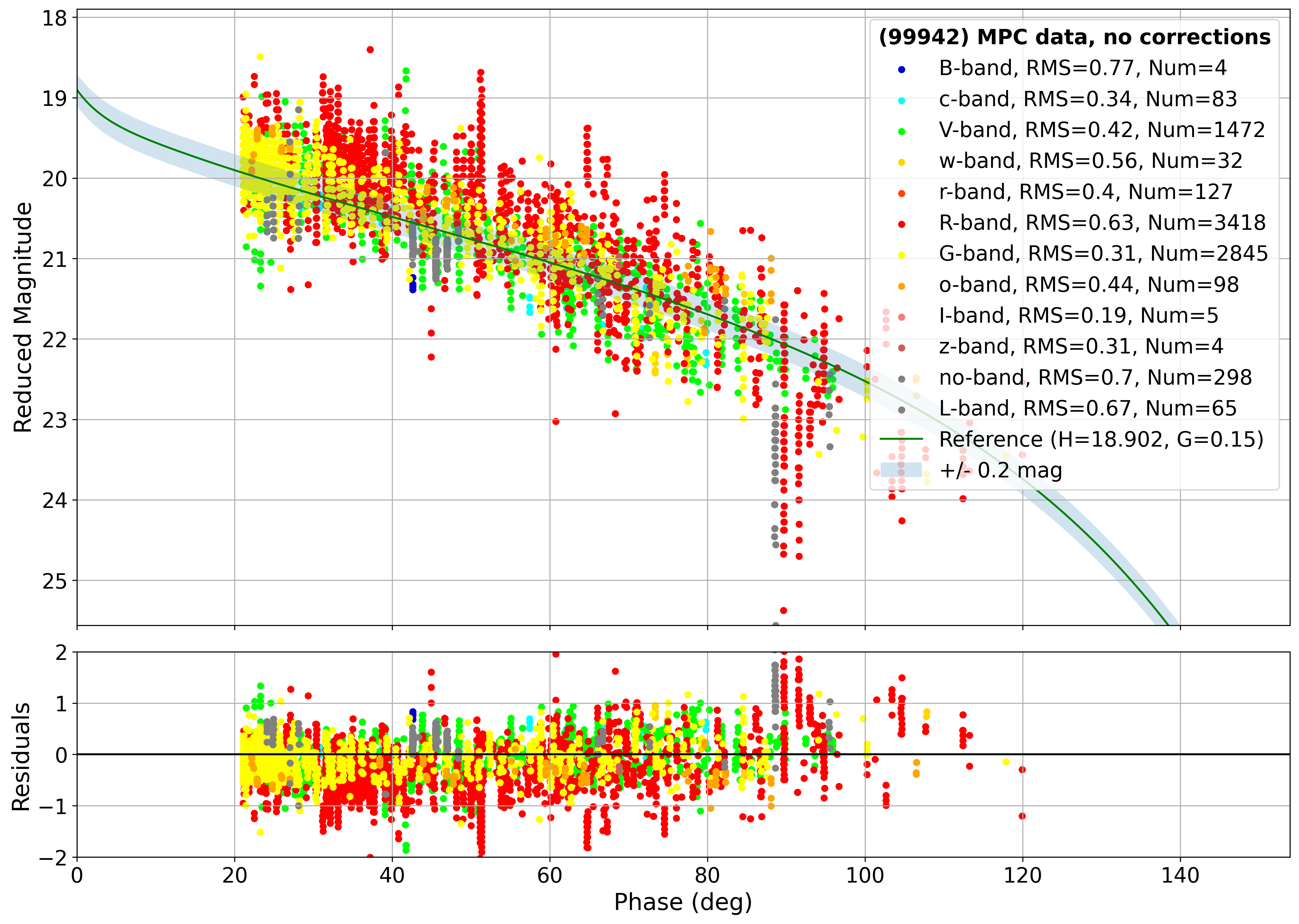} 
    \caption{$H$-$G$ phase function for asteroid (99942) Apophis ($H=18.945$, $G=0.15$; from NEOCC) with all available observations at \ac{MPC} in different color-bands, including the residuals below.}
    \label{fig:methods.phasefunction}
\end{figure}

\begin{figure*}[t]
    \centering
    \includegraphics[width=0.7\linewidth]{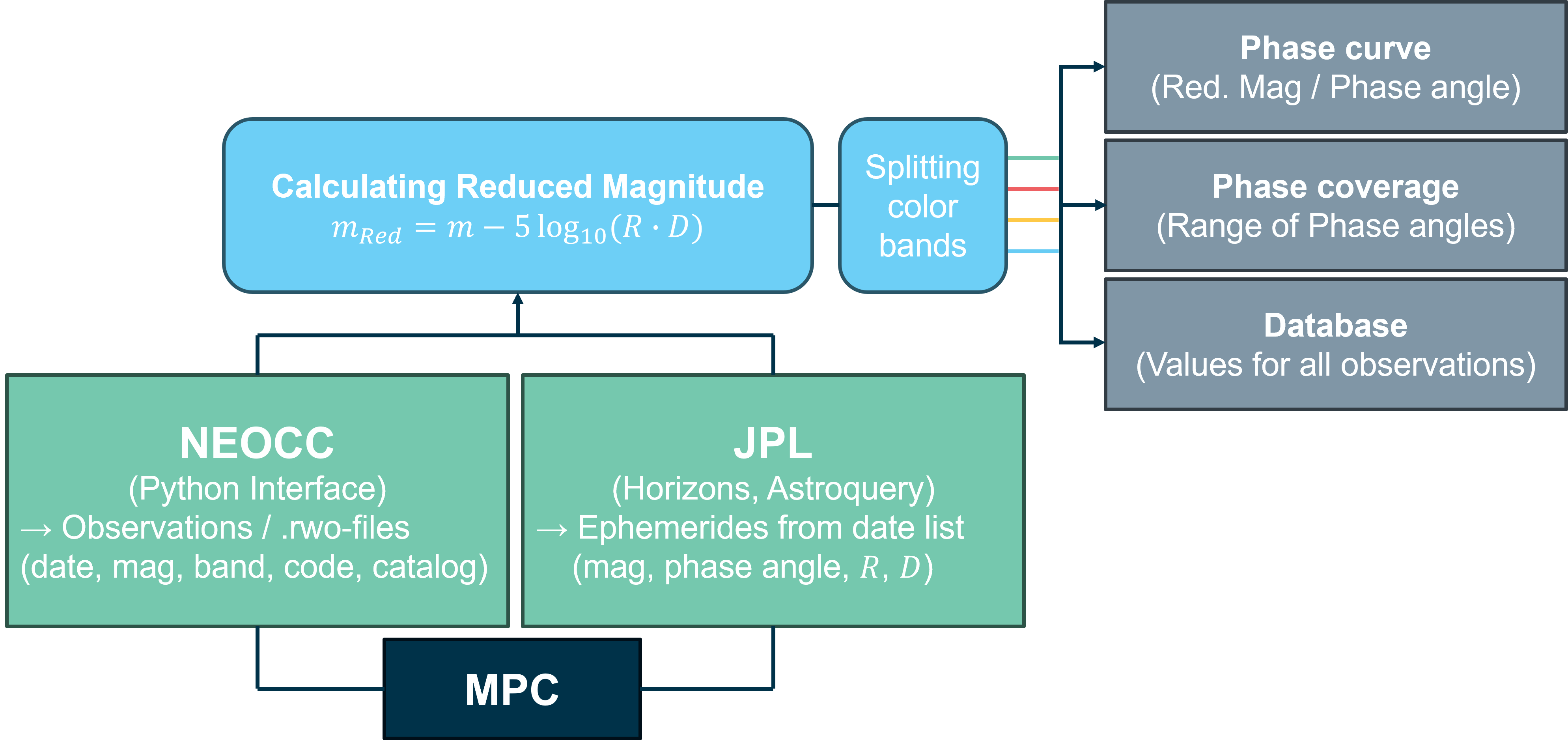} 
    \caption{Schematic representation of the photometric observation pre-processing pipeline with the data access from \ac{MPC} via \ac{NEOCC} and \ac{JPL}.}
    \label{fig:methods.pre.1}
\end{figure*}

The photometric observation pre-processing pipeline is schematically shown in Fig.~\ref{fig:methods.pre.1}.
Observational data is accessed via the \textit{\ac{NEOCC} Python interface}\footnote{\url{https://github.com/D-arioSpace/astroquery}}. 
For a given asteroid, all the observations with photometric measurements, which include the apparent visual magnitude $V_{\text{mag}}$ with the corresponding color bands and the meta-data (date, station code, catalog used), are downloaded and processed. On the other hand, orbital information from \textit{\acs{JPL} Horizons}\footnote{\url{https://ssd.jpl.nasa.gov/horizons/}} is used to obtain the asteroid's phase angle $\alpha$ and the heliocentric and geocentric distances, $r$ and $\Delta$ (in au), at the observation times. With that, the reduced magnitude $m_{red}$ is computed by 
\begin{align}
    m_{red} = V_{\text{mag}}-5\log_{10}(r\cdot \Delta)
\end{align}
in order to normalize the effects of distances on the brightness for asteroids \citep{Williams2013}.
The reduced magnitude is then used to visualize the asteroid's phase curve ($m_{red}$ over $\alpha$), separated for the different color bands that are used in the observations (cf. Fig.~\ref{fig:methods.phasefunction}). Additionally, plots of the range of phase angles covered in the observations can be generated for the asteroid, alongside with a data file that summarizes all the obtained values in the pre-processing pipeline. 

\subsection{Data analysis system}\label{sec:methods.analysis}
For a given dataset of accurately photometrically validated asteroids the observations from all observatory sites are statistically analyzed for biases. Therefore, we inspect the photometric residuals of the observations 
\begin{align}\label{eq:2.2-resid}
    \delta m_{red} = {[m_{red}]}_{H\text{-}G} - {[m_{red}]}_{obs.},
\end{align}
which are determined by the difference of the reduced observed magnitude ${[m_{red}]}_{obs.}$, derived by the pre-processing pipeline in Sec.~\ref{sec:methods.pre}, to the model value of the $H$-$G$ phase function ${[m_{red}]}_{H\text{-}G}$ defined by \citet{Bowell1989}:
\begin{align}\label{eq:phasecurve}
    m_{red}(\alpha;H,G) = H - 2.5\log_{10}\left((1-G)\phi_1(\alpha)+G\phi_2(\alpha)\right),
\end{align}
where for the functions $\phi_i$ with $i=1,2$ are defined by:
\begin{align}
    \phi_i(\alpha) &= W(\alpha)\phi_{iS}(\alpha)+\left(1-W(\alpha)\right)\phi_{iL}(\alpha),\\
    W(\alpha) &= \exp\left(-90.56\tan^2\frac{\alpha}{2}\right),\\
    \phi_{iS}(\alpha) &= 1-\frac{C_i \sin\alpha}{0.119+1.341\sin\alpha-0.754\sin^2\alpha},\\
    \phi_{iL}(\alpha) &= \exp\left(-A_i\left(\tan\frac{\alpha}{2}\right)^{B_i}\right), 
\end{align}
with parameters: $A_1=3.332$, $A_2=1.862$, $B_1=0.631$, $B_2=1.218$, $C_1=0.986$, $C_2=0.238$\ .

The fundamental parameters for this magnitude system, the absolute magnitude $H$ and slope parameter $G$, are extracted from the prepublished database from the \textit{Ondrejov Asteroid Photometry Project} updated as of 2023-08-11\footnote{\url{https://www.asu.cas.cz/~ppravec/newres.htm}} by \citet{Pravec2012}. This database contains results from asteroid rotational and binary light curve measurements, among others the $H$-$G$ parameters, for more than 1000 asteroids. The methodology of the measurements and their analysis is described by \citet{Pravec1996,Pravec1997}. Their derived asteroid absolute magnitudes $H$ correspond to the mean light (effectively an average over asteroid's rotation) in the Johnson V photometric system, calibrated with the \citet{Landolt1992} standard stars. Using calibrated measurements that consider the asteroids' light curves minimizes photometric errors in the $H$-$G$ parameters. Despite having a comparably low quantity of measurements, we consider this method to be effective in order to avoid intrinsic biases.

\begin{figure*}[t]
    \centering
    \includegraphics[width=0.8\linewidth]{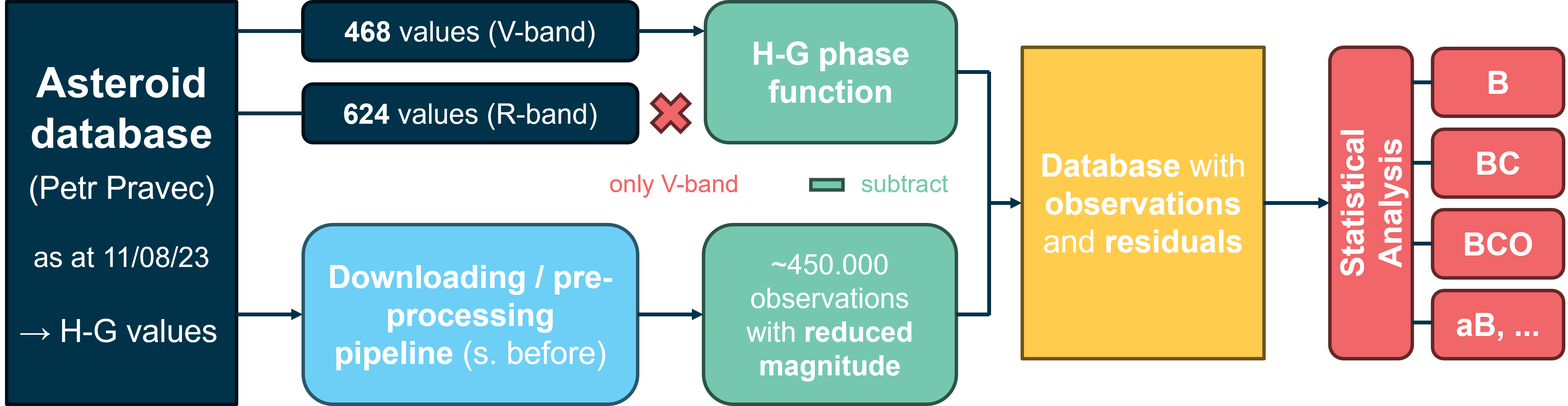} 
    \caption{Schematic representation of the bias analysis system with the implementation of the pre-processing pipeline from Fig.~\ref{fig:methods.pre.1}.}
    \label{fig:methods.ana.1}
\end{figure*}

The data analysis system as displayed in Fig.~\ref{fig:methods.ana.1} uses all asteroids in the database with $H$-$G$ magnitude system parameters in the V-band as a standard reference. In case of multiple measurements for the same asteroid the newer updated parameters are used. With that, we get a total of 468 different asteroids, including 327 \acp{NEO}, with their reference $H$-$G$ model parameters. For all these asteroids, all available observations from the \ac{MPC} are downloaded and pre-processed. The number of observations per object ranges from 51 up to 7059, with a median of 532 (cf. distribution in Fig.~\ref{fig:methods.numberobs} in the appendix). Each object has observations in at least 3 different color bands (maximum 19), with an median of 10 bands available.
Using the $H$ and $G$ parameter from \citet{Pravec2012} and the corresponding phase angles of the observations we can compute the $H$-$G$ phase function values. Comparing those values to the more than 450\,000 reduced magnitude measurements from observations, we get a database with observational residuals from all different observatories (up to 2023-08-11). This database not only includes the residuals, but also raw data (e.g., apparent magnitude measurement) and meta-data (e.g., color band, star catalog, observatory station, time).

Based on the meta-data for the residuals, we analyze the database with different criteria in order to obtain systematic biases for various observational parameters:
\begin{itemize}
    \item \textbf{Band (B) analysis:} Observations with the same color band are processed together and a statistical analysis is performed. This includes an evaluation of the distribution of the residuals with the average bias value $\overline{\delta_m}$ and its standard deviation $\sigma_{\delta_m}$ in all different color bands.
    \item \textbf{Band and Catalog (BC) analysis:} Observations with the same combination of color band and star catalog used are binned together. All combinations are then statically analyzed (like in the B-analysis) and the results for the same color band are compared.
    \item \textbf{Band, Catalog and Observatory (BCO) analysis:} Observations with the same combination of color band, star catalog used and observatory station (\ac{MPC} Code) are categorized. All combinations are then individually analyzed (like in the BC-analysis) and the results for the same color band are compared.
\end{itemize}

With this mechanism of subsequent grouping of the data, we can analyze the residual dataset and derive biases for different criteria, without neglecting the influence of previous dependencies. 

There might be an additional influence of the spectra of the individual asteroids, which would lead to deviating results. A high number of observations for these objects would dominate in the data sample and introduce a bias. Therefore, the statistics should be independent of the varying number of observations used from each asteroid (cf. Fig~\ref{fig:methods.numberobs}). For this reason, multiple observations of the same object in the same group (B, BC or BCO) should be summarized into one dataset by averaging the values. This can be achieved with the existing method by using the asteroid name as an additional grouping parameter and performing the analysis as for the previous cases. These additional \underline{\textbf{a}}steroid-dependent analyses are correspondingly named \textit{aB}, \textit{aBC} and \textit{\acsu{aBCO}} analysis. Before exporting specific biases and their errors, an average across all the different asteroids within the B, BC or BCO group is calculated. With this kind of analysis effectively all asteroids used in one group are getting weighted equally, thus minimizing the effect of intrinsic photometric properties of individual asteroids, especially in cases with a strongly deviating amount of observations per asteroid. The data analysis system hereby provides a statement about the bias for the average asteroid.

\subsection{Astro-photometric correction}

\subsubsection{Current correction systems}

\begin{table*}[ht]
\caption[]{Current astro-photometric correction offsets (in mag) to a reference V-band used by \ac{NEOCC} (before 2023-09-28)\footnotemark, \ac{MPC} and \ac{JPL} in their orbital determination systems \citep{Veres2018}. Values are sorted by the color band's effective wavelength $\lambda_{\text{eff}}$ and its bandwidth $\Delta\lambda_{\text{eff}}$.}
\centering
\begin{tabular}{clcclccc}\label{tab:methods.existing-corrections}
\textbf{Band} & \textbf{Note} & $\lambda_{\text{eff}}$ / nm & $\Delta\lambda_{\text{eff}}$ / nm & \textbf{Reference} & \textbf{\ac{NEOCC}} & \textbf{ \ac{MPC} } & \textbf{ \ac{JPL}  } \\ \hline
U & Johnson-Cousins & 366.3 & 65.0 & \citet{Bessell2005} & -1.30 & -1.30 & -1.30 \\ 
B & Johnson-Cousins & 436.1 & 89.0 & \citet{Bessell2005} & -0.80 & -0.80 & -0.80 \\ 
g & Sloan & 463.9 & 128.0 & \citet{Bessell2005} & -0.28 & -0.35 & -0.28 \\ 
c & \ac{ATLAS} cyan & 535.0 & 230.0 & \citet{Tonry2018} & -0.05 & -0.05 & -0.05 \\ 
V & Johnson-Cousins & 544.8 & 84.0 & \citet{Bessell2005} & 0.00 & 0.00 & 0.00 \\ 
v & equal to V & 544.8 & 84.0 & - & 0.00 & 0.00 & 0.00 \\ 
w & \ac{Pan-STARRS} & 608.0 & 382.0 & \citet{Tonry2012} & 0.16 & -0.13 & 0.16 \\ 
r & Sloan & 612.2 & 115.0 & \citet{Bessell2005} & 0.23 & 0.14 & 0.23 \\ 
R & Johnson-Cousins & 640.7 & 158.0 & \citet{Bessell2005} & 0.40 & 0.40 & 0.40 \\ 
G & Gaia Broadband & 673.0 & 440.0 & \citet{Jordi2010} & 0.28 & 0.28 & 0.24 \\ 
o & \ac{ATLAS} orange & 690.0 & 260.0 & \citet{Tonry2018} & 0.33 & 0.33 & 0.33 \\ 
i & Sloan & 743.9 & 123.0 & \citet{Bessell2005} & 0.39 & 0.32 & 0.39 \\ 
I & Johnson-Cousins & 798.0 & 154.0 & \citet{Bessell2005} & 0.80 & 0.80 & 0.80 \\ 
z & Sloan & 889.6 & 107.0 & \citet{Bessell2005} & 0.37 & 0.26 & 0.37 \\ 
y & \ac{Pan-STARRS} & 962.0 & 83.0 & \citet{Tonry2012} & 0.36 & 0.32 & 0.36 \\ 
Y & 1.035 micron band & 1020.0 & 120.0 & \citet{Binney1998} & 0.70 & 0.70 & 0.70 \\ 
J & 1.275 micron band & 1220.0 & 213.0 & \citet{Binney1998} & 1.20 & 1.20 & 1.20 \\ 
H & 1.662 micron band & 1630.0 & 307.0 & \citet{Binney1998} & 1.40 & 1.40 & 1.40 \\ 
K & 2.159 micron band & 2190.0 & 390.0 & \citet{Binney1998} & 1.70 & 1.70 & 1.70 \\ 
- & no band specified & - & - & - & -0.80 & -0.80 & -0.80 \\ 
u & unknown & - & - & [1] & -2.50 & 2.50 & 2.50 \\ 
L & unknown & - & - & [1] & 0.20 & 0.20 & 0.20 \\ 
C & ``clear'' formerly & - & - & [1],[2] & 0.40 & 0.40 & 0.40 \\ 
W & ``wide'' formerly & - & - & [1] & 0.40 & 0.40 & 0.40 \\ \hline
\end{tabular}\\
\raggedright
\hspace{8.5mm}\footnotesize{[1]\url{https://www.minorplanetcenter.net/iau/info/ADESFieldValues.html}}\\ 
\hspace{8.5mm}\footnotesize{[2]\url{https://www.minorplanetcenter.net/iau/info/OpticalObs.html}} 
\end{table*}

There are corrections in place at the \ac{MPC} to correct for photometric measurements calibrated in different bands. These standard corrections are used to convert the magnitude observed in different bands to a reference V-band. These correction offsets are necessary as the flux of the reflected sunlight on the asteroid's surface varies in the different wavelengths relative to the Sun. Default correction for the average asteroid taxonomic type can be computed by using photometric measurements \citep[cf.][]{Williams2013}. The current astro-photometric correction used by \ac{NEOCC}\footnotetext{\label{fn:1}Photometry correction used by NEOCC have been updated on 2023-09-28 based on the results of this study (\url{https://neo.ssa.esa.int/system-status}).}\footnotemark[4], \ac{MPC} and \ac{JPL} are listed in Tab.~\ref{tab:methods.existing-corrections}.

\subsubsection{Statistical correction algorithm}\label{sec:methods.correction}
We are using the results from the bias analysis system in Sec.~\ref{sec:methods.analysis} to develop an iterative algorithm that corrects the specific systematic biases in the apparent magnitude measurements for asteroids and especially \acp{NEO}, as well. For this, the average bias values $\overline{\delta_m}$ obtained by the \textit{aB}, \textit{aBC} and \textit{aBCO} analyses are used as an offset for the correction, which needs to be added on the observational apparent magnitude (cf. definition of the residuals $\delta m_{red}$ in Eq.~\eqref{eq:2.2-resid}).

In this study, we are applying many different corrections to astro-photometric observations, when certain criteria (e.g., band, catalog, observatory code) are fulfilled, not only the band correction. We propose a dictionary-based data format, where the keys indicate the parameter and the items contain the corresponding condition, when to apply the correction. A correction rule looks like the following: 

\begin{lstlisting}
    {'code': 'T08', 'band': 'o', 'type': None,
    'catalog': 'V', 'start': None, 'stop': None,
    'd': 0.326134}
\end{lstlisting}

where `code' indicates the \ac{MPC} observatory code (columns 78-80 in the \ac{MPC} 80-column format)\footnote{\url{https://www.minorplanetcenter.net/iau/lists/ObsCodes.html}}, `band' the magnitude band (column 71, cf. Tab.~\ref{tab:methods.existing-corrections}), `type' indicates how the observation was made (column 15)\footnote{\url{https://www.minorplanetcenter.net/iau/info/OpticalObs.html}}, `catalog' the one letter astrometric reference star catalog code (column 72)\footnote{\url{https://www.minorplanetcenter.net/iau/info/CatalogueCodes.html}}, `start' and `end' the boundaries for the date of observation in a ``YYYY MM DD.dddddd'' format and `d' the correction in magnitudes to be added on the observational apparent magnitude to convert to V-band.

\begin{figure}[htb]
    \centering
    \includegraphics[width=\linewidth]{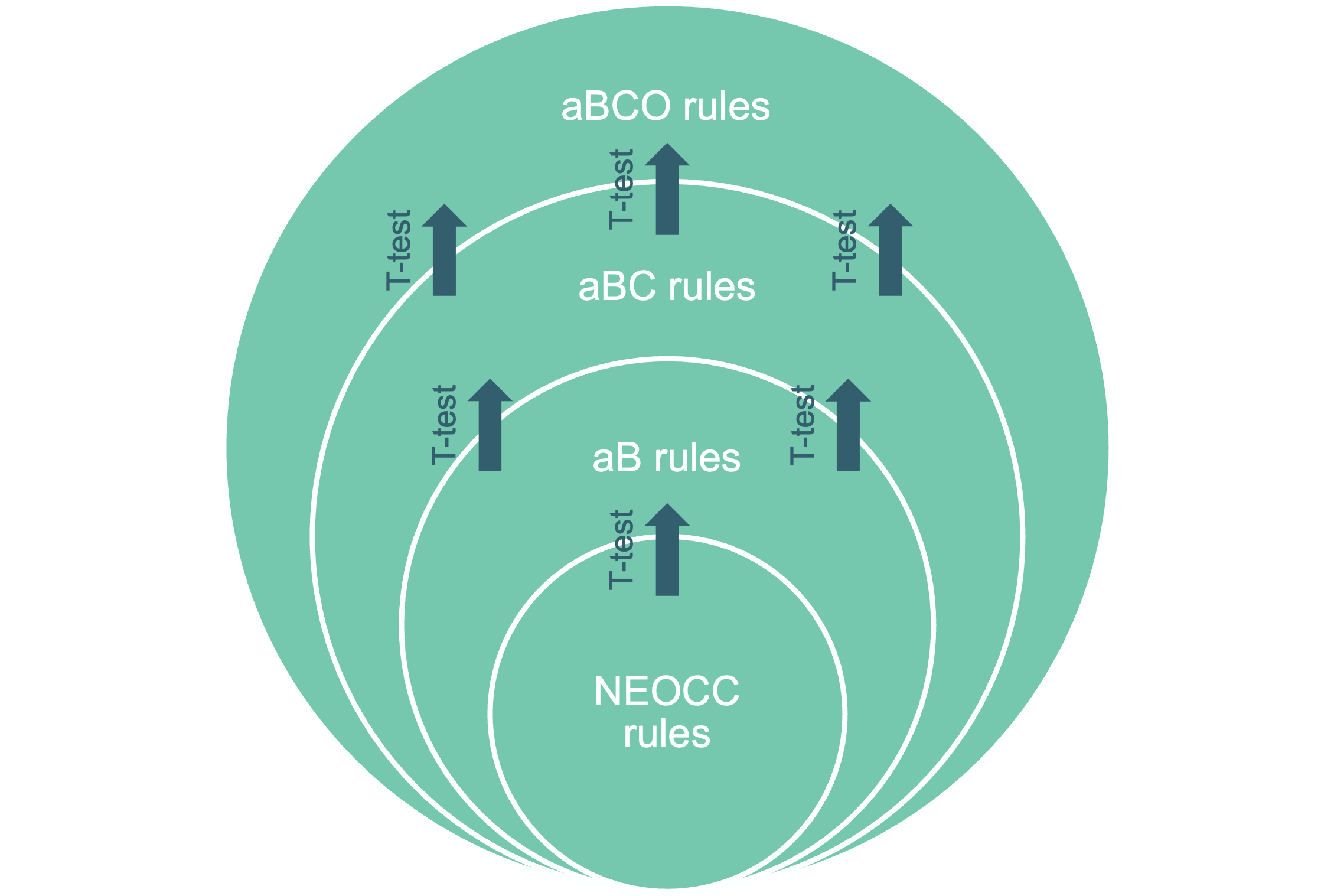} 
    \caption{Schematic representation of the correction system with NEOCC correction rules as reference and expanding successively with asteroid-weighted band (\textit{aB}), band-catalog (\textit{aBC}) and band-catalog-observatory (\textit{aBCO}) corrections if the t-test is fulfilled accordingly.}
    \label{fig:methods.corr.1}
\end{figure}

The corrections will be applied in an iterative approach, such that more specific corrections will overwrite general ones by appending them later to the list of corrections. This is shown in Fig.~\ref{fig:methods.corr.1}: We start with existing correction values from NEOCC (cf. Tab~\ref{tab:methods.existing-corrections}), appending the \textit{aB} and then the \textit{aBC} and \textit{aBCO} correction rules, but only in the case they are statistically significant. In other words, the correction $(\overline{\delta_m})_{ij}$, which is to be added, significantly deviates from the next more general correction $(\overline{\delta_m})_{i}$, i.e. an \textit{aBC} correction from the \textit{aB} correction with the same band. For \textit{aB} correction we compare with the corresponding NEOCC correction with the same band. The significance is checked with the statistical \textit{t-test}:

\begin{align}
    t = \sqrt{n}\cdot\frac{(\overline{\delta_m})_{ij}-(\overline{\delta_m})_{i}}{\sigma_{\delta_m}},
\end{align}

with the number $n$ of observations residuals $\delta_m$ in the new correction. A two-tailed test is then performed with the cumulative distribution function of the Student's t-distribution $F(t;\nu)$, with $\nu=n-1$ the degree of freedom, by checking if the two-tailed probability $p=2F(t;\nu)$ is larger than or equal to a given significance level $\alpha$ (null hypothesis $H_0$: $p\geq \alpha$. If this is not the case, the new correction is deviating significantly (rejection of the null hypothesis $H_1$: $p<\alpha$) and therefore gets appended to the list of correction rules. By repeating this algorithm for all possible corrections step-wise, the final correction system is defined, with all corrections having a confidence $\gamma=1-\alpha$ larger than a given level.

Currently, the presented statistical correction algorithm is not using all possible parameters, such as `type', the observation date or other additional ones. These could be added in further studies.

\subsubsection{Light curve effects}\label{sec:methods.lightcurve}
Since most astrometric observers only make a few observations of an object during the night, with rough estimated astro-photometry, the observational coverage does not account for light curve effects that are due to rotation and the shape of the asteroid, which typically happens on a period longer than the duration of the observation for main-belt asteroids -- for \acp{NEO} rotation periods can significantly shorter, with some already observed as brief as $2.5$ seconds \citep{Devogele_2024,Thirouin_2016}. Overall, these effects create a periodic variation in brightness, thus introducing natural deviations from the $H$-$G$ system phase curve. 

Analyzing this effect for the 468 asteroids used in the analysis from Sec.~\ref{sec:methods.analysis}, we get an average amplitude of $0.4042$ mag. This is also representative for other asteroids: The \ac{LCDB} by \citet{Warner2009}, containing 32\,455 asteroids with amplitude information, gives us an average amplitude of $0.4008$ mag. 

Assuming the variation of light curve has a sinusoidal shape, the expected RMS of the observations due to the rotational effects is given by $1/2\sqrt{2}$ times the peak-to-trough amplitude, which leads us to an average effect on the RMS for asteroids of $(0.1417 - 0.1429)$ mag. Including the typical numerical precision of astro-photometrical observations of $0.10$ mag, a minimum RMS of roughly $0.25$ mag is expected as a physical boundary, if we do not consider and compensate for light curve effects.\footnote{Also other factors influence the errors in the $H$ and $G$ parameters, e.g. the phase angle coverage, leading to a further increased expected RMS.} As shown in other studies, this value is also assumed to be a typical error for the $H$ parameter \citep{Mainzer2012}.

\section{Results}
\subsection{Bias analysis}\label{sec:res.bias}
We use the methods from Sec.~\ref{sec:methods.analysis} to analyze the distribution of specific biases in color bands (\ref{sec:res.bias.band}), catalogs (\ref{sec:res.bias.catalog}) and observatories (\ref{sec:res.bias.obs}). From the \textit{Ondrejov
Asteroid Photometry Project} database 394 out of the 468 asteroids ($\approx84\%$) with V-band measurements for the \textit{H} and \textit{G} parameter were used for the analysis. These are selected by having an export date of the measurements earlier than 2021-10-21. The subset of the remaining 74 asteroids with more recent measurements will be used for the validation of the correction system (Sec.~\ref{sec:res.pred}).

\subsubsection{Band bias results}\label{sec:res.bias.band}
\begin{figure}[ht]
    \centering
    \includegraphics[width=0.95\linewidth,trim=0mm 1mm 0mm 0mm,clip]{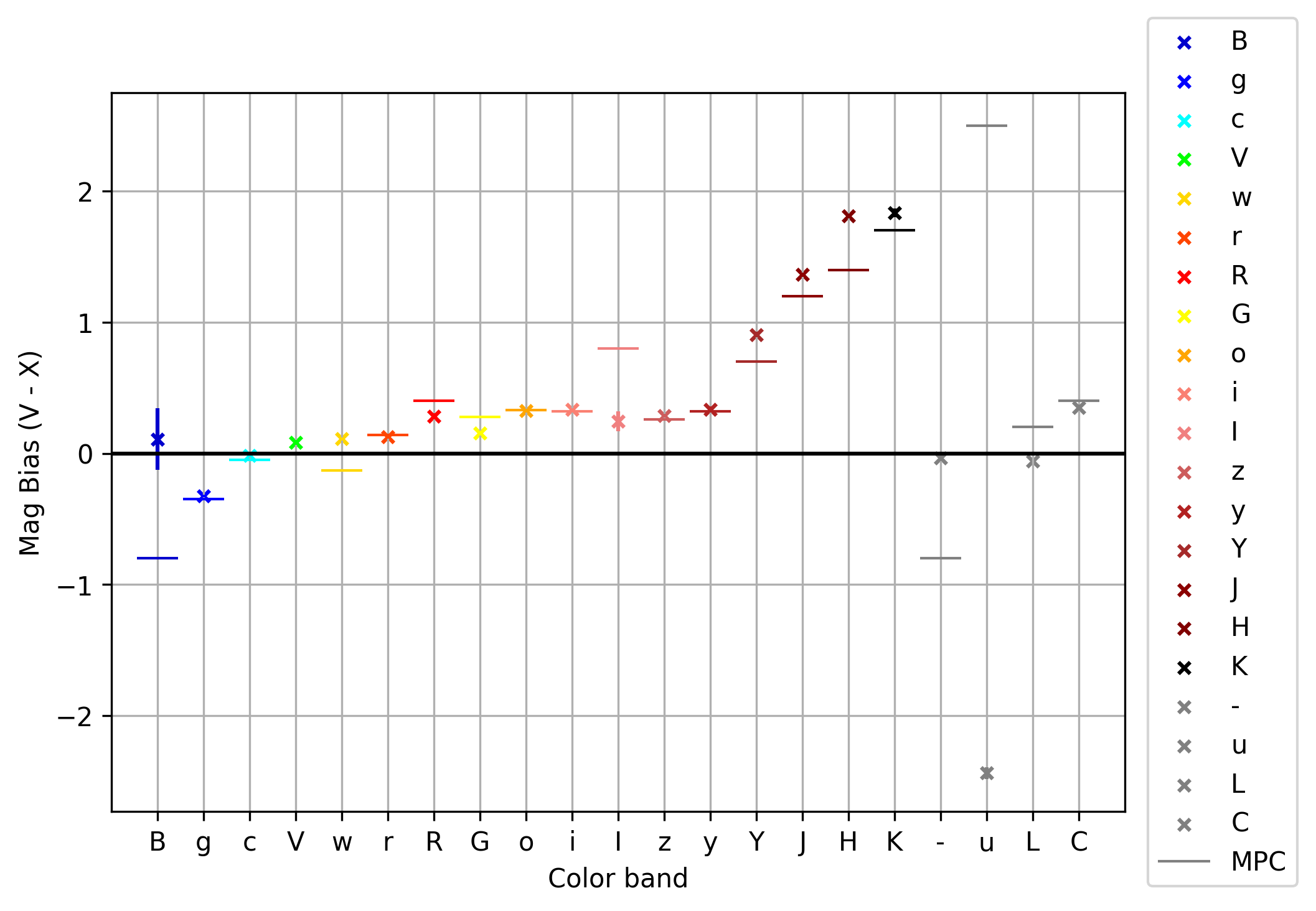}
    \vspace{-3mm}
    \caption{Graphic representation of the average color band bias results (crosses) from \textit{aB} analysis in Tab.~\ref{tab:res.aB} in comparison to the current \ac{MPC} band correction (bars).}
    \label{fig:res.aB.1}
\end{figure}

In Tab.~\ref{tab:res.aB} the average color band bias results from the \textit{aB} analysis and the number of observations and asteroids used can be seen. In contrast, current \ac{MPC} correction values are also shown for comparison. A visual representation of the color band biases can be seen in Fig.~\ref{fig:res.aB.1}.

\begin{table}[htb]
    \caption{Results from \textit{aB} bias analysis with data from 394 out of 468 asteroids, with the average bias and its standard error in the different color bands compared to the current \ac{MPC} correction from Tab.~\ref{tab:methods.existing-corrections}, and with the number of observations and different asteroids available for the analysis.}
    \centering 
    \begin{tabular}{crrrr}\label{tab:res.aB}
         \textbf{Band} &  \begin{tabular}[c]{@{}c@{}}\textbf{Num.}\\ \textbf{Obs.}\end{tabular}&\begin{tabular}[c]{@{}c@{}}\textbf{Num.}\\ \textbf{Ast.}\end{tabular}&  \begin{tabular}[c]{@{}c@{}}\textbf{Mean Bias}\\ / mag\end{tabular}&\begin{tabular}[c]{@{}c@{}}\textbf{\ac{MPC}}\\ / mag \end{tabular} \\ \hline
                  B&  227&  2&  $0.11 \pm  0.23$& -0.80\\
                  g&  4\,600&  233&  $-0.325 \pm  0.010$& -0.35
\\
                  c&  18\,980&  297&  $-0.017 \pm  0.012$
& -0.05
\\
                  V&  63\,520&  391&  $0.085 \pm  0.020$& 0.00
\\
                  w&  25\,582&  371&  $0.111 \pm  0.012$
& -0.13
\\
                  r&  20\,661&  295&  $0.126 \pm  0.014$
& 0.14
\\
                  R&  52\,887&  392&  $0.282 \pm  0.023$
& 0.40
\\
                  G&  102\,697&  376&  $0.154 \pm  0.018$
& 0.28
\\
                  o&  55\,716&  342&  $0.325 \pm  0.013$
& 0.33
\\
                  i&  5\,178&  301&  $0.334 \pm  0.010$& 0.32
\\
                  I&  137&  17&  $0.246 \pm  0.077$
& 0.80
\\
                  z&  453&  95&  $0.287 \pm  0.016$
& 0.26
\\
                  y&  202&  49&  $0.336 \pm  0.024$
& 0.32
\\
                  Y&  36&  15&  $0.906 \pm  0.021$
& 0.70
\\
                  J&  69&  32&  $1.362 \pm  0.018$
& 1.20
\\
                  H&  20&  10&  $1.810 \pm  0.017$& 1.40
\\
                  K&  39&  18&  $1.835 \pm  0.034$
& 1.70
\\
                  -&  50\,399&  322&  $-0.037 \pm  0.024$
& -0.80
\\
                  u&  20&  4&  $-2.436 \pm  0.045$
& 2.50
\\
                  C&  976&  113&  $0.351 \pm  0.023$& 0.40\\  \hline
    \end{tabular}
\end{table}

For some commonly used color bands, like V, G, R and r or the \ac{ATLAS} c- and o-bands and \ac{Pan-STARRS} w-band, a significant amount of observations (nearly more than 20\,000 each) was used in the analysis. These measurements also originate from a broad sample of asteroids (around more than 300 each), allowing more significant statistics. Also, there are more than 50\,000 observations that do not have any color band specified, the so-called `blank'-band. As the color bands were not provided in old, mostly photographic measurements, these `blank'-band are formally handled, so far, as if they were blue-sensitive B-band measurements, thus having a correction of $-0.80$ mag assigned (cf. Tab.~\ref{tab:methods.existing-corrections}). The average bias for the observations in this ``band'' in the present study, including also modern CCD or CMOS measurements, with $-0.037$ mag deviates noticeably: this is likely the consequence of recent users misinterpreting the absence of an explicit band indication as a way to submit unfiltered observations, in contrast with the formal interpretation of blank corresponding to B. There is also a noticeable discrepancy in the \ac{Pan-STARRS} w-band, with a computed correction of  $+0.111$ mag, significantly different to the $-0.13$ mag of the current MPC correction. Here, the analysis results rather match with the correction of $+0.16$ mag used by \ac{NEOCC} and \ac{JPL} (cf. Tab.~\ref{tab:methods.existing-corrections}).

Looking at the V-, G- and R-bands, we can see that the bias results deviate by more than $3\sigma$ (standard deviations) from current correction, but all with a comparably small difference in absolute terms of less than $0.15$ mag. Nevertheless, such biases can lead to relevant effects. Most striking here is the finding that the V-band is biased by around $0.085$ mag on average.

There are also deviations of more than $3\sigma$ for the blue B-band and the infrared I-, Y-, J-, H- and K-bands, but those have less than 1000 observations in the database of this study. Also, the `unknown' u-band has a large difference, better seen effectively as a sign change, from $+2.50$ mag (\ac{MPC}) to $-2.436$ mag (\textit{aB} analysis).

On the other hand, the \textit{aB} bias analysis matches well within $3\sigma$ for the Sloan g-, r- and i-bands, and especially the \ac{ATLAS} c- and o-bands, with absolute differences below $0.05$ mag. Those modern and well-defined photometric bands are typically used by observers that more accurately determine their photometry and thus tend to have lower biases in their measurements. This shows that, for bands where good accuracy in the starting dataset is expected, the debiasing methodology is indeed giving the expected results. Other bands, like the z-, y- and C-band are also giving results matching the current correction, but there are less than 1000 observations per color band, which makes these correction not as relevant in a statistical sense. 

\begin{figure}[htb]
    \centering
    \includegraphics[width=0.75\linewidth, trim=0mm 4mm 0mm 10mm,clip]{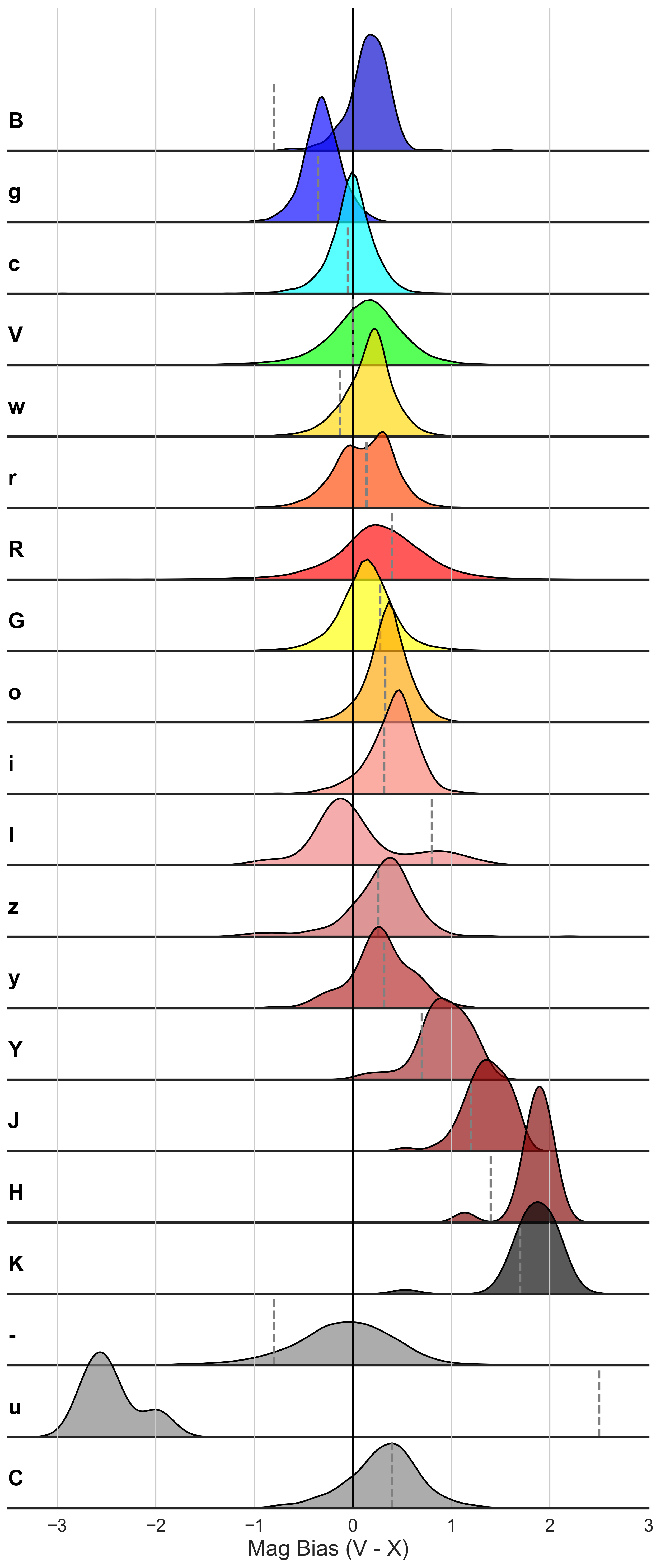} 
    \caption{Distribution of bias values in different bands from \textit{aB} analysis with current MPC correction in dashed lines.}
    \label{fig:res.aB.2}
\end{figure}

In Fig.~\ref{fig:res.aB.2} the obtained biases are also represented in a graphical way by showing the distribution of the bias values in each color band. Most of them have symmetrical peak-shapes, with some having narrow widths like g, c, w, o or i and others having broader widths like V, R, G or `blank', where the latter are the ones with noticeable deviations to the current correction, as described before. Most interesting are the color bands where there are asymmetric or double peaked shapes, like the r- and I-band. A bi-modality test by \citet{Hartigan1985, Hartigan1985b} and implemented by \textit{diptest}\footnote{\url{https://github.com/RUrlus/diptest/releases/tag/v0.8.1}} shows a $p$-value of the dip statistics of 0.0086 for r-band and 0.9822 for I-band, meaning r-band is most likely having a bi-modality. In both cases, one portion of the observations is having one specific bias, conspicuously around $0$ mag, and one portion is having another bias, around the value of the correction which would be expected. A bias of $0$ mag means, that there is no bias and no correction is needed. As there should be bias expected by the intrinsic color properties of the light reflected by the asteroid as derived by \citet{Williams2013}, it can be suspected that there are observers or measurers who provide the wrong color band in their reports, or even already correct their photometry to V with the correction values by themselves (while still reporting the original band in the MPC submission). Those would be corrected a second time by the \ac{MPC}, thus aggravating the results. This underlines the importance of investigating the biases of individual observatories.

\subsubsection{Catalog bias results}\label{sec:res.bias.catalog}
\begin{figure*}[tp]
    \centering
    \begin{subfigure}[b]{0.47\textwidth}
        \centering
        \includegraphics[width=\textwidth, trim=0mm 7mm 0mm 2mm,clip]{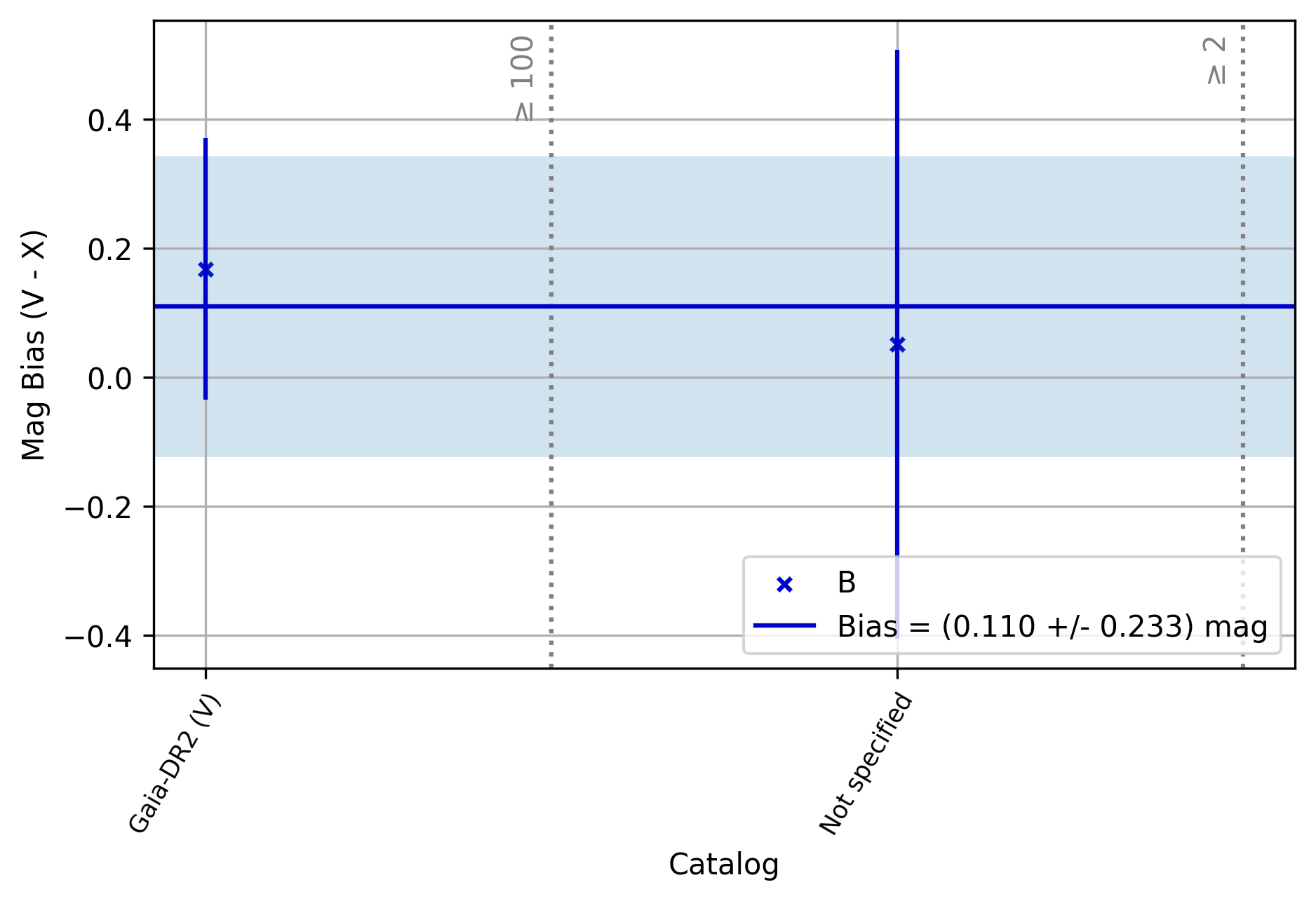}
    \end{subfigure}
    \hfill
    \begin{subfigure}[b]{0.47\textwidth}
        \centering
        \includegraphics[width=\textwidth, trim=0mm 7mm 0mm 2mm,clip]{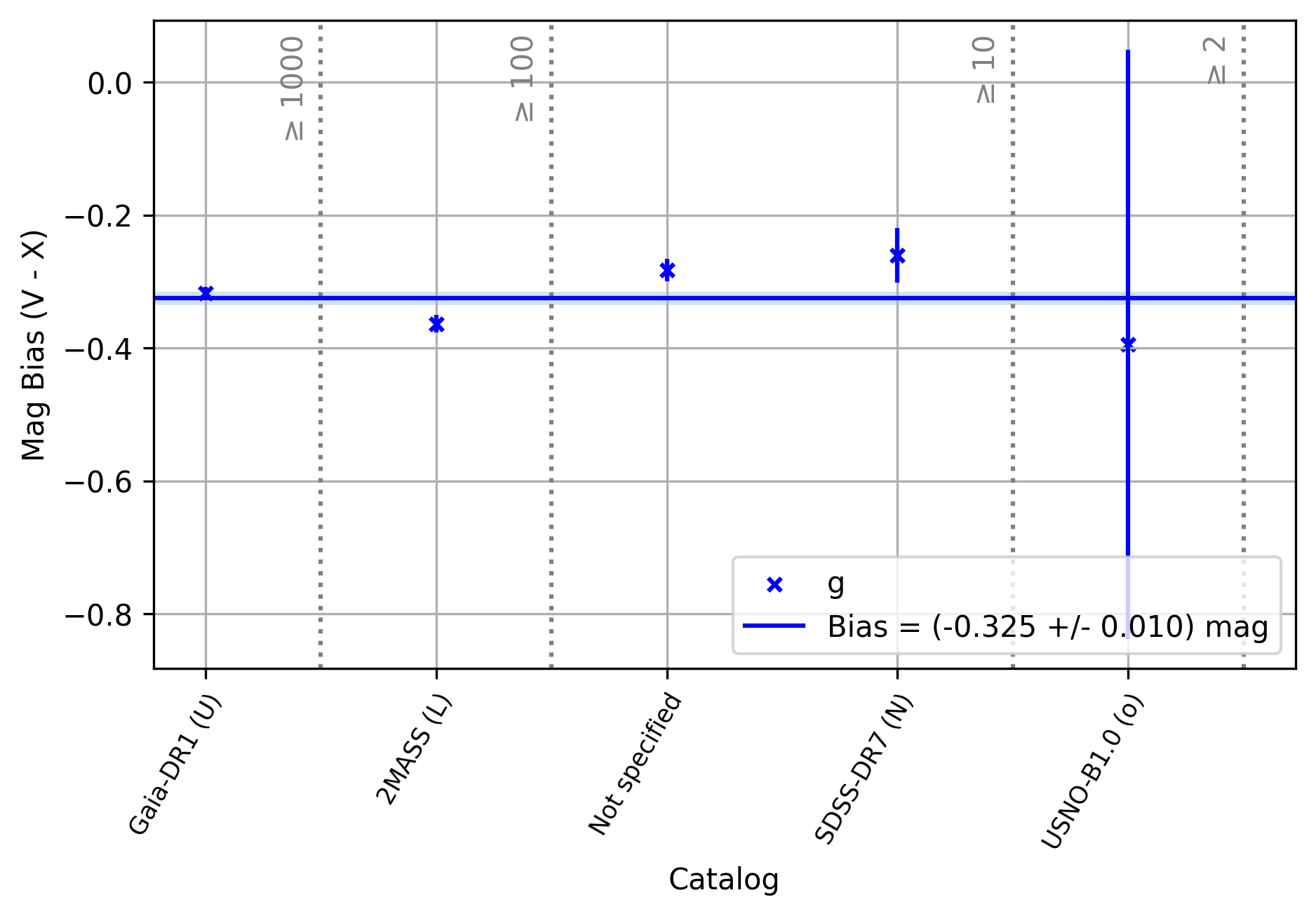}
    \end{subfigure}

    \begin{subfigure}[b]{0.47\textwidth}
        \centering
        \includegraphics[width=\textwidth, trim=0mm 7mm 0mm 2mm,clip]{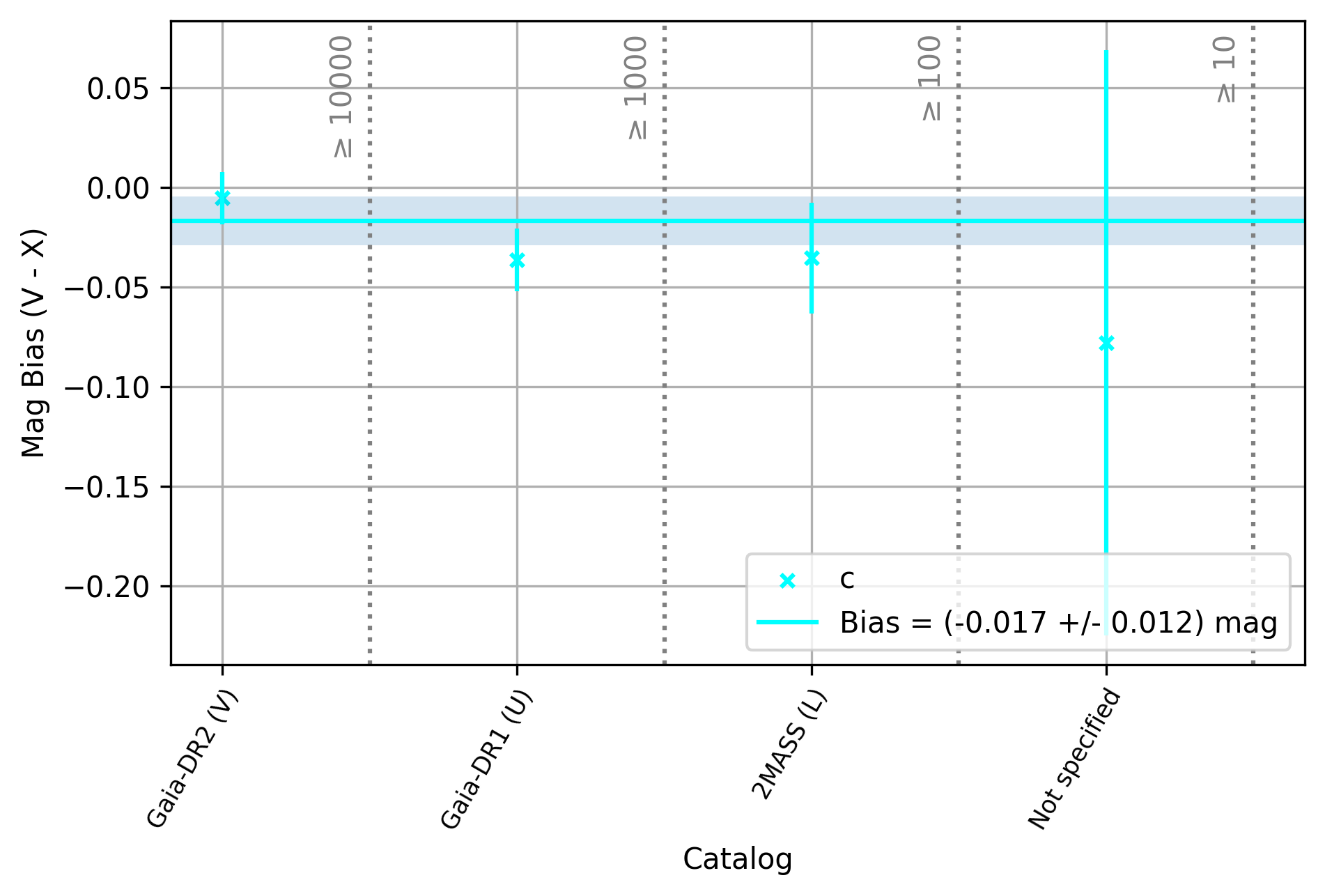}
    \end{subfigure}
    \hfill
    \begin{subfigure}[b]{0.47\textwidth}
        \centering
        \includegraphics[width=\textwidth, trim=0mm 7mm 0mm 2mm,clip]{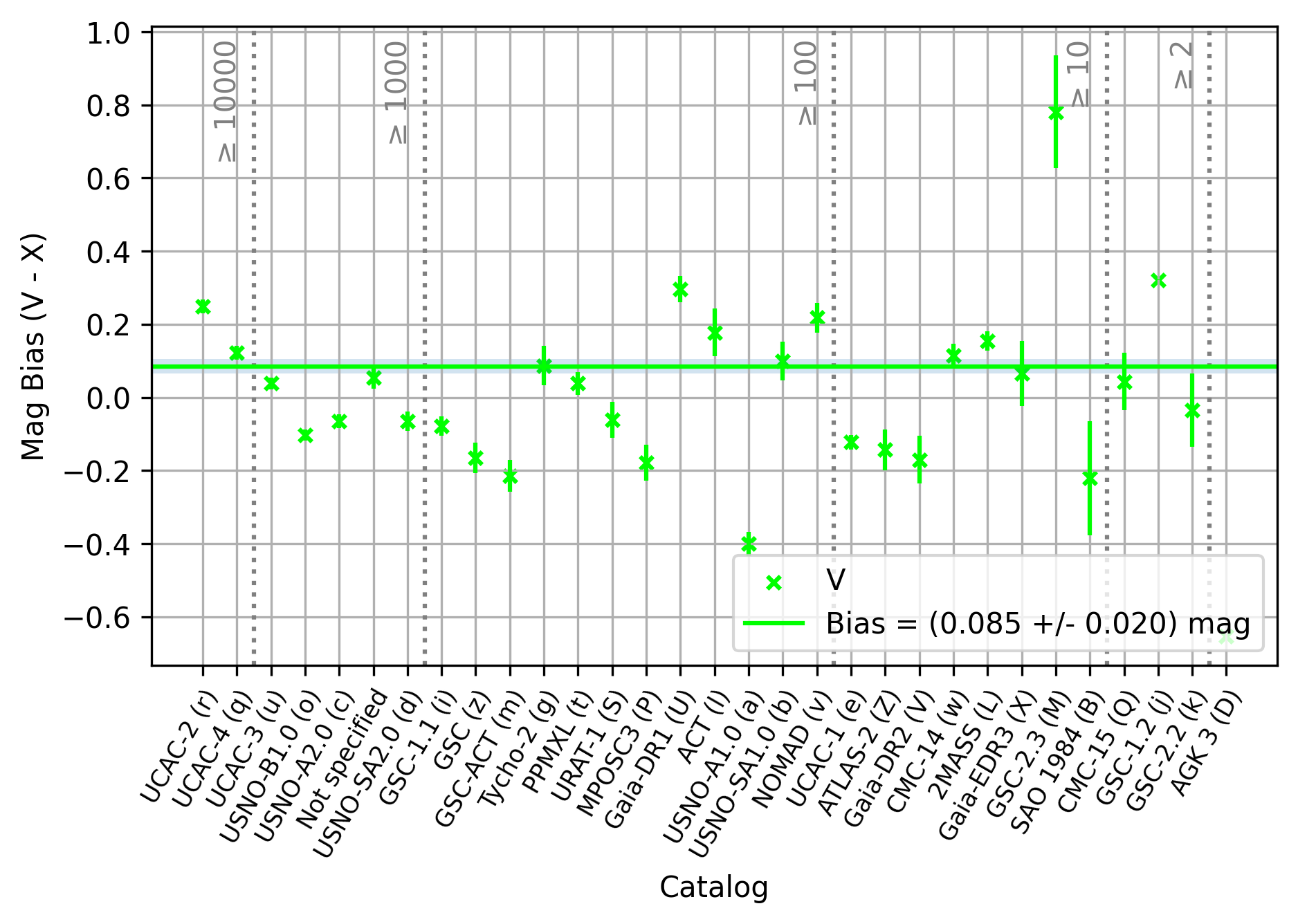}
    \end{subfigure}
    \vspace{2mm}

    \begin{subfigure}[b]{0.47\textwidth}
        \centering
        \includegraphics[width=\textwidth, trim=0mm 7mm 0mm 2mm,clip]{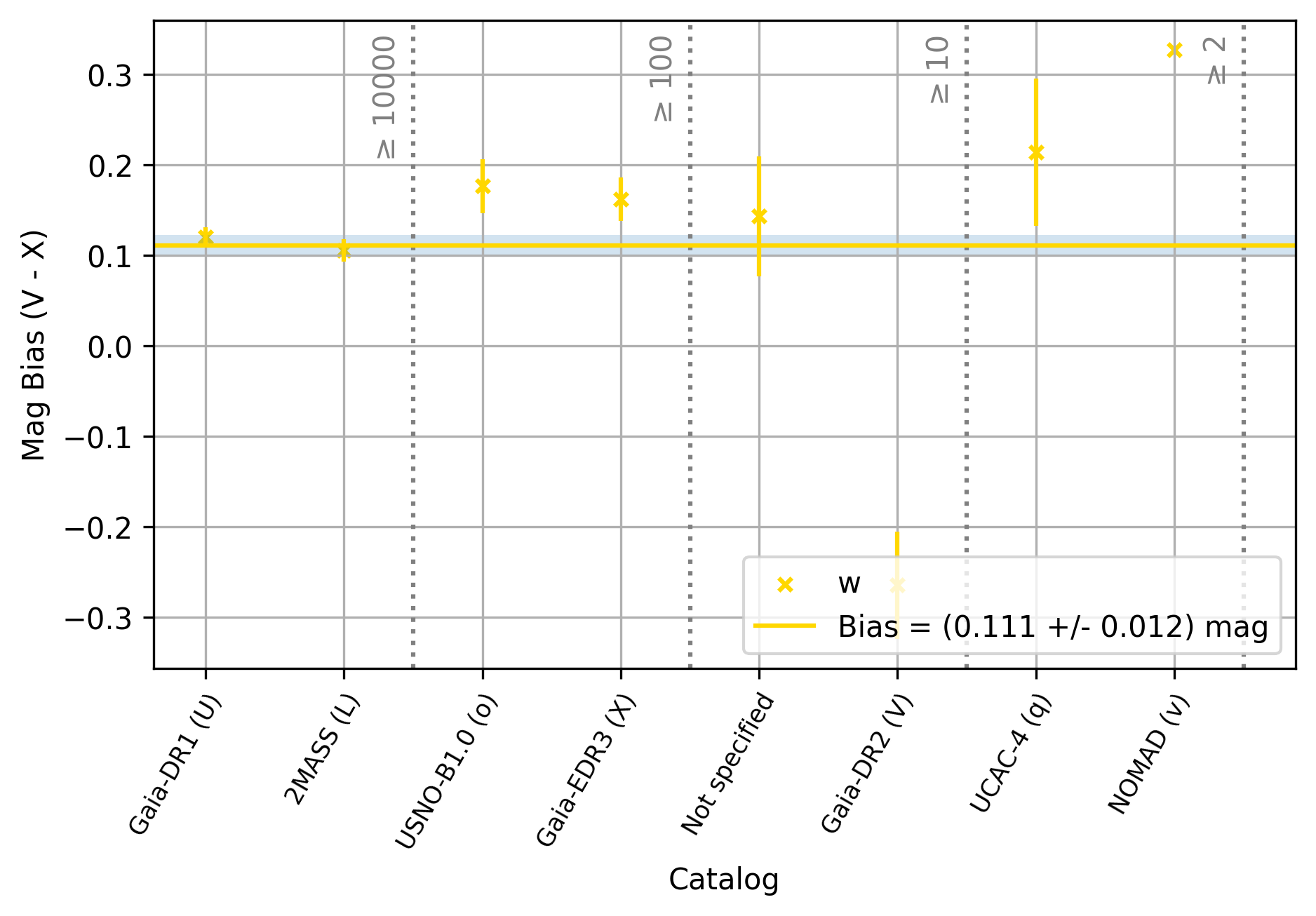}
    \end{subfigure}
    \hfill
    \begin{subfigure}[b]{0.47\textwidth}
        \centering
        \includegraphics[width=\textwidth, trim=0mm 7mm 0mm 2mm,clip]{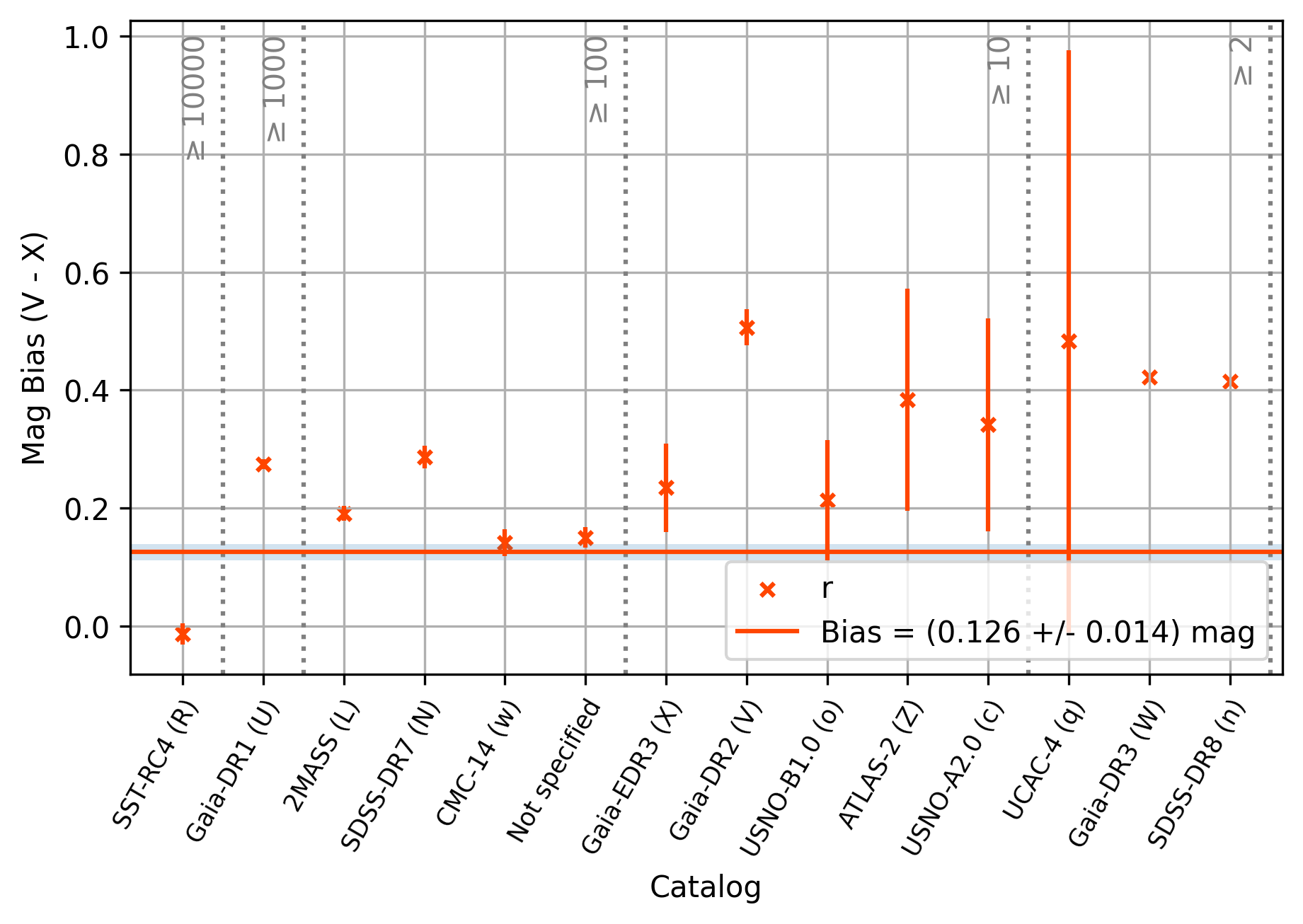}
    \end{subfigure}
    
    \begin{subfigure}[b]{0.47\textwidth}
        \centering
        \includegraphics[width=\textwidth, trim=0mm 7mm 0mm 2mm,clip]{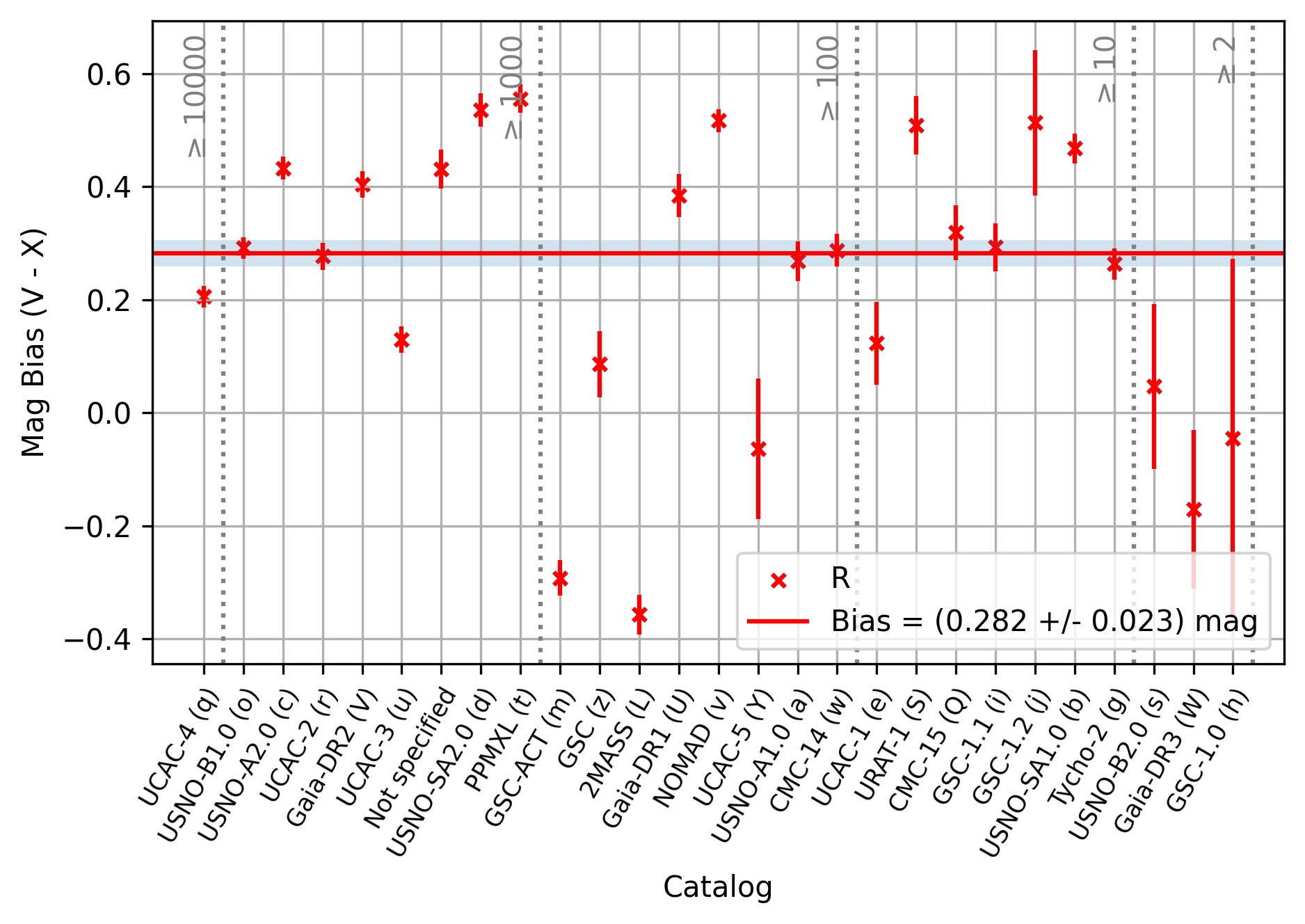}
    \end{subfigure}
    \hfill
    \begin{subfigure}[b]{0.47\textwidth}
        \centering
        \includegraphics[width=\textwidth, trim=0mm 7mm 0mm 2mm,clip]{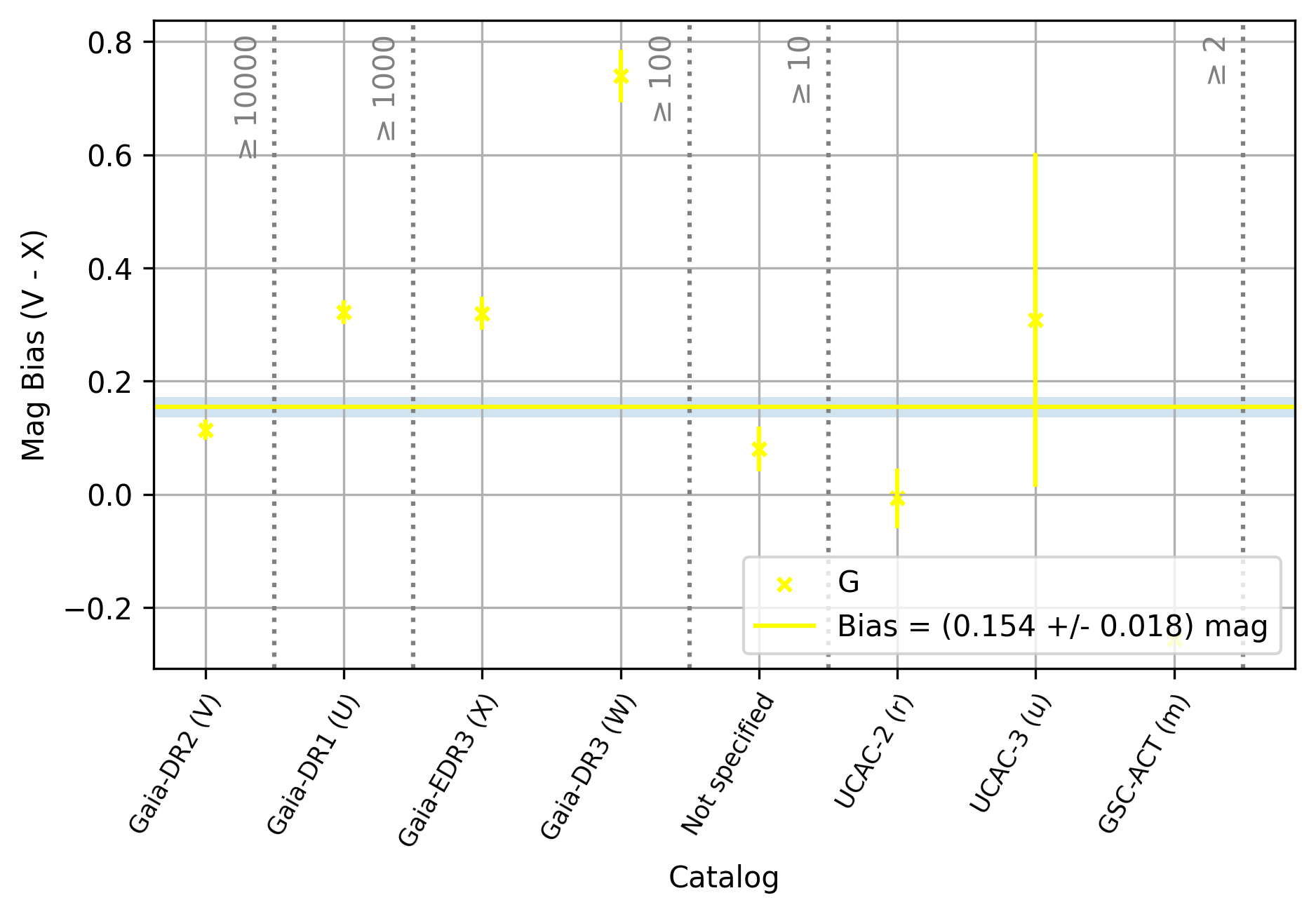}
    \end{subfigure}

\raggedleft (continued on next page)
\end{figure*}%
\begin{figure*}[tp]\ContinuedFloat 
    \centering
    \begin{subfigure}[b]{0.47\textwidth}
        \centering
        \includegraphics[width=\textwidth, trim=0mm 7mm 0mm 2mm,clip]{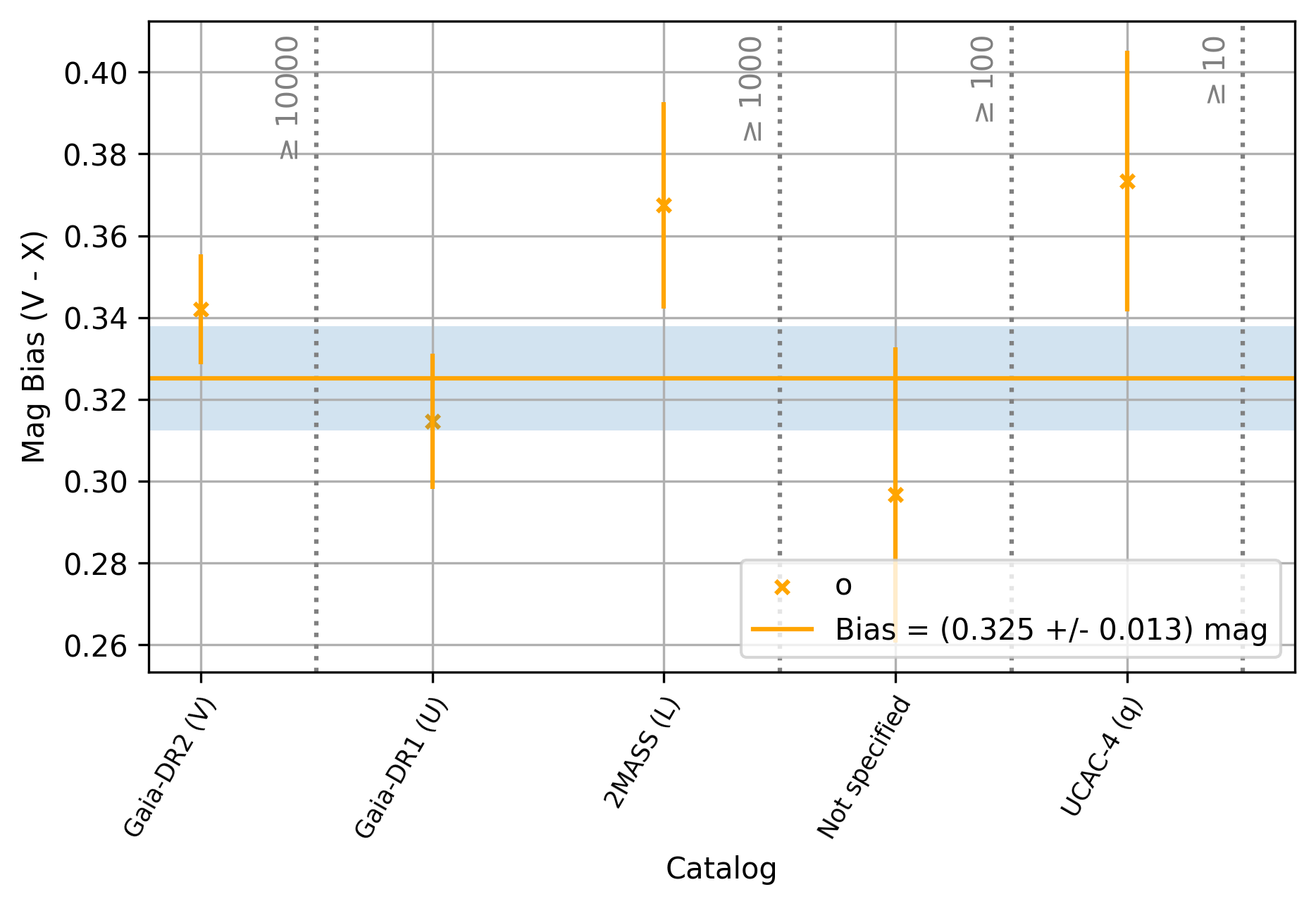}
    \end{subfigure}
    \hfill
    \begin{subfigure}[b]{0.47\textwidth}
        \centering
        \includegraphics[width=\textwidth, trim=0mm 7mm 0mm 2mm,clip]{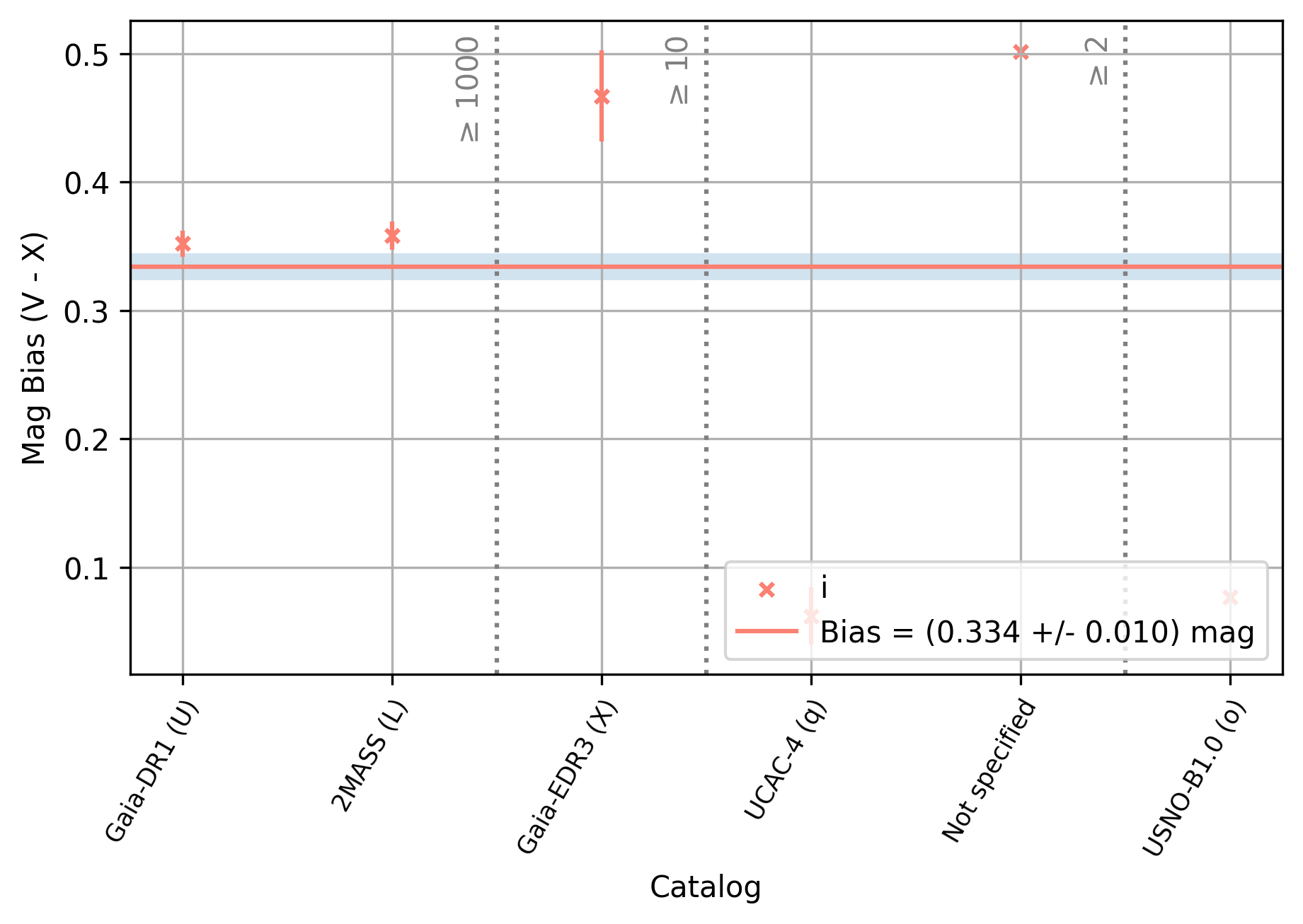}
    \end{subfigure}

    \begin{subfigure}[b]{0.47\textwidth}
        \centering
        \includegraphics[width=\textwidth, trim=0mm 7mm 0mm 2mm,clip]{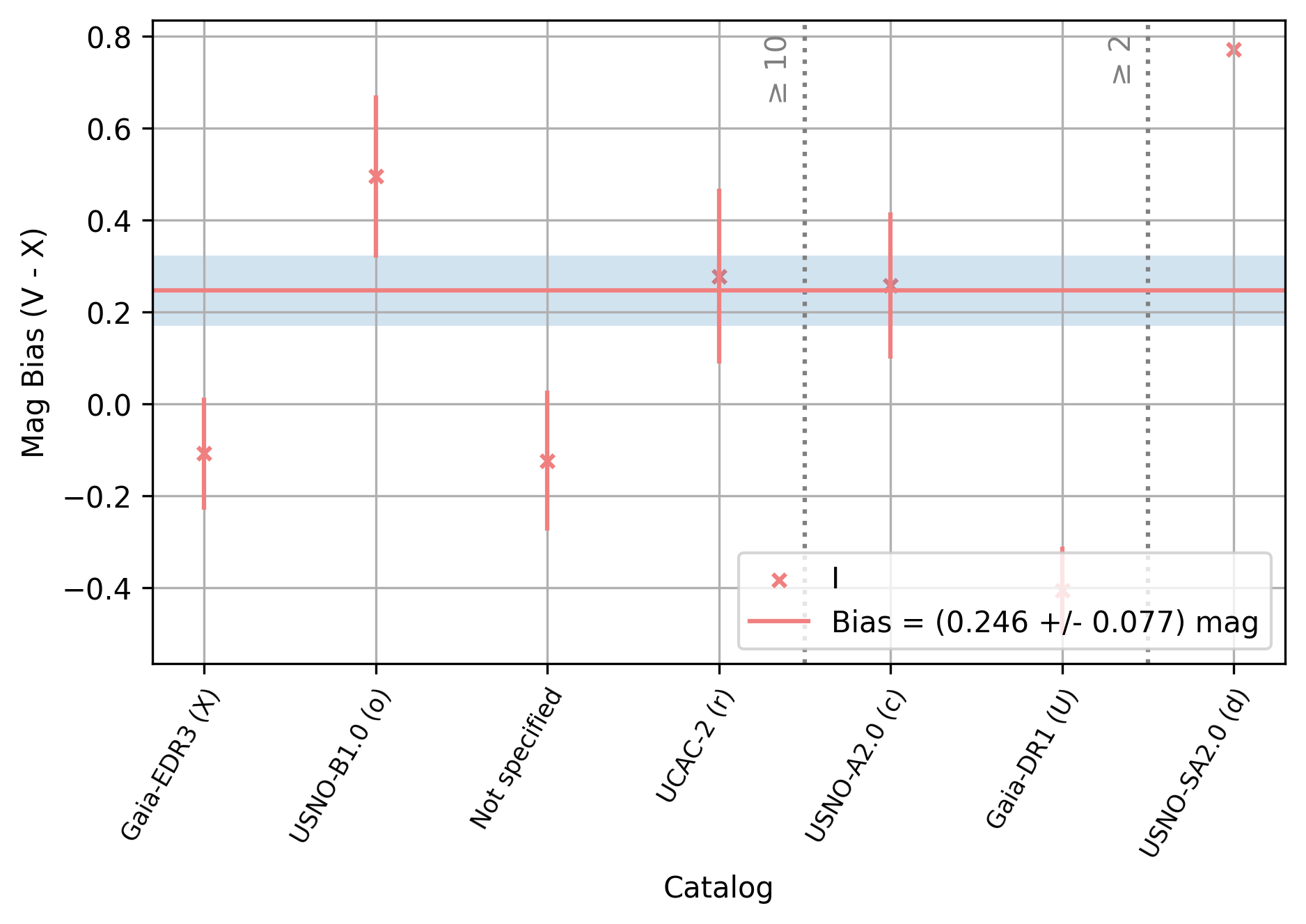}
    \end{subfigure}
    \hfill
    \begin{subfigure}[b]{0.47\textwidth}
        \centering
        \includegraphics[width=\textwidth, trim=0mm 7mm 0mm 2mm,clip]{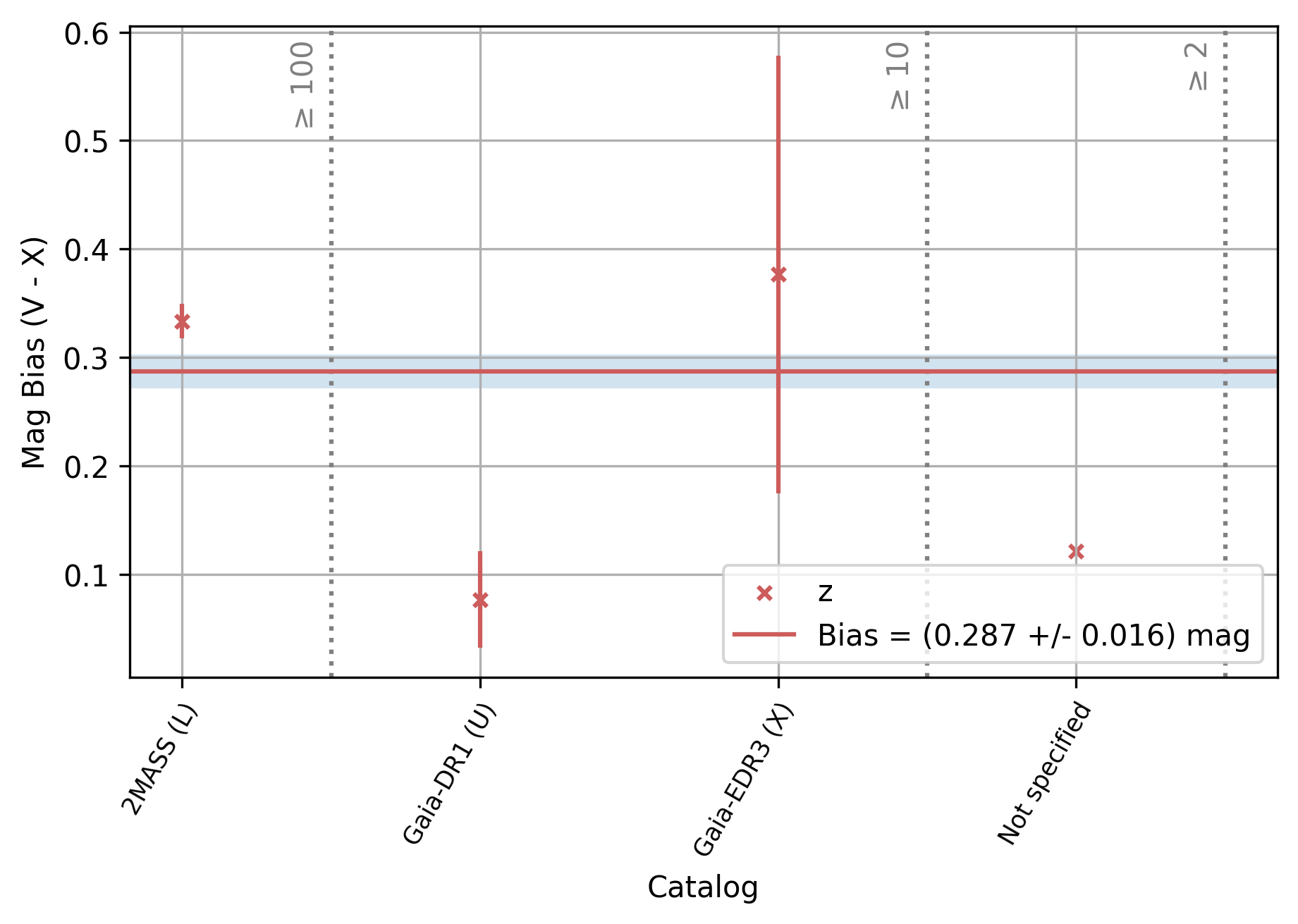}
    \end{subfigure}
    
    \begin{subfigure}[b]{0.47\textwidth}
        \centering
        \includegraphics[width=\textwidth, trim=0mm 7mm 0mm 2mm,clip]{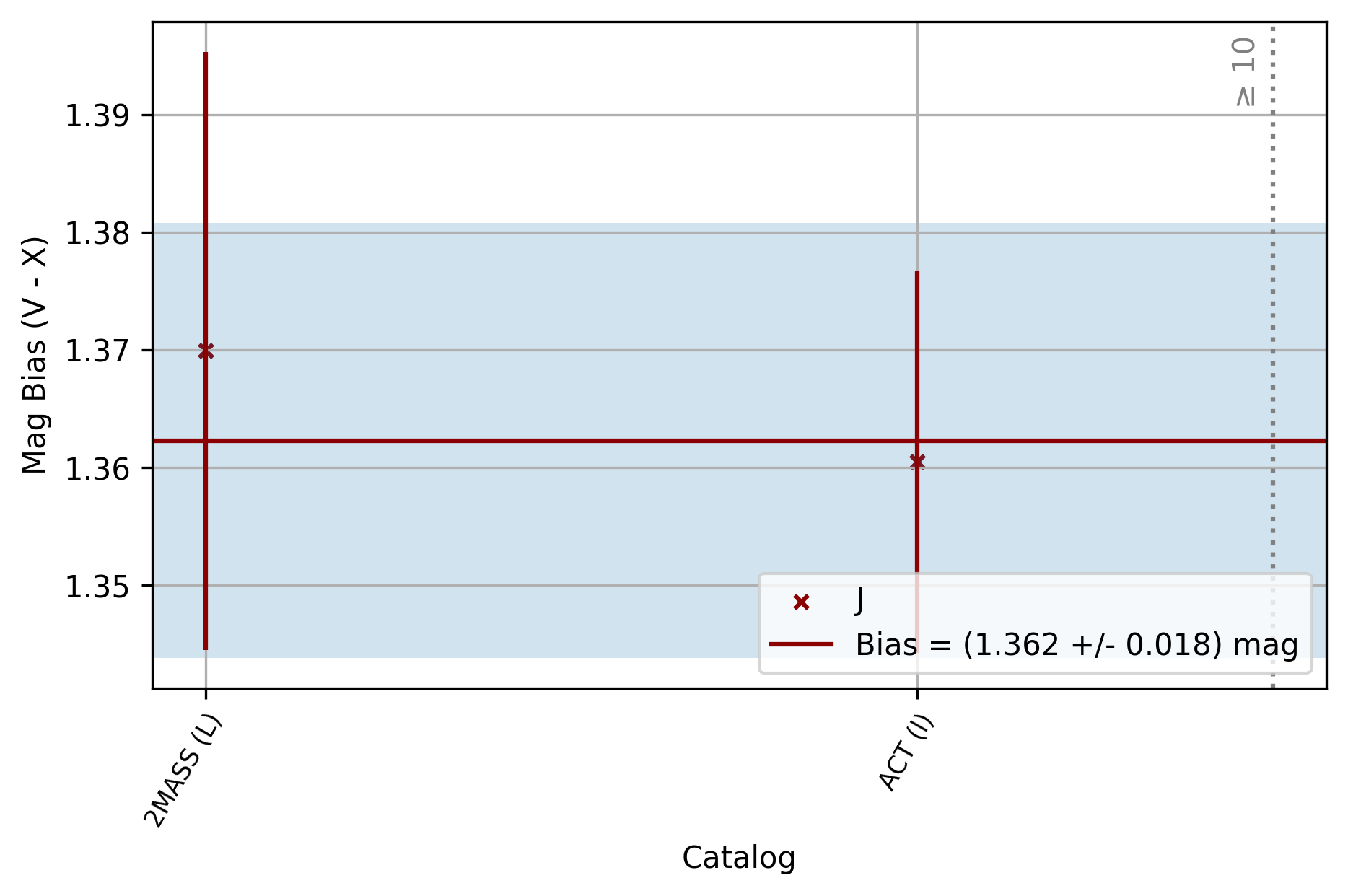}
    \end{subfigure}
    \hfill
    \begin{subfigure}[b]{0.47\textwidth}
        \centering
        \includegraphics[width=\textwidth, trim=0mm 7mm 0mm 2mm,clip]{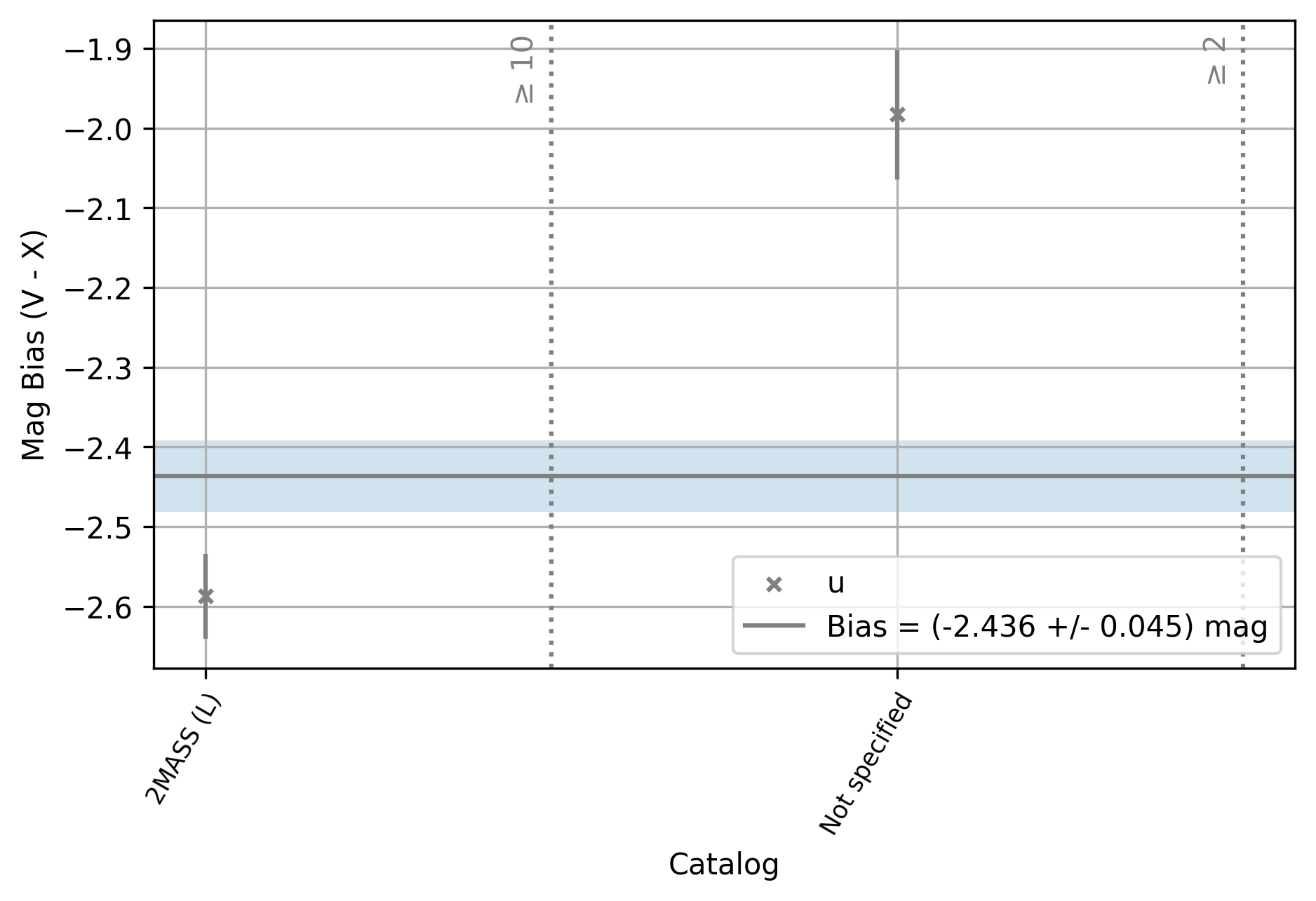}
    \end{subfigure}

    \begin{subfigure}[b]{0.47\textwidth}
        \centering
        \includegraphics[width=\textwidth, trim=0mm 7mm 0mm 2mm,clip]{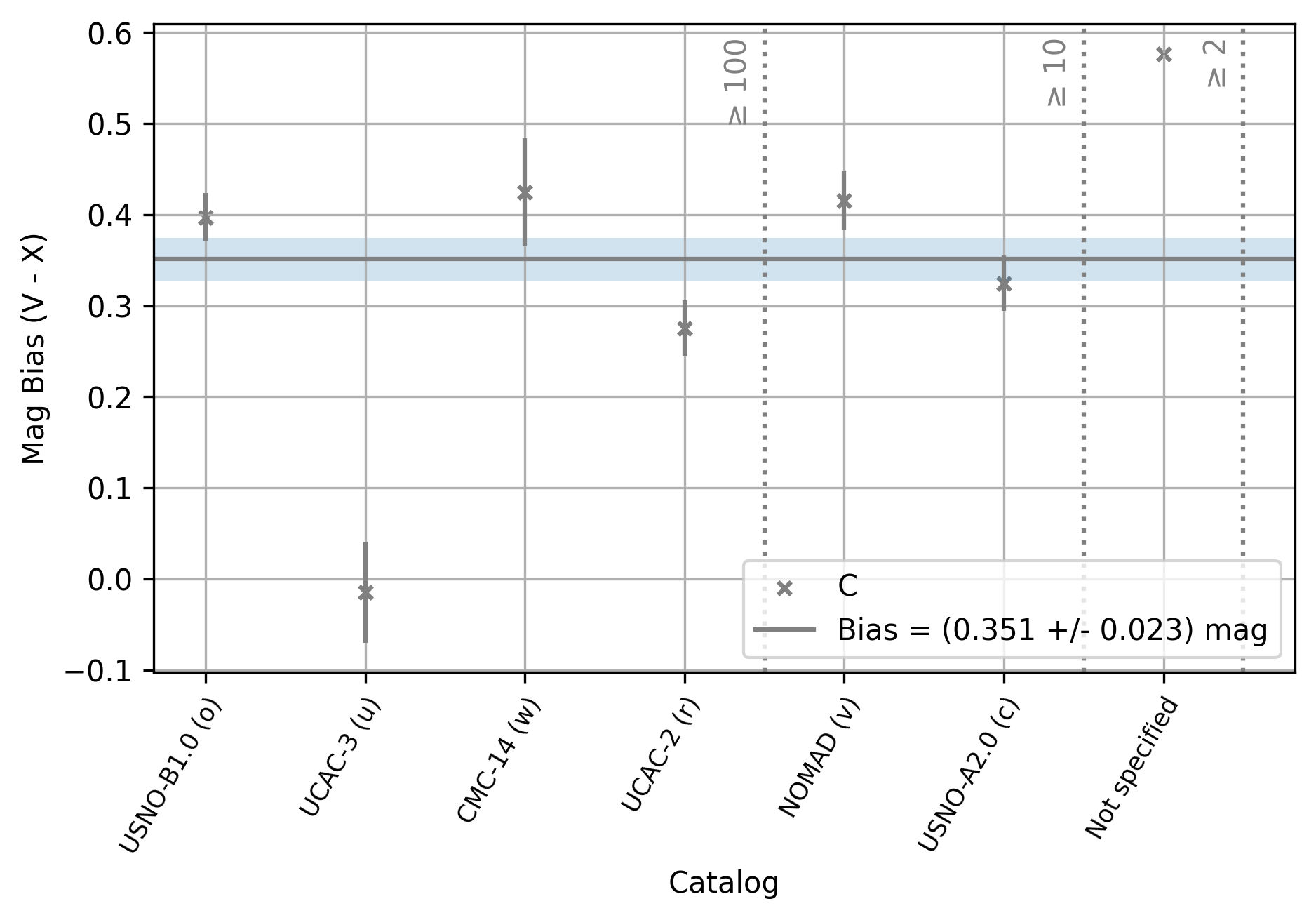}
    \end{subfigure}
    \hfill
    \begin{subfigure}[b]{0.47\textwidth}
        \centering
        \includegraphics[width=\textwidth, trim=0mm 7mm 0mm 2mm,clip]{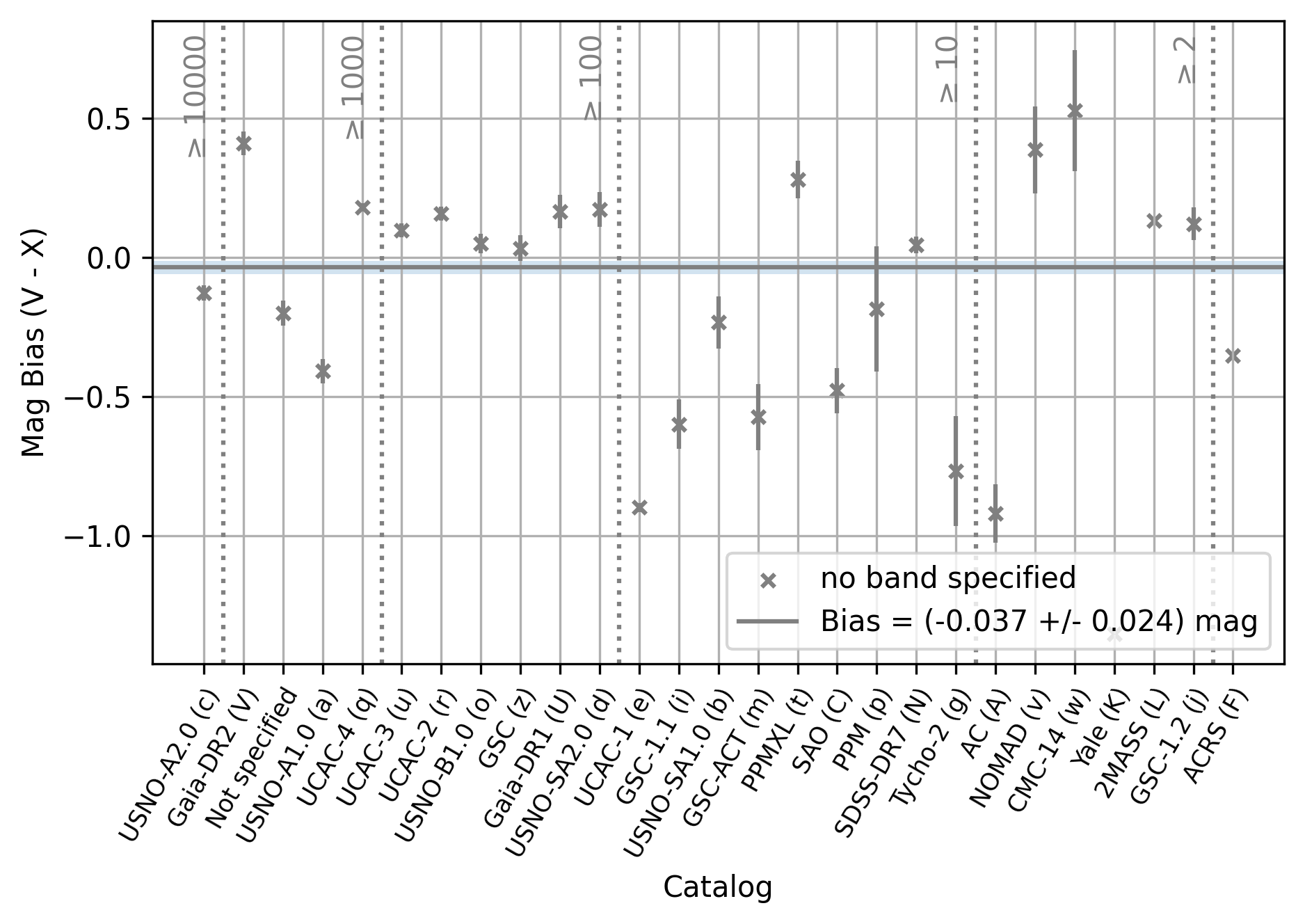}
    \end{subfigure}

    \caption[Results aBC]{Results from \textit{aBC} bias analysis with data from 394 out of 468 asteroids, with the average bias and its standard error in the different color bands and astrometric catalogs (one letter format\footnotemark in brackets), sorted in decreasing order of the number of observations, shown per band with comparison band bias (colored line) including error margins (blue area) from the \textit{aB} analysis (Tab.~\ref{tab:res.aB}), leaving out the following bands with just one catalog: y, Y, H, K (only 2MASS) and L (only Gaia DR2)}
    \label{fig:res.aBC}
\end{figure*}

Binning the observations in each color band by the reference catalog used, we can analyze systematic magnitude biases by catalogs, independent of the color band bias. The catalog refers to the astrometric reference star catalog used to derive the measurements, which is encoded by a single character\footnotetext{\url{https://www.minorplanetcenter.net/iau/info/CatalogueCodes.html}} in column 72 of the observations in the \ac{MPC} 80-column format, which has been introduced by \citet{Williams2013}. More recent observations with the \ac{ADES} would allow the report of a separate photometric catalog, but as this option was not that extensively used until very recently, we assume that there is no major deviation between astrometric and photometric reference star catalog. 

There are in total 159 band-catalog combinations. The results for the average catalog bias from the \textit{aBC} analysis are visually shown, separated by color band, in Fig.~\ref{fig:res.aBC} in comparison with the color band bias. 

We can see in the analysis that there is a significant influence of the reference star catalog used on the bias and accuracy of the photometry. In some color bands like V, R, G and when no band is specified (`blank') the results are varying by up $0.5$ mag, with only a few catalogs lying in the uncertainty margins of the overall color band bias. For the `blank'-band this behavior is expected as the observations might be taken in fact in various color bands that correspond with certain catalogs (e.g., Gaia catalogs with the Gaia G-band), which were just not provided in the report. 

Looking at the `blank'-band results, we can identify certain accumulations of bias values: there is a group of values around $+0.4$ to $+0.2$ mag, corresponding to the G- or R-band, another group around $0.0$ mag, corresponding to V-band, and a third group with negative biases from $-0.2$ to $-0.8$ mag, corresponding to the B-band. The catalogs with the ``blueish'' negative biases are all older catalogs, mostly based on photographic plate measurements, from the 1990s until the early 2000s. Recent catalogs like the Gaia data releases or \ac{UCAC} are based on CCD data and used in more recent observations taken with CCD/CMOS cameras, thus having a rather red-sensitive signal with positive bias. This broad range of biases with older observations (bias up to $-1.0$ mag) and newer ones (bias up to $+0.5$ mag) shows the necessity to further distinguish observations within the color band, especially the `blank'-band, before applying a correction. 

Moreover, in other color bands we see strong deviations from the color bias results in many catalogs. In the Gaia G-band there is a quite noticeable difference among the Gaia data release versions. Here, Gaia-DR3 with a bias of around $+0.7$ mag is a clear outlier, that originates from the high proportion of polluted observations from one observatory (N86), which has strongly biased data. But still, Gaia-DR1, DR2 and EDR3 vary by around $0.2$ mag. Interesting to see is that there are a few observations in G-band with \ac{UCAC} catalogs, which do not contain G-band photometry at all. This suggests that there is a mismatch between color band of the photometric catalog and the observed (or reported) band.

In the case of the visual V-band, we have the same phenomenon, but with the opposite bands. There are Gaia catalog observations in the V-band (e.g., Gaia-DR1 with a bias of $+0.30$ mag) that would rather match with the G-band bias. More prominent even is the bias of UCAC-2 and UCAC-4 catalog, each having around 20\,000 observations, with $+0.25$ and $+0.12$ mag. Again, there are negative biases by older catalogs like the USNO as already seen for the `blank'-band. 

Similar results as for the V-band can be seen in the Cousins R-band, with a shift to a higher bias around $+0.3$ mag, as expected. There are two accumulations of catalog biases, one around $+0.3$ mag and one around $+0.5$ mag, with some outliers towards $0.0$ mag or even negative biases (GSC, 2MASS and UCAC-5). In the Sloan r-band, most biases are localized around $+0.2$ mag with a prominent outlier of $-0.01$ mag from the SST-RC4 catalog, which is only used by one observatory (G45), but has more than 10\,000 observations, thus over-weighting the average. Comparing this with the results from the distribution of biases in the r-band in Fig.~\ref{fig:res.aB.2}, we can see that this is exactly the reason for the second peak in the distribution around $0$ mag. This observatory seems to already correct its measurements to V-band in an accurate way, which we already discussed in Sec.~\ref{sec:res.bias.band}.

On the other side, there are color bands (like g-, c-, o- and w-band), where most biases of catalogs with many observations lie within the uncertainty regions of the \textit{aB} analysis without having high fluctuations. As already mentioned in Sec.~\ref{sec:res.bias.band}, these are the most accurate photometric bands, so we expect a constancy with less specific catalog bias and mostly just color-intrinsic biases as offset here. This again underlines the validity of this method.

\subsubsection{Observatory bias results}\label{sec:res.bias.obs}
We can additionally separate the results by the observatory code that made the measurements. With that, we get in total 2287 band-catalog-observatory combinations. The results for the average observatory bias from the \textit{aBCO} analysis are shown in the appendix (Fig.~\ref{fig:res.aBCO}).

In the \ac{ATLAS} c- and o-band the spread among the observatories is again quite low as expected and within $3\sigma$ of the standard error. There is only a slight deviation in c-band of around $0.03$ mag of T05 in Gaia-DR2 (V) compared to the other observatories. Also in o-band there is a similar deviation, but also for M22 and especially W68 in Gaia-DR2. This shows slight discrepencies among the ATLAS telescopes.

For the Sloan filters g, r and i we can see that there is mostly only one major observatory per band and catalog, so the conclusions from Sec.~\ref{sec:res.bias.catalog} also apply to the \textit{aBCO} results. As already mentioned, the significant deviation in r-band is caused by the deviation (and possible calibration) of G45. In i-band there is a slight effect of a $0.03$ mag difference for the \ac{Pan-STARRS} telescopes F51 and F52 in Gaia-DR1 (U). This difference can also be seen in the \ac{Pan-STARRS} w-band and for other catalogs (L and X), with F52 always having a higher bias than F51, but the amount of the discrepancy is varying among the bands and catalogs. Additionally, there are some T08 and C41 measurements in the w-band, which deviate from the other observatories.

In Gaia G-band there are nearly 300 observatory combinations. Most biases for observatories with many observations vary around $0.2$ mag, with the already mentioned outlier of N86 in Gaia-DR3 (W). It is striking that there are several observatories (including again G45) with biases around $0.0$ mag, mostly with Gaia-DR2 (V), but there are also other observatories with Gaia-DR2 at around $0.2$ mag. There seems to be a systematical difference within the measurements with this catalog. 
There is a large spread for the observations that do not have a band specified (blank). In some cases, the measurements within one observatory code vary a lot, such as for code 247 (Roving Observer) and A84; however, code 247 is a placeholder used for stations without an assigned observatory code, and it is therefore expected that observations submitted with that code should show much broader inconsistencies, since they originate from a variety of different instruments and observers. Observatory code 704 (the now inactive LINEAR survey), with more than 10\,000 observations in this band with USNO-A2.0 catalog (c), on the other hand shows a quite stable bias of $-0.16$ mag.

The majority of observatories use the R-band, where the biases are also quite spread, especially with some clear outliers deviating more than $0.5$ mag. A comparable spread is also noticeable for V-band observations. It is conspicuous that most major observatories reporting in the V-band have a positive bias of around $0.25$ mag, among which there are the telescopes with codes 703 and G96 from the \ac{CSS}. For those observations, they all used the UCAC-2 (r) or UCAC-4 (q) catalog, which have the same bias as already discussed (\ref{sec:res.bias.catalog}). As other observatories, especially the ones with the UCAC-2 catalog, have the same bias, this might be caused by intrinsic biases of the catalogs and not of the observatories. However, there are also other observatories with same catalogs that have zero or negative biases, like C97 and H21. This shows that distinguishing by specific observatory codes can lead to an improved correction.

\subsection{Correction systems}\label{sec:res.correction}
With the algorithm defined in Sec.~\ref{sec:methods.correction} we generate different correction systems based on the bias results with different levels of confidence $\gamma = 1-\alpha$. These \textit{\ac{DePhOCUS}} corrections are then compared to the results of previous corrections that are used. 

The bias analysis results contain 20 band (B), 159 band-catalog (BC) and 2287 band-catalog-observatory (BCO) combinations, which we use as correction values. In the statistical correction algorithm, the standardized t-test is applied to all these combinations with confidence levels $\gamma$ of $0$ (all), $0.68$ ($1\sigma$), $0.90$, $0.95$ ($2\sigma$), $0.99$ and $0.997$ ($3\sigma$), reducing the number of corrections with increasing confidence. This approach keeps the highly statistical relevant corrections for high confidence levels. In Tab.~\ref{tab:res.corr.comp} we can see the decrease of correction rules from nearly 2500 to roughly 150 with increasing $\gamma$.

For the validation subset, consisting of 74 out of the 468 asteroids from the asteroid photometry database\footnote{\url{https://www.asu.cas.cz/~ppravec/newres.htm}} by \citet{Pravec2012} that were not used for obtaining the biases, we can compute the RMS of the reduced magnitudes of all observations for each asteroid before and after applying the correction. Again, in order to avoid any effect of specific asteroids with comparatively many observations, the RMS is calculated for each asteroid and we look at the average RMS among all asteroids. The results for the average RMS for the different correction system (including existing corrections by \ac{MPC}, \ac{NEOCC} and \ac{JPL}) are shown in Tab.~\ref{tab:res.corr.comp}. 

\begin{table}
    \caption{Summary of the results of the different \textit{\ac{DePhOCUS}} correction systems with confidence level $\gamma$ for the number of correction rules in band (B), band-catalog (BC) and band-catalog-observatory (BCO) combinations, and their effect on the RMS of observations from 74 independent validation objects with their relative change in RMS (assuming boundary of $0.25$ mag, cf. Sec.~\ref{sec:methods.lightcurve}).}
    \centering
    \begin{tabular}{l|rrrrr}
        \textbf{Correction}&  \textbf{B}&  \textbf{BC}&  \textbf{BCO}& \begin{tabular}[c]{@{}c@{}}\textbf{RMS}\\ / mag\end{tabular} & \begin{tabular}[c]{@{}c@{}}\textbf{Relative}\\ \textbf{Change} \end{tabular}\\\hline
        None&  0&  0&  0&  0.4251& 0.0\%\\
        MPC&  24&  0&  0&  0.4321& +4.0\%\\
        JPL&  24&  0&  0&  0.4206& -2.6\%\\
        NEOCC$^{[1]}$&  24&  0&  0&  0.4244& -0.4\%\\
        NEOCC$^{[2]}$& *17& 0& 0& 0.3768&-27.6\%\\\hline
        DePhOCUS & & & & &\\ \hline
        $\gamma=0$& *20& 159& 2287& 0.3720&-30.3\%\\
        $\gamma=0.68$& *17& 90& 701& 0.3621&-35.9\%\\
        $\gamma=0.90$& *17& 61& 339& 0.3613&-36.4\%\\
        $\gamma=0.95$& *16& 53& 241& 0.3622&-35.9\%\\
        $\gamma=0.99$& *12& 41& 128& 0.3691&-32.0\%\\
        $\gamma=0.997$& *10& 34& 91& 0.3694&-31.8\%\\\hline
    \end{tabular}\\
    \raggedright
    \hspace{3mm}\footnotesize{[1] Used by \ac{NEOCC} before 2023-09-28 (cf. Tab.~\ref{tab:methods.existing-corrections})}\\
    \hspace{3mm}\footnotesize{[2] Used by \ac{NEOCC} since 2023-09-28 (cf. Tab.~\ref{tab:res.corr.neocc})}\\
    \hspace{3mm}\footnotesize{\ *\  Additional to previous NEOCC correction}
    \label{tab:res.corr.comp}
    \vspace{-4mm}
\end{table}

Without any correction, we see an average RMS of $0.4251$ mag. Using the \ac{MPC} correction (cf. Tab.~\ref{tab:methods.existing-corrections}) the RMS surprisingly increases slightly to $0.4321$ mag. Further analysis show that even though the \ac{MPC} correction reduces the RMS in most bands, bands with many observations, like the c- and w-band, and especially where there is no band specified, have higher RMS than without correction. This is also the case for the B-band, but there is not enough data to make meaningful statements. 

For the \ac{NEOCC}  correction that was used prior to 2023-09-28 there are also similar issues, but the different correction in the i- and especially w-band reduce uncertainties, compensating the worsening and leading to an almost zero change compared to no correction. On the other hand, the current \ac{NEOCC} correction, in use since 2023-09-28, includes the \textit{aB} bias analysis results of the present study. Band corrections that fulfill the $\gamma=0.9$ criterion are allowed to replace previous correction (cf. Tab.~\ref{tab:res.corr.neocc}), which happens to be the case for 17 out of the 24 different bands. This correction, that is just using band correction, already reduces the overall RMS to $0.3768$ mag, a decrease of $27.6\%$ compared to $0.25$ mag estimated correction limit (cf. Sec.~\ref{sec:methods.lightcurve}). The \ac{JPL} correction, using the same band correction as the prior 2023-09-28 \ac{NEOCC}-version, except for G-band, is also reducing the RMS by $4.0\%$, showing the impact of minor correction changes for commonly-used bands.

\begin{table} 
    \caption{Current NEOCC correction values (used since 2023-09-28), only using \textit{aB} bias analysis results in correction algorithm ($\gamma=0.9$). The results for the values with two decimal places do not fulfill the significance criteria, therefore the previous values are being used there.}
    \centering
    \begin{tabular}{cl|cl}
        \textbf{Band} & \begin{tabular}[c]{@{}c@{}}\textbf{NEOCC}\\ / mag\end{tabular} & \textbf{Band} & \begin{tabular}[c]{@{}c@{}}\textbf{NEOCC}\\ / mag\end{tabular}\\ \hline
    U & -1.30     & I & 0.246276    \\
    B & 0.109632  & z & 0.286980    \\
    g & -0.325416 & y & 0.36        \\
    c & -0.016834 & Y & 0.905773    \\
    V & 0.084846  & J & 1.362321    \\
    v & 0.00      & H & 1.810332    \\
    w & 0.111135  & K & 1.834697    \\
    r & 0.125577  &   & -0.036556   \\
    R & 0.282052  & u &-2.50        \\
    G & 0.154115  & L & 0.20        \\
    o & 0.33      & C & 0.350984    \\
    i & 0.333879  & W & 0.40       
    \end{tabular}
    \label{tab:res.corr.neocc}
    \vspace{-4mm}
\end{table}

Introducing the full \textit{\ac{DePhOCUS}} correction, that is combining all the band, catalog and observatory bias results, we can see that the RMS is reducing by more than $30\%$ for all confidence levels, having an optimum RMS of $0.3613$ mag ($-36.4\%$) for a confidence level $\gamma=0.90$. This already shows the large potential of catalog and observatory correction in addition to band correction. Using more or even all corrections (smaller $\gamma$) increases the uncertainties, as corrections might not be statistically solid enough. Using less corrections (higher $\gamma$) selects the more statistically relevant corrections, but also removes corrections that are improving the results. A confidence level of $\gamma=0.90$ is therefore our empirical optimum for the correction algorithm. This set of corrections can be found in the appendix.

\subsection{Validation and prediction of correction model}\label{sec:res.pred}

\begin{figure*}[ht]
    \centering
    \hfill
    \begin{subfigure}[b]{0.49\textwidth}
        \centering
        \includegraphics[width=\textwidth, trim=0mm 0mm 0mm 0mm,clip]{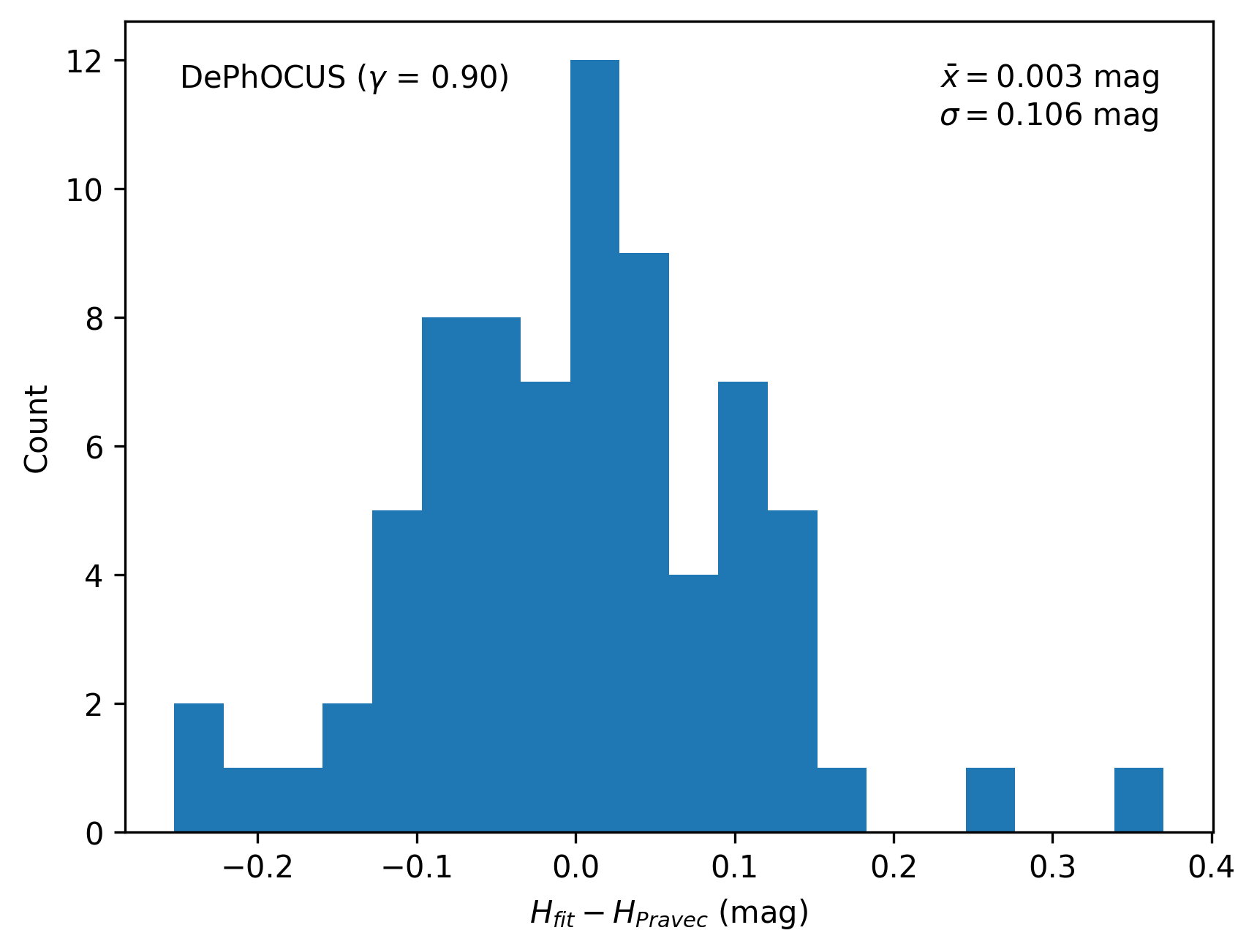}
    \end{subfigure}
    \hfill
    \begin{subfigure}[b]{0.49\textwidth}
        \centering
        \includegraphics[width=\textwidth, trim=0mm 0mm 0mm 0mm,clip]{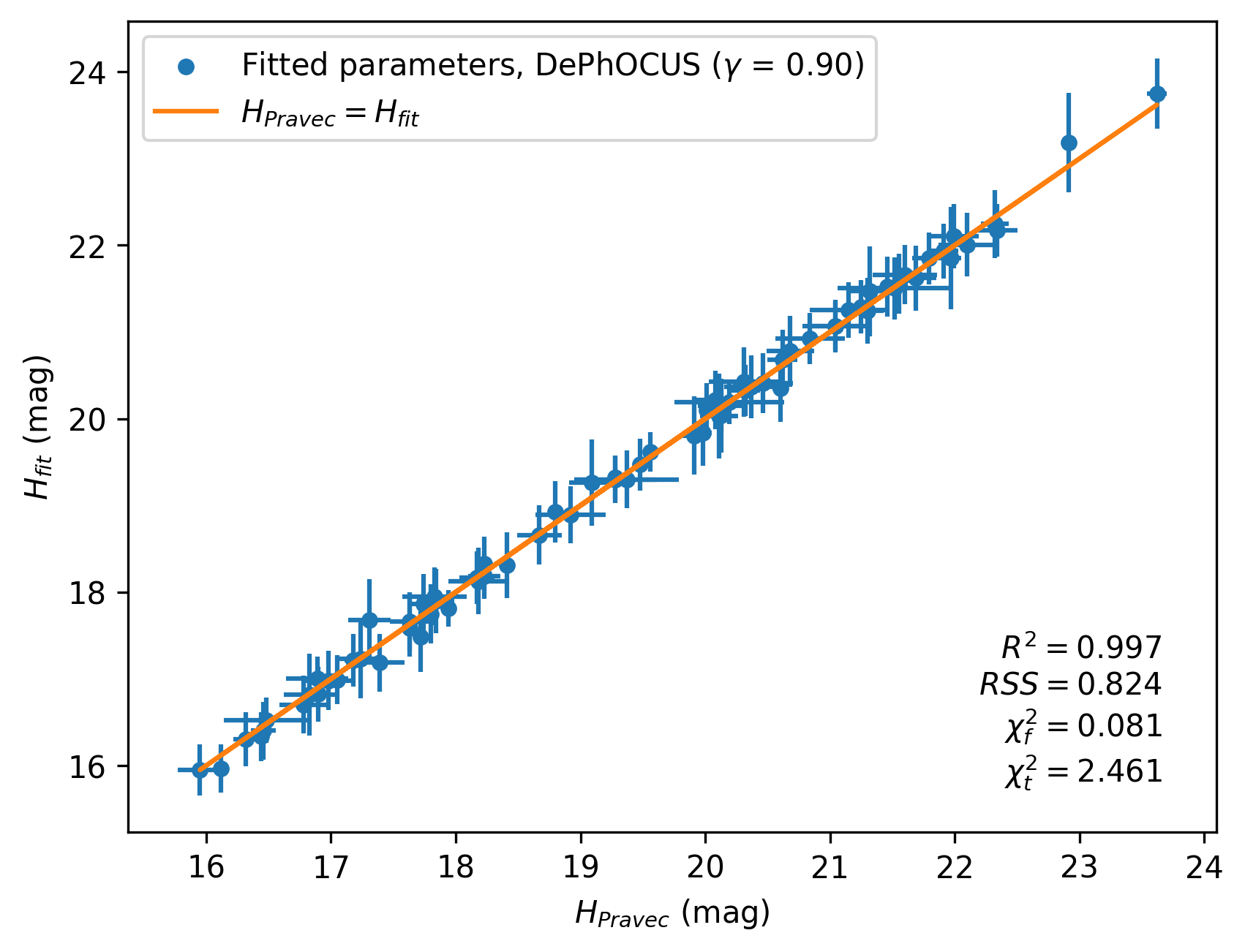}
    \end{subfigure}
    \vspace{2mm}

    \centering
    \hfill
    \begin{subfigure}[b]{0.49\textwidth}
        \centering
        \includegraphics[width=\textwidth, trim=0mm 0mm 0mm 0mm,clip]{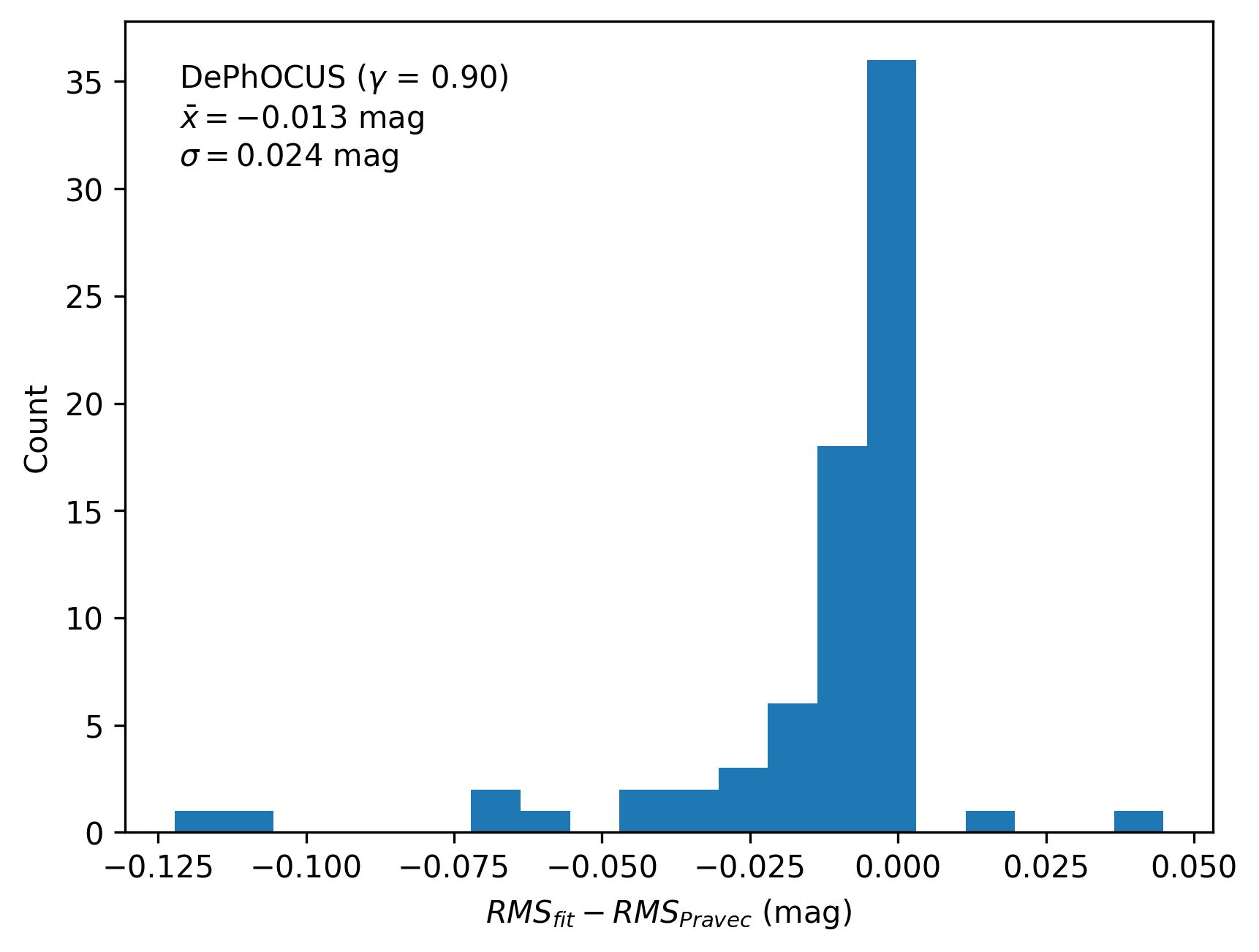}
    \end{subfigure}
    \hfill
    \begin{subfigure}[b]{0.49\textwidth}
        \centering
        \includegraphics[width=\textwidth, trim=0mm 0mm 0mm 0mm,clip]{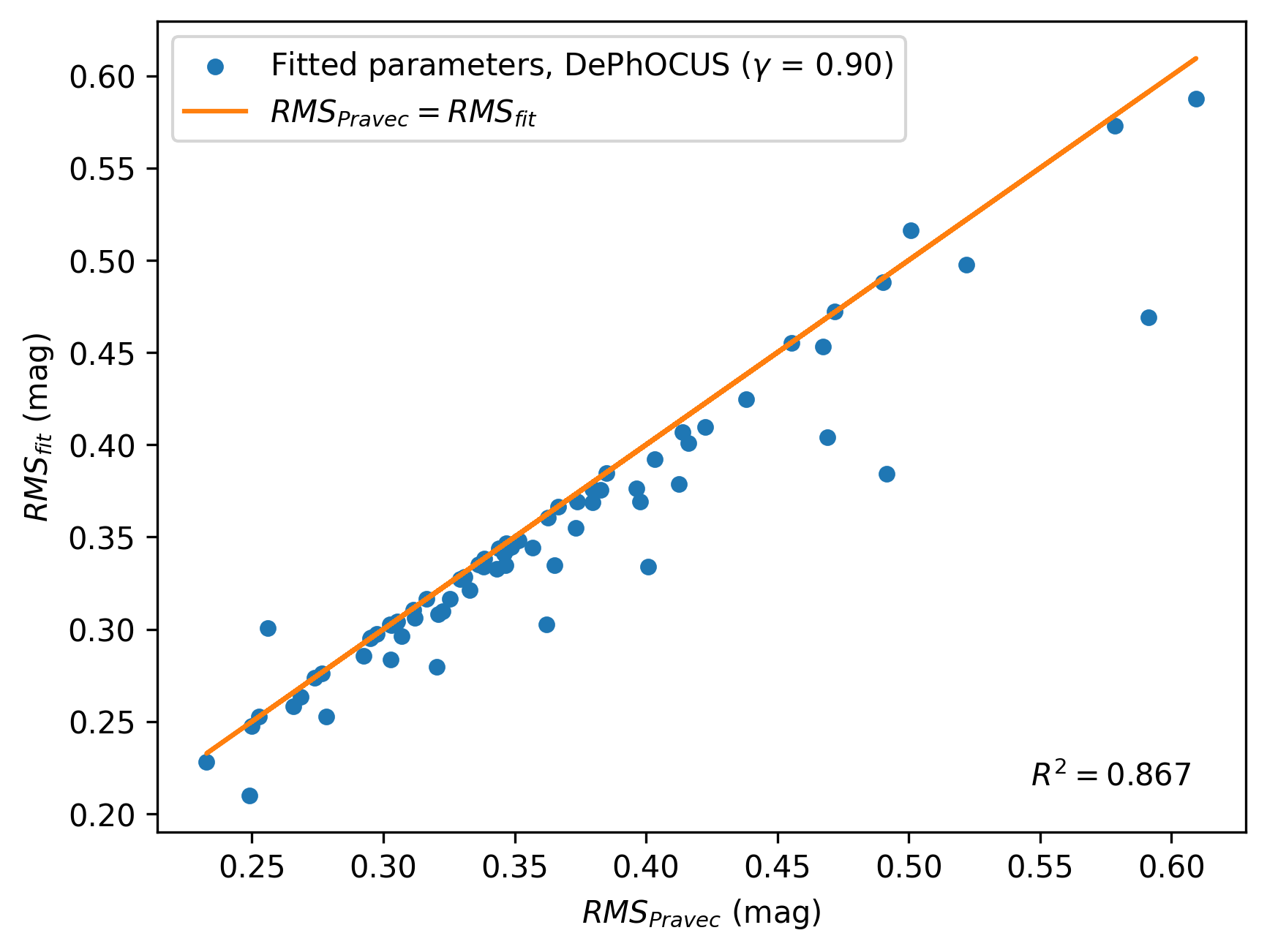}
    \end{subfigure}
    
    \caption{Histogram of difference between fitted $H$ parameter, after applying \textit{\ac{DePhOCUS}} correction ($\gamma=0.9$), and reference values $H_{\text{Pravec}}$ by Pravec et al. (left) with their correlation (right), when keeping $G=G_{\text{Pravec}}$ for 74 asteroids as validation. Also comparing the $RMS$ of measurements by using the fitted parameters or the reference values.}
    \label{fig:res.valid.onlyH}
\end{figure*}

We use the subset of 74 asteroids to validate the results for the developed correction system and see if this leads to a better prediction for the $H$-$G$ magnitude system. We achieve this by blindly fitting the corrected astro-photometric observations to the $H$-$G$ phase curve and comparing the fitted parameters to the reference $H$-$G$ parameters by \citet{Pravec2012}. In order to avoid specific biases from the fitting of the $G$ parameter, which is not well understood and rarely done up to this point, we constrain the $G$ parameter in the fit to the reference value $G_{\text{Pravec}}$ and do only optimize the phase curve function for $H_{\text{fit}}$. The fitting procedure is using a non-linear least squares method with Eq.~\eqref{eq:phasecurve} as fitting function, weighting the observations by their photometric precision\footnote{As used by \ac{NEOCC}: $\sigma=(0.5, 0.7, 1.0)$ mag for 2, 1 and 0 decimal places.}.

Fig.~\ref{fig:res.valid.onlyH} shows the results for the fitted $H_{\text{fit}}$ and the resulting $RMS_{\text{fit}}$ in comparison to the reference $H_{\text{Pravec}}$, where we have a deviation $RMS_{\text{Pravec}}$ in the observation with the applied \textit{\ac{DePhOCUS}} correction ($\gamma=0.9$). The results show that the $H_{\text{fit}}$ predicts well the reference values within a standard deviation of $0.106$ mag, having no significant systematic deviation or bias ($0.003$ mag), which shows there is no selection bias\footnote{Could have been due to the selection by time (after 2021-10-21).} of the reference asteroids. The results also show high agreement for the correlation of fitted and reference $H$ parameter ($R^2=0.997$) with a reduced $\chi^2$ value of $0.081$ for the fitting uncertainties and $2.461$ for the reference error estimates. We are aware that the fitted $H$ parameter uncertainties are underestimating the true uncertainties. As they are based on the RMS of the observation, the small fitting errors do not consider the intrinsic fluctuations of the parameter (e.g., due to rotational effects). We see that the RMS is decreasing on average by $0.013$ mag with the fitted parameters, with minor exceptions, as the fitted parameters should minimize deviations, leading to an expected low correlation factor $R^2=0.867$.

The results for the $H$ parameter prediction with the \textit{\ac{DePhOCUS}} correction are significantly reducing uncertainties compared to previous corrections; the deviation of the fitted $H$ parameter to the reference values with \ac{MPC} correction is $0.136$ mag for the same objects (cf. Appendix, Fig. \ref{fig:res.valid.MPConlyH}), resulting in a $28\%$ reduction of uncertainties with the new method. 
While the coefficient of correlation is already nominally high for the \ac{MPC} method ($R^2=0.995$), we especially see improvements in the reduced $\chi^2$ value with a $21\%$ and $94\%$ increase in goodness of fit for the fitting uncertainties and the reference error estimates respectively.\footnote{For \ac{MPC} correction: $\chi^2_f=0.098$ and $\chi^2_t=4.770$.}

\begin{figure}[tb]
    \centering
    \includegraphics[width=\linewidth, trim=0mm 0mm 0mm 0mm,clip]{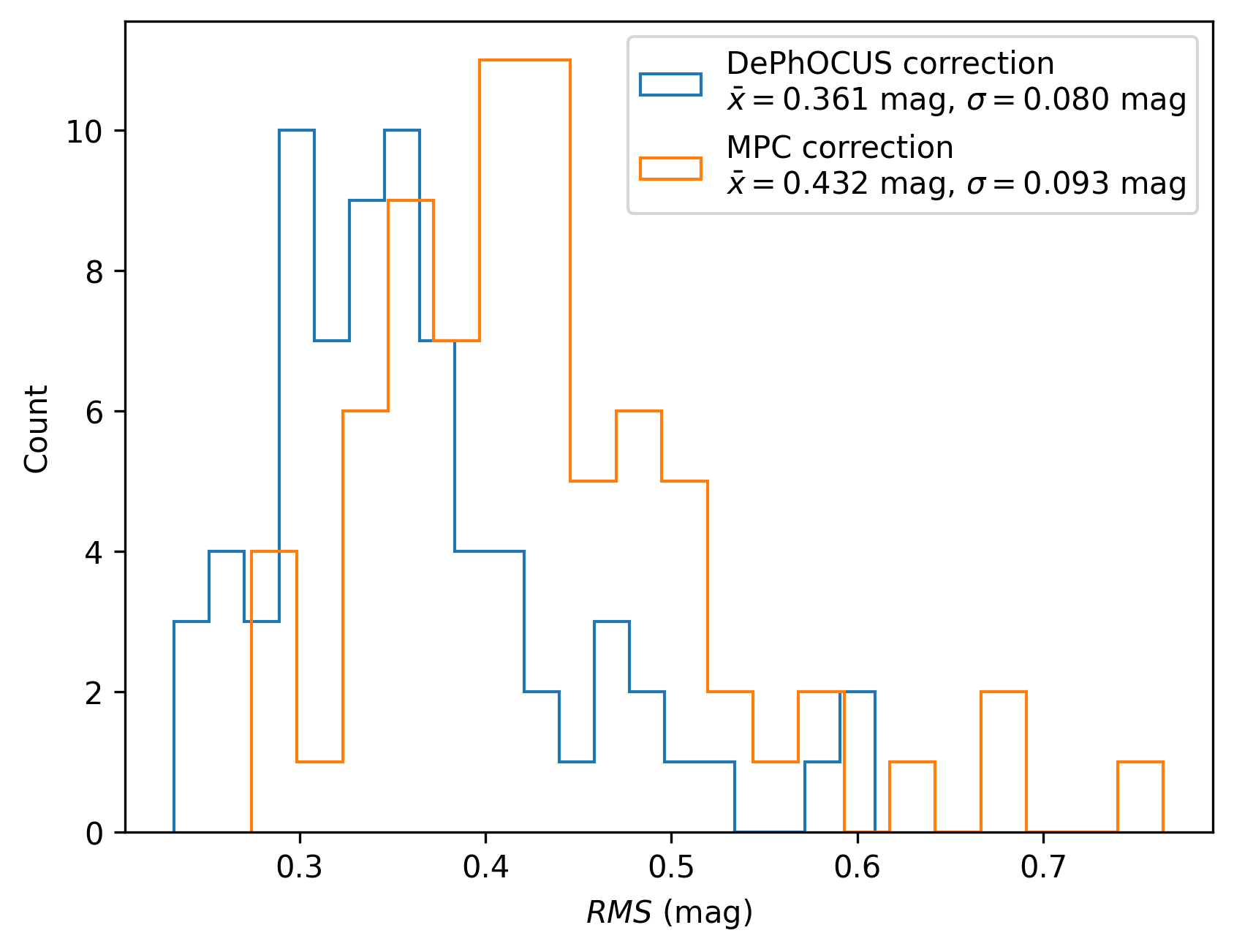} 
    \caption{RMS distribution for 74 validation asteroids with \textit{\ac{DePhOCUS}} correction ($\gamma=0.9$) and \ac{MPC} correction.}
    \label{fig:res.dephocus_rmsdiff}
\end{figure}

\begin{figure}[tb]
    \centering
    \includegraphics[width=\linewidth, trim=0mm 0mm 0mm 0mm,clip]{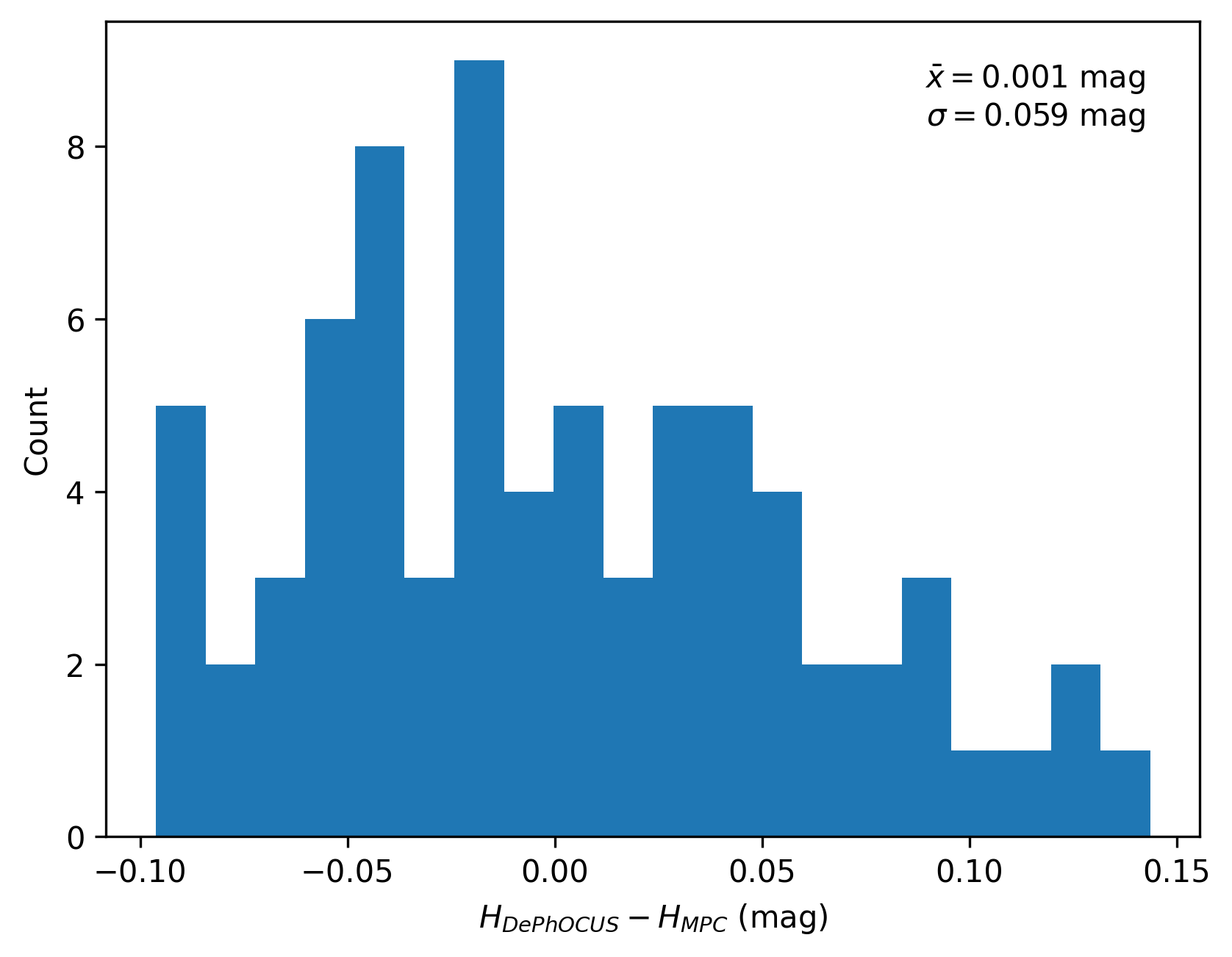} 
    \caption{Distribution of difference in optimized $H$ for 74 validation asteroids with \textit{\ac{DePhOCUS}} correction ($\gamma=0.9$) and \ac{MPC} correction.}
    \label{fig:res.dephocus_Hdiff}
\end{figure}

Looking at Fig.~\ref{fig:res.dephocus_rmsdiff}-\ref{fig:res.dephocus_Hdiff}, we can also compare the distribution of RMS and difference of the $H$ parameter for the 74 asteroid phase curves with the novel correction method in contrast to the current \ac{MPC} correction (cf. Fig.~\ref{fig:res.valid.MPConlyH}). The $H$ parameter changes in a range of $-0.10$ to $0.15$ mag (standard deviation $0.059$ mag) with the new method. On average, the $H$ values do not tend to change in one specific direction (average difference $0.001$ mag). But however, for the RMS it is noticeable that the distribution shifts to lower RMS values, on average by $-0.071$ mag, reaching a peak at around $0.35$ mag with the new method due to change and improvement of the individual $H$ parameters. The reduction is consistent at all levels of RMS. The distribution also gets narrower with the \textit{\ac{DePhOCUS}} method, especially reducing the number of high-RMS outliers. This leads to a reduction in the standard deviation from $\sigma_{\text{MPC}}=0.093$ down to $\sigma_{\text{DePhOCUS}}=0.080$.

\section{Discussion}
\subsection{Advanced bias analysis results}\label{sec:disc.adv}
The seven largest stations with the most observations in our study are: C57 (TESS, 61\,233 observations), 703 (\ac{CSS}, 45\,185 observations), T08 (\ac{ATLAS}-MLO, 39\,522 observations), T05 (\ac{ATLAS}-HKO, 37\,934 observations), 704 (Lincoln Laboratory ETS, 36\,993 observations), G96 (Mt. Lemmon Survey, 30\,134 observations) and F51 (\ac{Pan-STARRS} 1, 21\,792 observations). To determine the limits of our correction model, we have a look at the dependence of different parameters that could have an effect on the photometric bias for these stations. For that, we visualize the biases obtained from the observations of all 468 asteroids in the analysis system graphically. As both \ac{ATLAS} telescopes T05 and T08 are technically equivalent, we show the results for T05 as reference here (Fig.~\ref{fig:dis.advanced.T05c}); the other results can be found in the appendix (Fig.~\ref{fig:dis.advanced.mag}-\ref{fig:dis.advanced.time}). We have a look at:
\begin{itemize}
    \item Apparent magnitude (\acs{SNR}),
    \item Apparent motion (trailing effects),
    \item Astrometric residuals (measuring errors),
    \item Absolute magnitude (size estimate),
    \item Galactic latitude (density of surrounding stars),
    \item Time (long-term effects),
\end{itemize}
in comparison to the obtained correction values for the observatory stations. We achieve this by binning the data points of each station in a hexagonal grid (spacing of $0.1$ mag for the bias) and aggregate all observations that fall in each segment and compute its number of included points. This visualizes the density of observations to be located at a certain parameter- and bias-area. 

\begin{figure*}[htbp]
    \centering
    \begin{subfigure}[b]{0.49\linewidth}
        \centering
        \includegraphics[width=\textwidth, trim=0mm 0mm 0mm 2mm,clip]{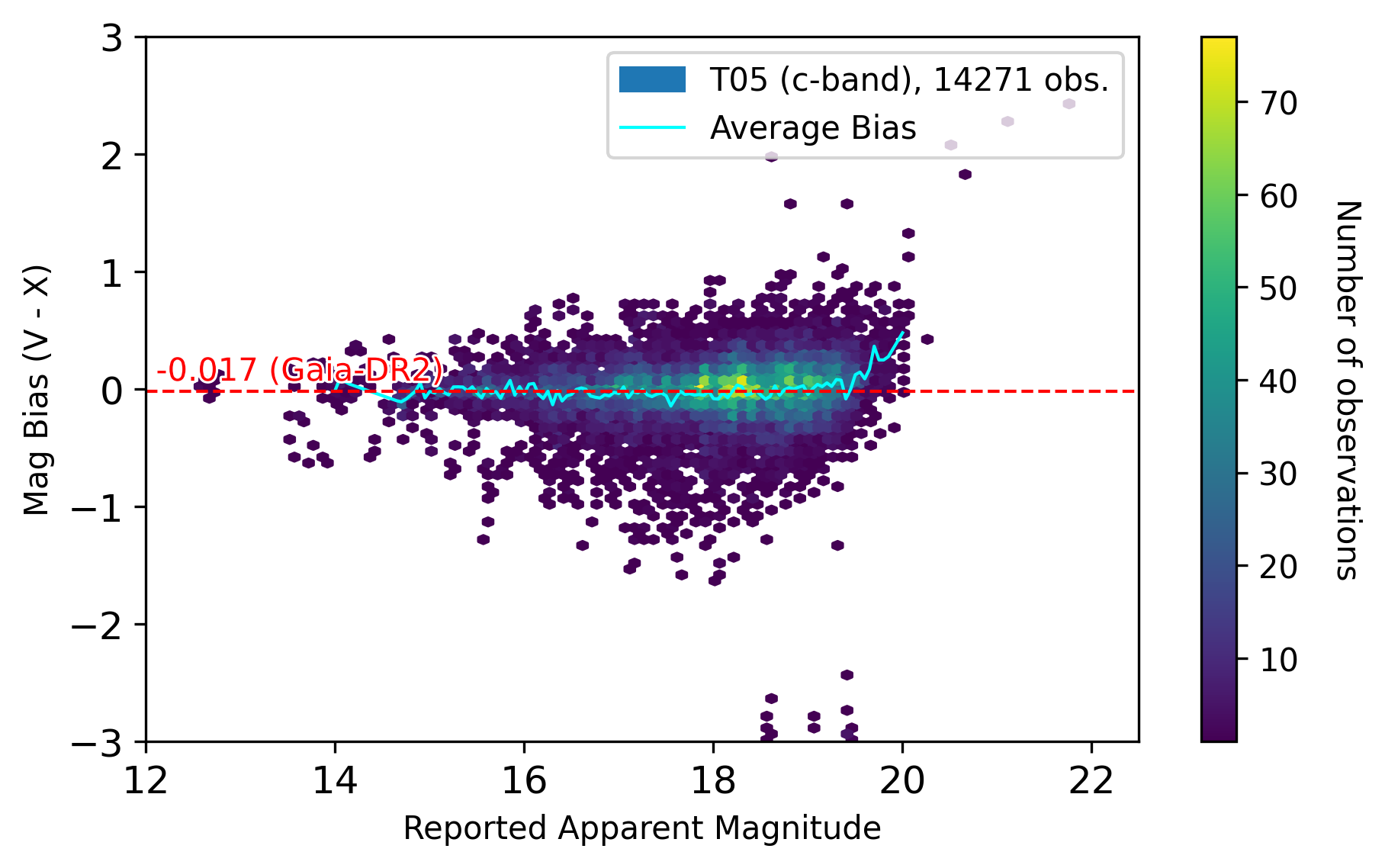}
    \end{subfigure}
    \hfill
    \begin{subfigure}[b]{0.49\linewidth}
        \centering
        \includegraphics[width=\textwidth, trim=0mm 0mm 0mm 2mm,clip]{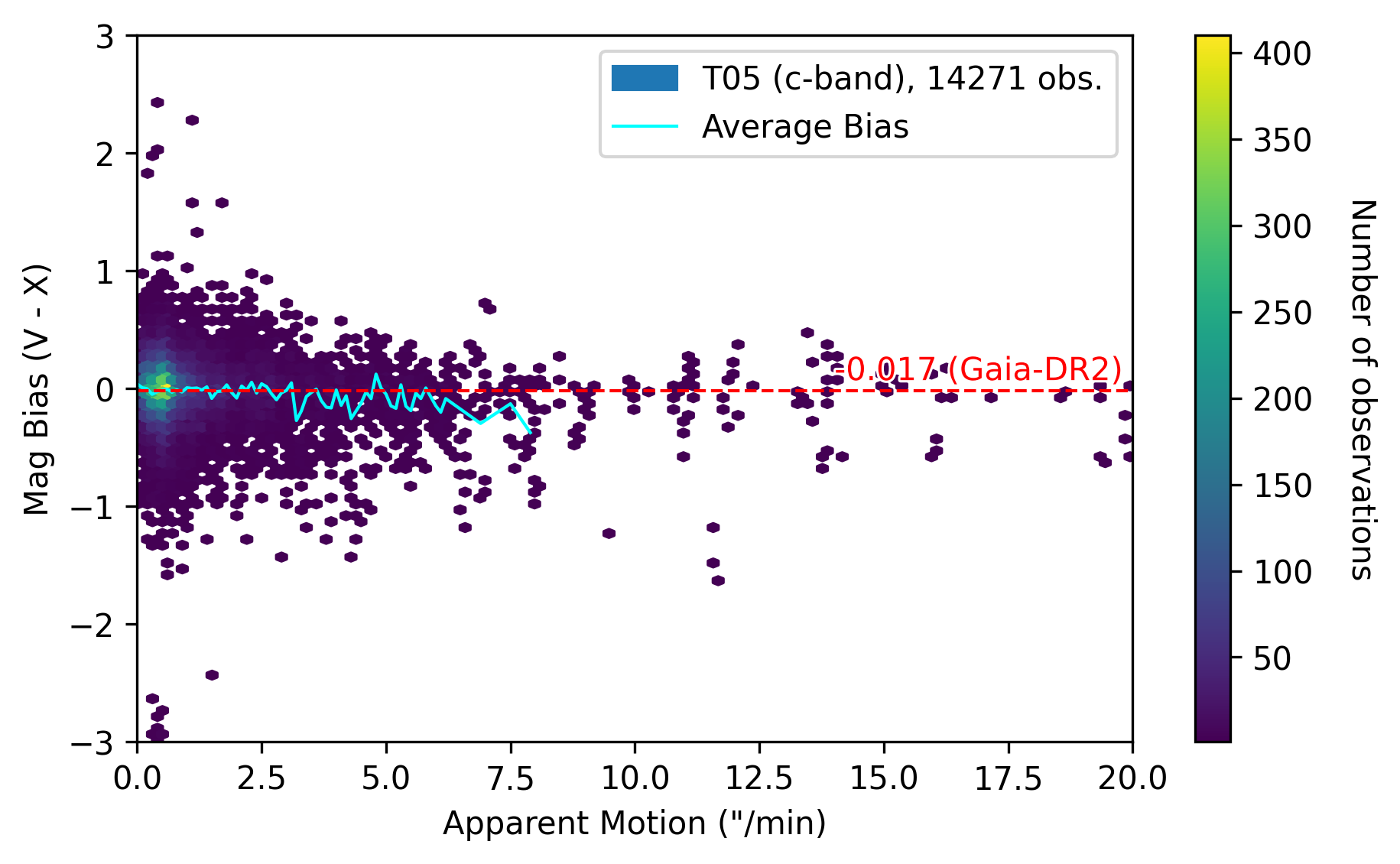}
    \end{subfigure}

    \begin{subfigure}[b]{0.49\linewidth}
        \centering
        \includegraphics[width=\textwidth, trim=0mm 0mm 0mm 2mm,clip]{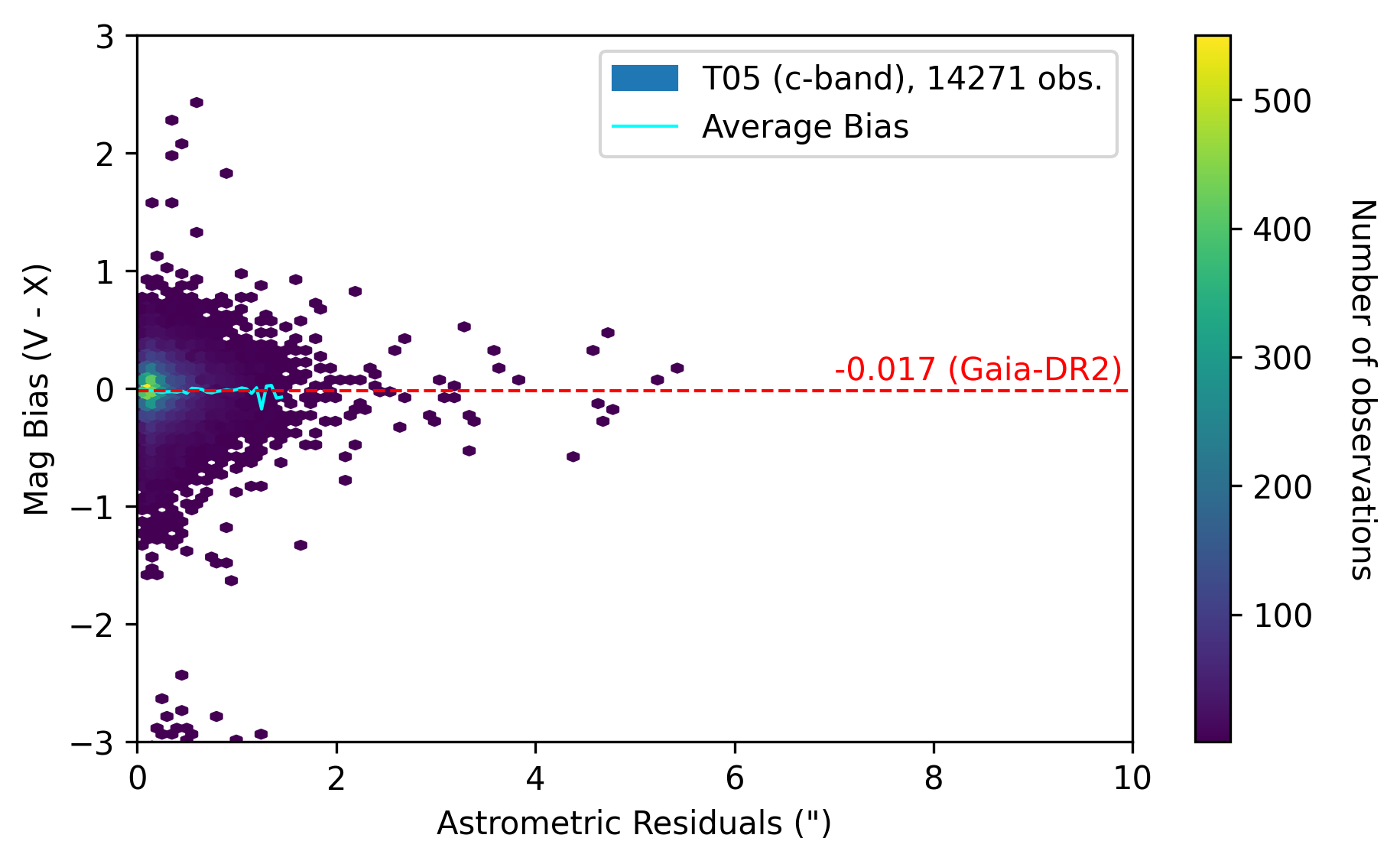}
    \end{subfigure}
    \hfill
    \begin{subfigure}[b]{0.49\linewidth}
        \centering
        \includegraphics[width=\textwidth, trim=0mm 0mm 0mm 2mm,clip]{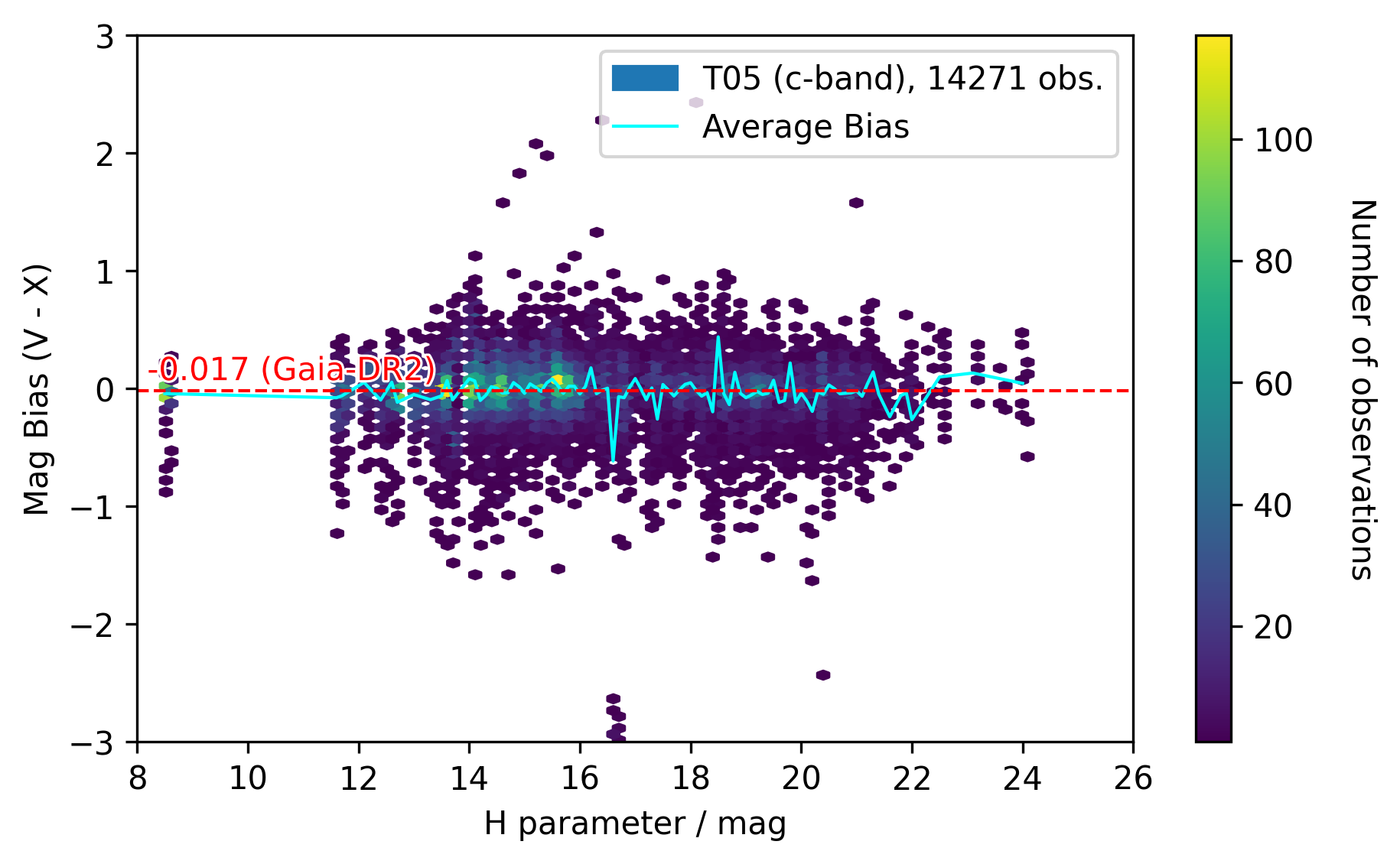}
    \end{subfigure}

    \begin{subfigure}[b]{0.49\linewidth}
        \centering
        \includegraphics[width=\textwidth, trim=0mm 0mm 0mm 2mm,clip]{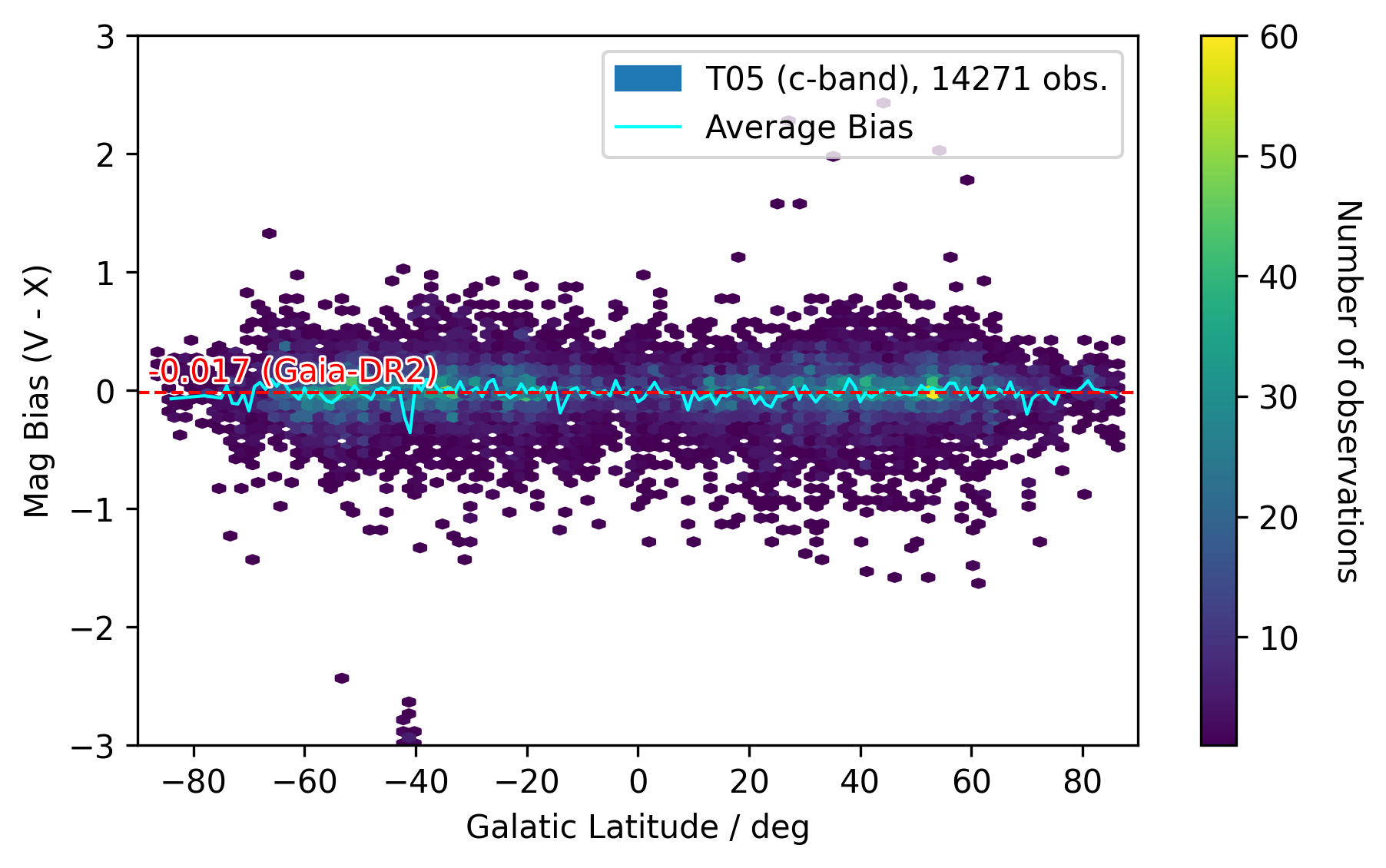}
    \end{subfigure}
    \hfill
    \begin{subfigure}[b]{0.49\linewidth}
        \centering
        \includegraphics[width=\textwidth, trim=0mm 0mm 0mm 2mm,clip]{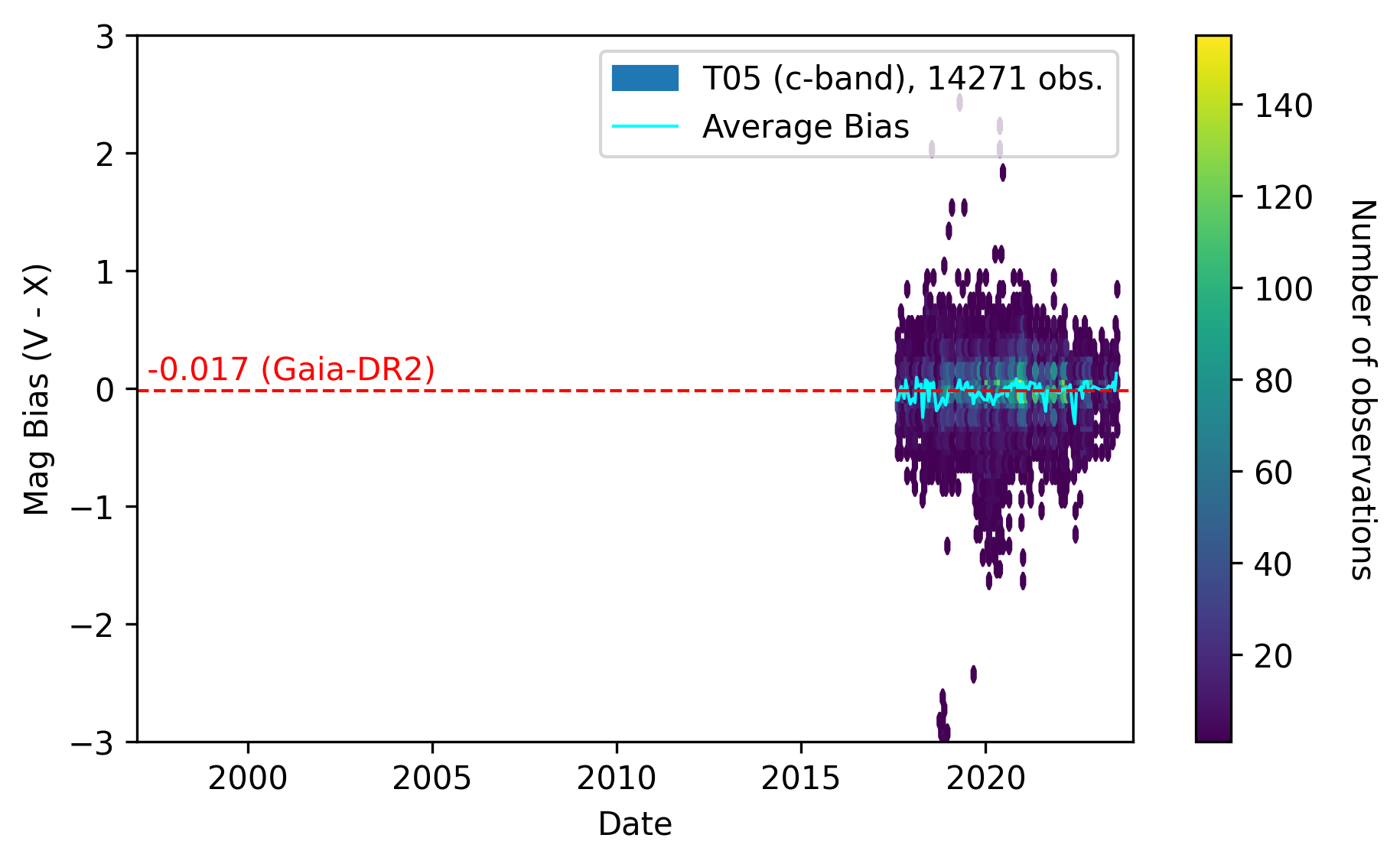}
    \end{subfigure}

    \caption{Dependence of the apparent magnitude and motion, astrometric residuals, absolute magnitude, galactic latitude and time on the photometric bias for T05 in c-band.}
    \label{fig:dis.advanced.T05c}
\end{figure*}

The \textbf{apparent magnitude} is a proxy on how well the observatory could measure the object; a high apparent magnitude corresponds to a low \ac{SNR} of the measurements and vice versa. In Fig.~\ref{fig:dis.advanced.mag} we can see that for observatory code 704 there is an increase of bias with increasing apparent magnitude. Code F51 has no such correlation and the results also match well with the derived \textit{aBCO} correction. The same can be seen for codes T05 and T08, however, for apparent magnitudes larger than 19 there is an increase of bias, meaning that low \ac{SNR} observations have a higher bias with the objects being reported brighter than they should be. This can be explained by the selection effect by \citet{Jedicke2002}, as a selection bias is introduced due to better observability towards rotation amplitude maxima near the limiting magnitude of a telescope. This might result in a systematic $H$ error in cases where most observations were obtained near the technical limit. 

The \textbf{apparent motion} of an object is a metric for potential effects of trailing on the images as exposure times could have been too high in order to have the asteroid as a point in the image, therefore potential issues could occur in the brightness measurement. In general, high fluctuations can be seen in Fig.~\ref{fig:dis.advanced.motion} for most observatories as most observations are in the region of up to 1''/min, except for code C57, where there is also a large proportion of observations up to 3''/min. Also other stations, like T08, do not have large dependencies of bias on the apparent motion. Code F51 on the other hand has the tendency to have an increasingly negative bias for high motions, thus objects being reported too faint. This might indicate trailing issues for some observations; if an ordinary point-spread-function is assumed for astro-photometric derivation, the function would cut out parts of the elongated trail of an object in the image, which leads to an alleged decrease of brightness. However, there are recent methods using aperture photometry to reduce errors in the measurement of trailed images \citep{Devogele_2024}.

The \textbf{astrometric residuals} of the observations are a metric for potential general mistakes in the measurement. This could be also due to trailing or an inaccurate measurement. In Fig.~\ref{fig:dis.advanced.resid} we can see that many of the observatories have a tendency for an increasingly negative bias for higher astrometric residuals, especially codes 704 and F51, which match well with the trends from the apparent motion discussion and could be due to trailing. The results from codes T05, T08, G96 and C57 show no effect on the bias, even though the latter has a quite large range of astrometric errors compared to the others. Overall, it can be seen that code C57, despite having significant errors in the astrometry, does not seem to have systematic deviations in the astro-photometry\footnote{apart from a small proportion of large outliers around $-3$ mag bias.}. 

The \textbf{absolute magnitude} of an object corresponds (inversely) to the size of an object and could therefore be a metric to indicate whether the object's sizes have an influence on the photometric bias. Among all analyzed observatory sites in Fig.~\ref{fig:dis.advanced.size} there is no significant effect visible, but it is notable that there is a lack of data especially for small sized objects ($H>22$).

The \textbf{galactic latitude} of an observation is a metric for the density of background stars. In the center of the Milky Way (low galactic latitude between $\pm10^\circ$) there is a large density of stars, which impedes measurements of asteroids. Fig.~\ref{fig:dis.advanced.glx} shows that most observatories do not have many observations towards the galactic center, even some stations (codes 703 and F51) do not observe in these regions of the sky at all. Code 704 does have some observations in the galactic center, but it is noticeable that there is a change in bias comparing both galactic hemispheres - the upper hemisphere has a bias around $0$ mag, while the lower hemisphere has a bias around $-0.5$ mag. Between both hemispheres at the galactic center, there is a slight gradient visible. The shift of bias for both hemispheres could indicate a systematic error influenced by the photometric catalog that was used, in this case the USNO-A1.0 and A2.0 \citep{Monet1996,Monet1998}, which both surveys utilized two separate calibrations for the northern and southern sky. For other sites, e.g. for code G96, T05 or T08, which are mostly using the more recent Gaia catalog, no such effect is discernible there.

The display of the \textbf{evolution} of the bias \textbf{over time} as shown in Fig.~\ref{fig:dis.advanced.time} is a measure to monitor long-term effects, but also abrupt changes. For codes 703 and G96 there is a significant step visible happening during the summer of 2018, where the bias drops by more than $0.3$/$0.1$ mag respectively. This drop perfectly corresponds to the switch from Gaia DR1 to DR2 as reference star catalog and matches well with the corresponding \textit{aBCO} corrections obtained for the sites. This shows that effects that result from different catalogs being used are well covered by our algorithm.
However, the increase in the bias for code G96 starting after the summer of 2022, even though still using Gaia DR2, is not included in the corrections. Also, more gradual effects like the steep increase of bias for code 704 in its first years of measurement up to the year 2000 and the consequent shallow decrease up to the year 2006 can be seen in the analysis. Also code F51 had high fluctuations in its first years of measurement, having a rather negative bias then, but afterwards it has, as well as code 704, a quite constant bias.\footnote{with the exception of down-times during 2018 and 2022.} For code T05 and T08 there is no change of bias visible over time.\footnote{with the exception of a lack of measurements during 2022.}

\subsection{Additional points}
The analysis in Sec.~\ref{sec:res.bias} shows that multiple of the mainly used color bands in the field of asteroid observation are systematically biased (e.g., V-band by around $0.1$ mag). A bias of this magnitude leads to a bias in the size of an asteroid by factor of $5\%$, using 
\begin{align}
    D = 1.329 \cdot 10^6 \cdot p_{\text{V}}^{-1/2} \cdot 10^{-0.2H}
\end{align}
by \citet{Harris2002} as an estimate for the size $D$ (in meters) based on the albedo in V-band $p_{\text{V}}$ and the absolute magnitude $H$. However, the uncertainties of the $H$ parameter (and other factors such as the albedo) are larger. Nevertheless, we achieve a $28\%$ reduction in the $H$ parameter uncertainties with the new correction system, leading to a better size estimate.

We showed that the method used is valid for determining biases as they appear for bands, catalogs and observatories known for photometric errors, while expectedly good observations match well with our results (cf. Sec.~\ref{sec:res.bias}). 

The correction algorithm with the statistical t-test in a successive approach turns out to minimize the number of necessary corrections and select the most relevant ones (cf. Sec.~\ref{sec:res.correction}). For example, multiple observatory corrections can be summarized in one catalog or even one band correction. With that, only biases deviating from the average results receive their own correction.

It turns out that there are some highly significant bias results with much underlying data and high deviations to the previous correction. These have the largest individual impact on the results. On the other hand, there are many lower significant results with less data and only slight differences. These might be giving questionable results or not having large individual impact, but there could still be reasonable results among them, where we might miss their impact, when not including them. Finding the optimum point for the minimum RMS based on the different confidence levels gives us the opportunity to include all relevant correction, but not to loose too many reasonable results at the same time.

Overall, the whole potential of the dataset is not currently entirely being used; there are other factors (as shown in Sec.~\ref{sec:disc.adv}) that have an influence on the bias and there is also still the option in the correction format to include further parameters than the band-catalog-observatory criteria (cf. Sec.~\ref{sec:methods.correction}). Also, the debiasing would further improve over time and with a larger database for accurate $H$-$G$ model parameters, leading to more observations to analyze and thus allowing more refined statistics and more specific parameters to analyze for future works \citep{Veres2015,Waszczak2015,Colazo2021,Mahlke2021}.

\section{Conclusion}
In this paper we determined significant biases for the major color bands used for minor planet observations, showing that there is a large variation in accuracy in the field of astro-photometry. We were able to quantify the systematic deviations in different color band, reference star catalogs and for specific observatories and give reasons for some of the occurring issues. Therefore, we propose to use band-, catalog- and observatory specific bias corrections. Based on these bias results, we were able to develop a novel statistical correction algorithm and generated a correction system in order to debias all the astro-photometric observations. This led to a $36\%$ decrease in the RMS of the asteroids' phase curve compared to current methods, shown for an independent set of objects. These were also used to validate our method for deriving the absolute magnitude $H$. With the newly corrected observations we were able to improve the goodness of fit statistics by $94\%$ and reduce the uncertainties for $H$ by $28\%$ and thus improving size estimates. These results can be further improved in future studies by including other more specific corrections and using a photometric weighting system based on the deviations of observatories obtained in this study.
All in all, there is only a small proportion of objects in the solar system with accurately measured photometry, therefore astro-photometry is the best guess for most objects, especially in time-critical domains, as planetary defense with imminent impactors, or large data models in Solar System Science. Therefore, the new method can be beneficial in estimating the physical properties and their accuracy of almost all minor planets. 

\section*{Acknowledgements}
TH thanks the European Space Agency for the possibility of a student internship, its financial support, computing infrastructure and the OPS-SP team for technical support and feedback. TH gratefully acknowledges financial support from the Erasmus+ program, co-funded by the European Union, and the Cusanuswerk Bischöfliche Studienförderung, funded by the German Federal Ministry of Education and Research. 
Part of this research was conducted at the Jet Propulsion Laboratory, California Institute of Technology, under a contract with the National Aeronautics and Space Administration (80NM0018D0004).
The work at Ond\v{r}ejov Observatory was supported by the Grant Agency of the Czech Republic, Grant 23-04946S.
This research has made use of data and/or services provided by the International Astronomical Union's Minor Planet Center. 
\section*{Author Contributions}
\textbf{Tobias Hoffmann}: Methodology, Software, Formal analysis, Investigation, Writing - Original Draft, Visualization 
\textbf{Marco Micheli}: Conceptualization, Resources, Writing - Review \& Editing, Supervision
\textbf{Juan Luis Cano}: Conceptualization, Writing - Review \& Editing, Supervision, Project administration
\textbf{Maxime Devogèle}: Validation, Resources, Writing - Review \& Editing
\textbf{Davide Farnocchia}: Conceptualization, Methodology, Validation, Writing - Review \& Editing, Supervision
\textbf{Petr Pravec}: Conceptualization, Resources, Writing - Review \& Editing,
\textbf{Peter Vereš}: Resources, Writing - Review \& Editing
\textbf{Björn Poppe}: Writing - Review \& Editing, Supervision, Project administration, Funding acquisition

\section*{Competing Interests}
The authors declare that they have no competing financial interests.
\section*{Data Availability}
This work makes use of the following publicly available data:
Photometry database from the Ondrejov NEO Program \citep{Pravec2012}, Data from the \acs{IAU} Minor Planet Center (\url{https://www.minorplanetcenter.net/data}), \acs{ESA}'s NEO Coordination Centre (\url{https://neo.ssa.esa.int/}) and NASA's \acl{JPL} (\url{https://ssd.jpl.nasa.gov/horizons/}).

\section*{Code Availability}
This work makes use of the following publicly available codes: Pandas \citep{reback2020pandas,mckinney-proc-scipy-2010}, Matplotlib \citep{Hunter:2007}, NumPy \citep{harris2020array}, SciPy \citep{2020SciPy-NMeth}, Astropy \citep{2013A&A...558A..33A, 2018AJ....156..123A}, Astroquery \citep{2019AJ....157...98G}, diptest \citep{Hartigan1985b}.


\bibliographystyle{elsarticle-harv} 
\bibliography{Literature}

\newpage
\appendix
\onecolumn
\section{Correction files}
Correction system based on NEOCC, aB, aBC and aBCO corrections with confidence level $\gamma$ can be found as separate data files, named ``corr\_DePhOCUS\_g[XXX].txt'', with $\gamma=0.90$ being the optimum correction according to the results.

\section{Additional Figures}
\label{app1}

\begin{figure}[htb]
    \centering
    \hfill
    \begin{subfigure}[b]{0.49\textwidth}
        \centering
        \includegraphics[width=\textwidth, trim=0mm 0mm 0mm 0mm,clip]{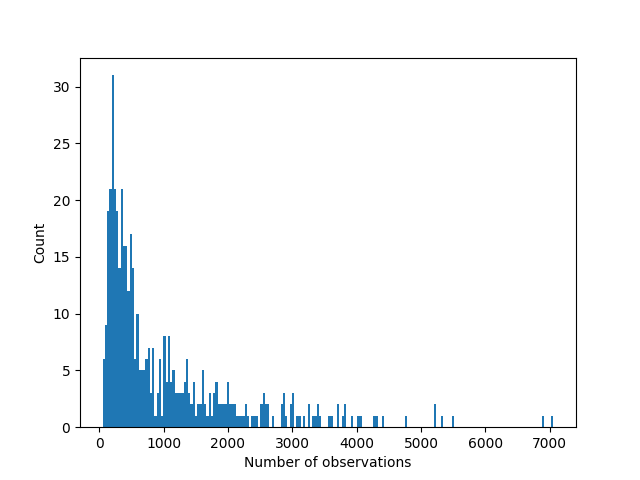}
    \end{subfigure}
    \hfill
    \begin{subfigure}[b]{0.49\textwidth}
        \centering
        \includegraphics[width=\textwidth, trim=0mm 0mm 0mm 0mm,clip]{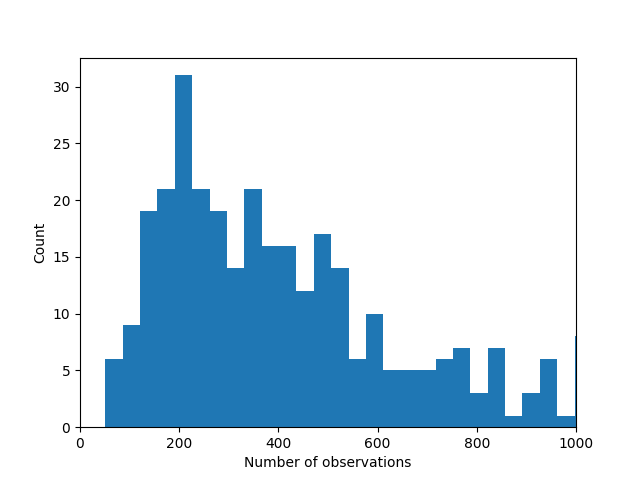}
    \end{subfigure}
    \hfill
    
    \caption{Histogram for the number of observations for each of the 468 asteroids in full range (left) and cut-out up to 1000 observations (right).}
    \label{fig:methods.numberobs}
\end{figure}

\begin{figure}[htb]
    \centering
    \hfill
    \begin{subfigure}[b]{0.49\textwidth}
        \centering
        \includegraphics[width=\textwidth, trim=0mm 0mm 0mm 0mm,clip]{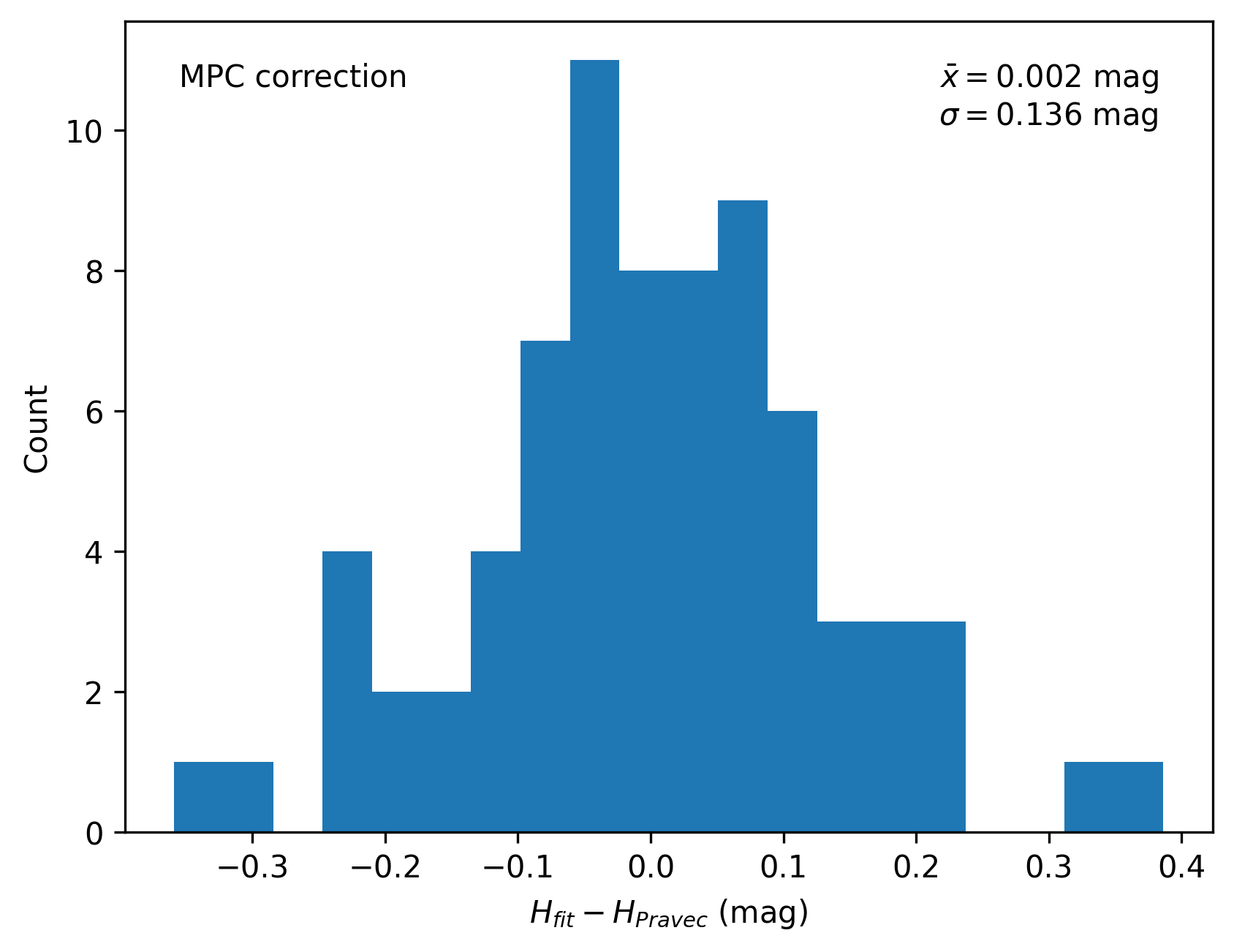}
    \end{subfigure}
    \hfill
    \begin{subfigure}[b]{0.49\textwidth}
        \centering
        \includegraphics[width=\textwidth, trim=0mm 0mm 0mm 0mm,clip]{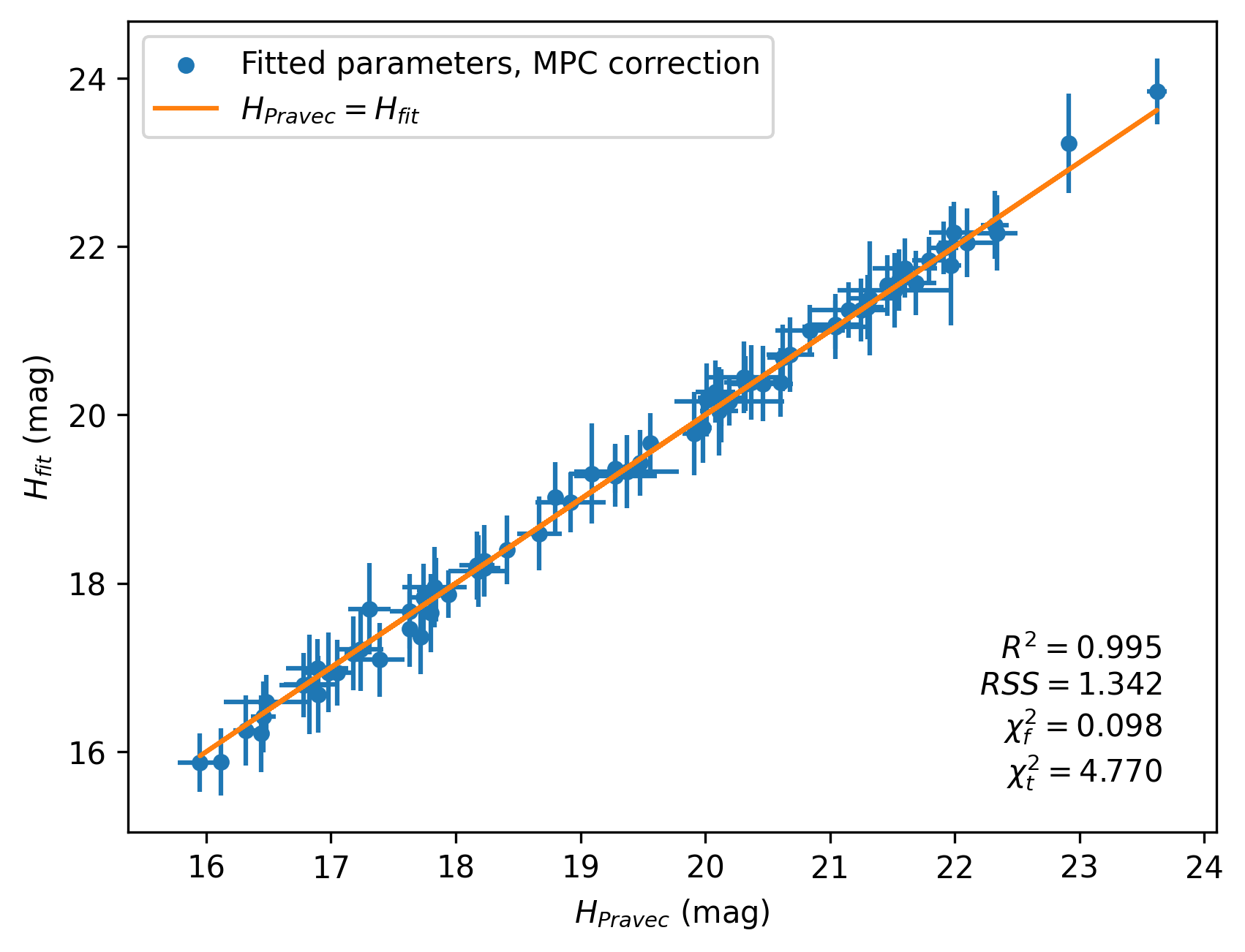}
    \end{subfigure}
    \hfill

    \hfill
    \begin{subfigure}[b]{0.49\textwidth}
        \centering
        \includegraphics[width=\textwidth, trim=0mm 0mm 0mm 0mm,clip]{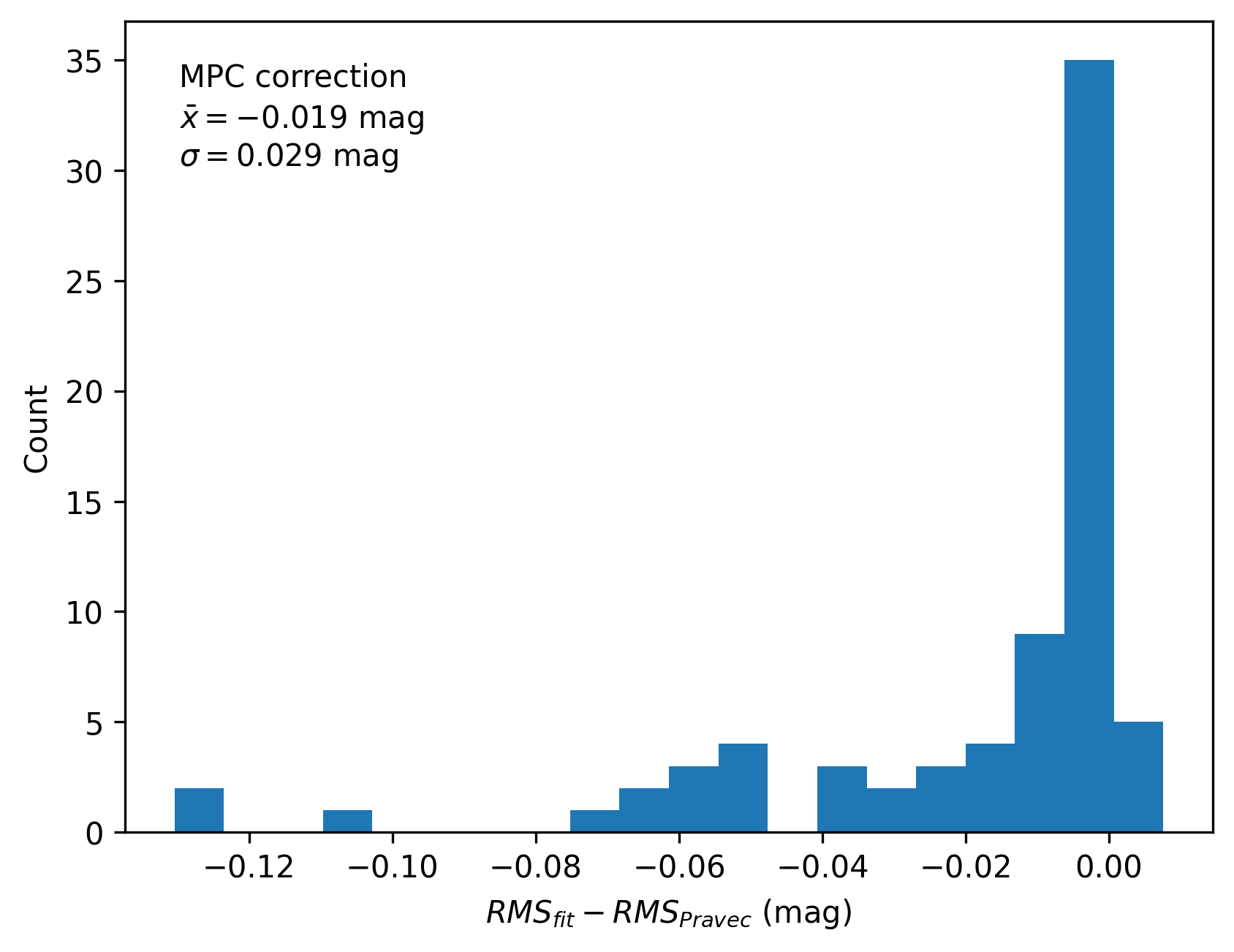}
    \end{subfigure}
    \hfill
    \begin{subfigure}[b]{0.49\textwidth}
        \centering
        \includegraphics[width=\textwidth, trim=0mm 0mm 0mm 0mm,clip]{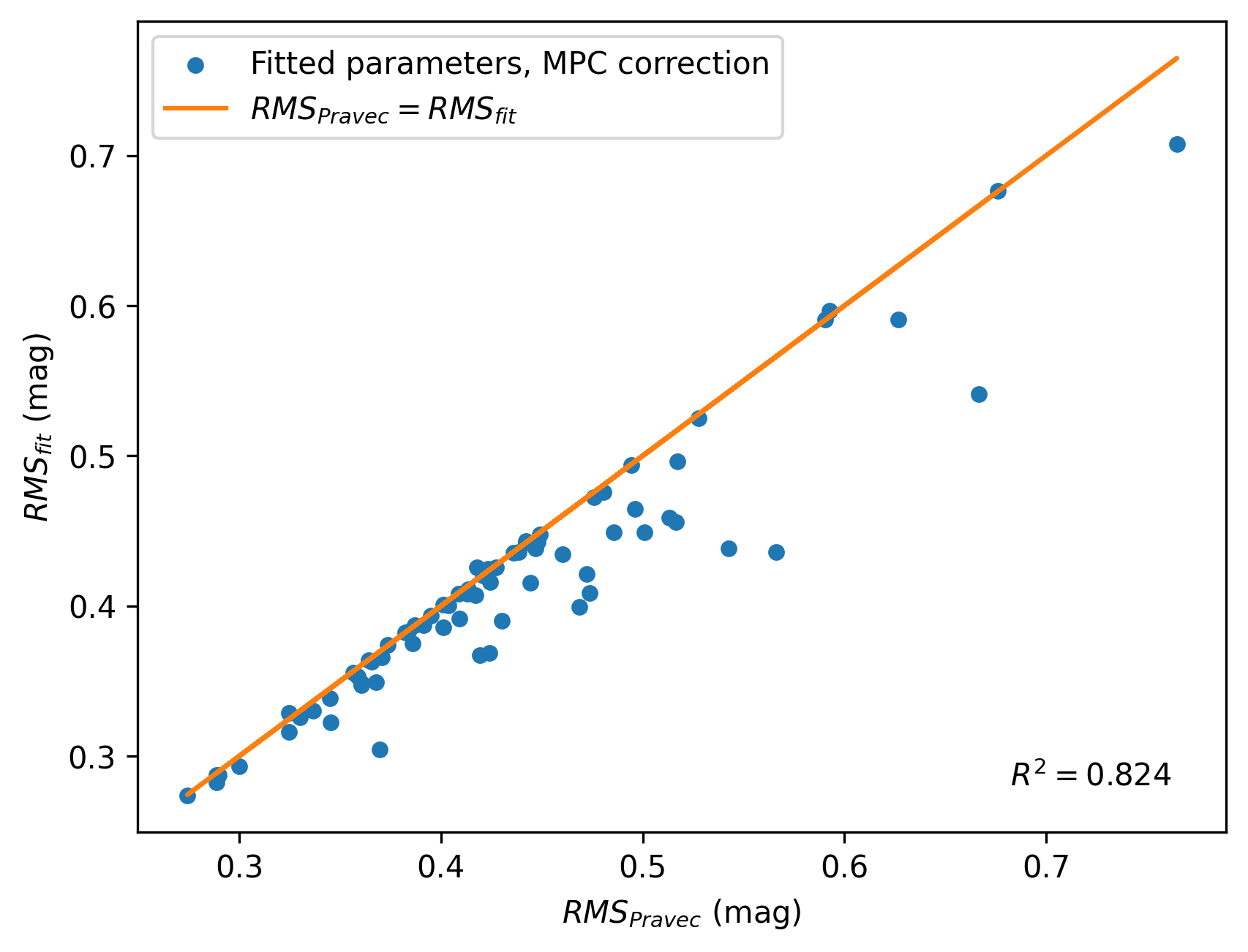}
    \end{subfigure}
    \hfill

    \caption{Histogram of difference between fitted $H$ parameter, after applying MPC correction, and reference values $H_{\text{Pravec}}$ by Pravec et al. (left) with their correlation (right), when keeping $G=G_{\text{Pravec}}$ for 74 asteroids as validation. Also comparing the $RMS$ of measurements by using the fitted parameters or the reference values.}
    \label{fig:res.valid.MPConlyH}
\end{figure}

\begin{figure*}[tp]
    \centering

    \begin{subfigure}[b]{0.32\textwidth}
        \centering
        \includegraphics[width=\textheight, trim=0mm 7mm 0mm 2mm,clip, angle=90]{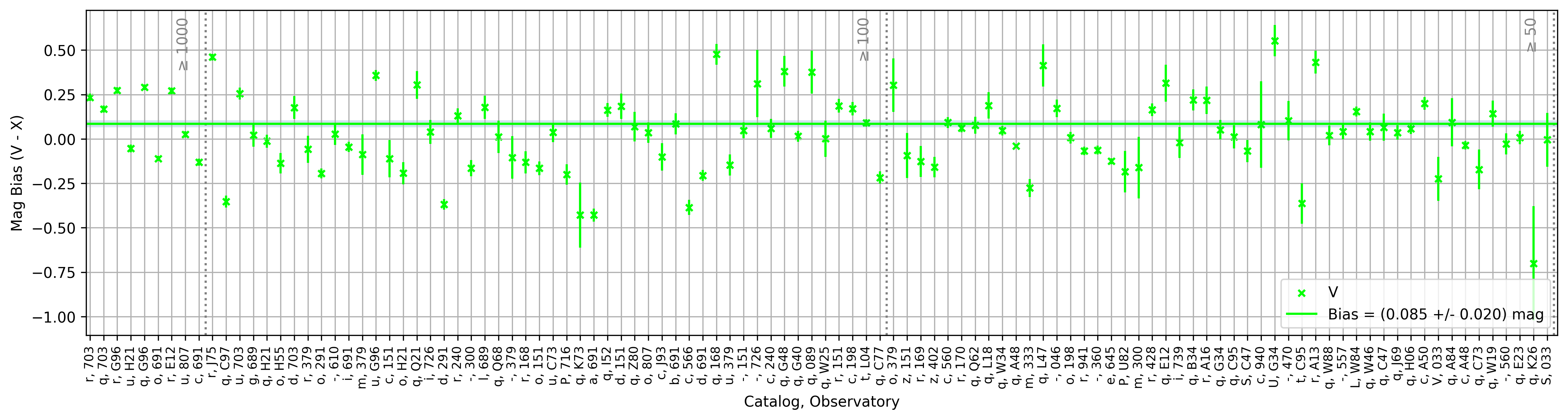}
    \end{subfigure}
    \hfill
    \begin{subfigure}[b]{0.32\textwidth}
        \centering
        \includegraphics[width=\textheight, trim=0mm 7mm 0mm 2mm,clip, angle=90]{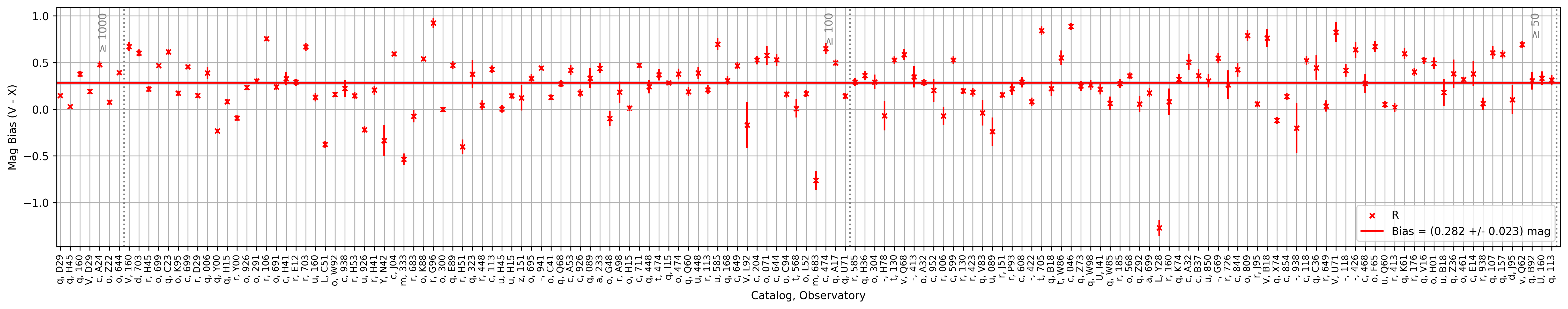}
    \end{subfigure}
    \hfill
    \begin{subfigure}[b]{0.32\textwidth}
        \centering
	\includegraphics[width=0.37\textheight, trim=0mm 7mm 0mm 2mm,clip, angle=90]{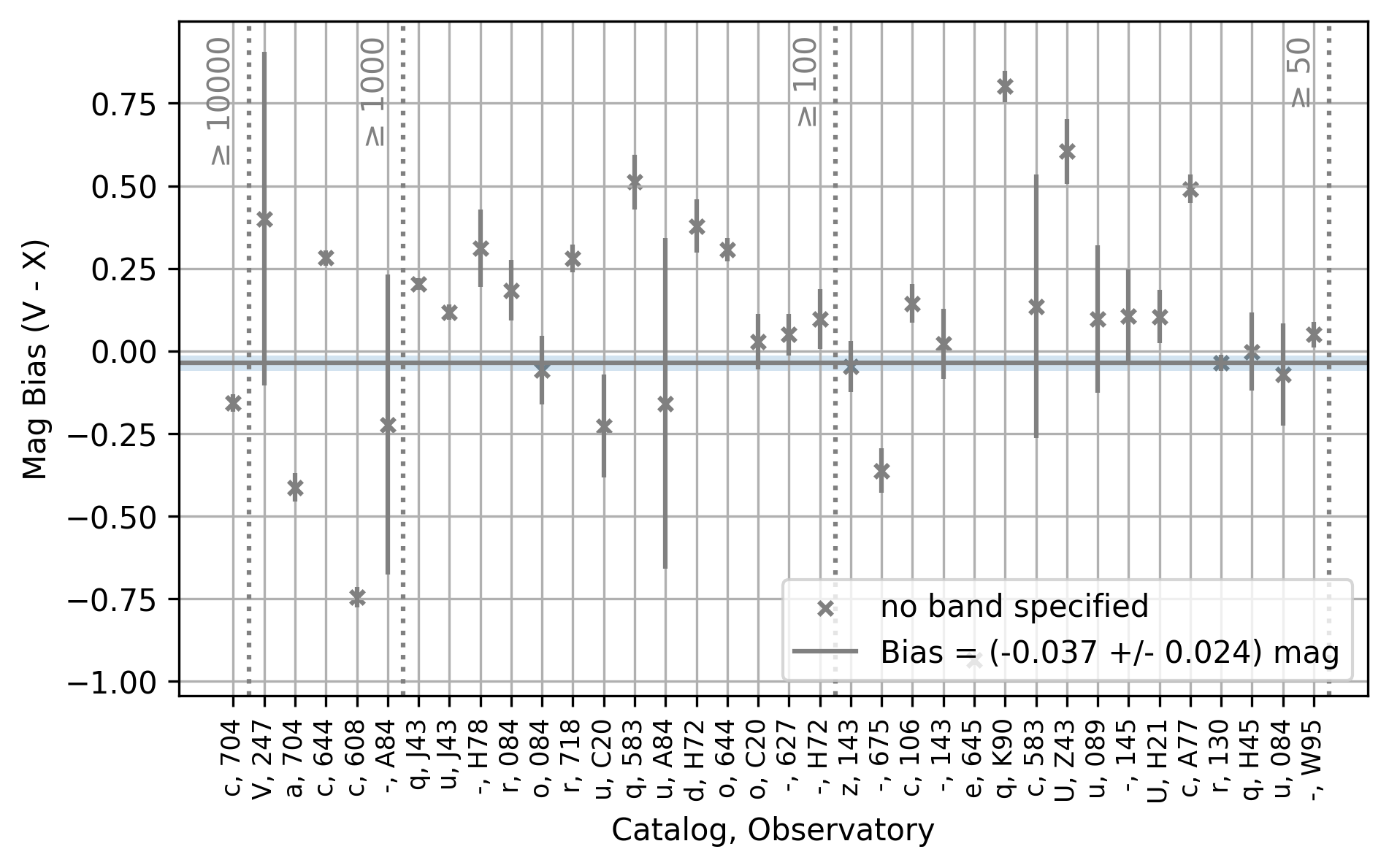}
        \includegraphics[width=0.63\textheight, trim=0mm 7mm 0mm 2mm,clip, angle=90]{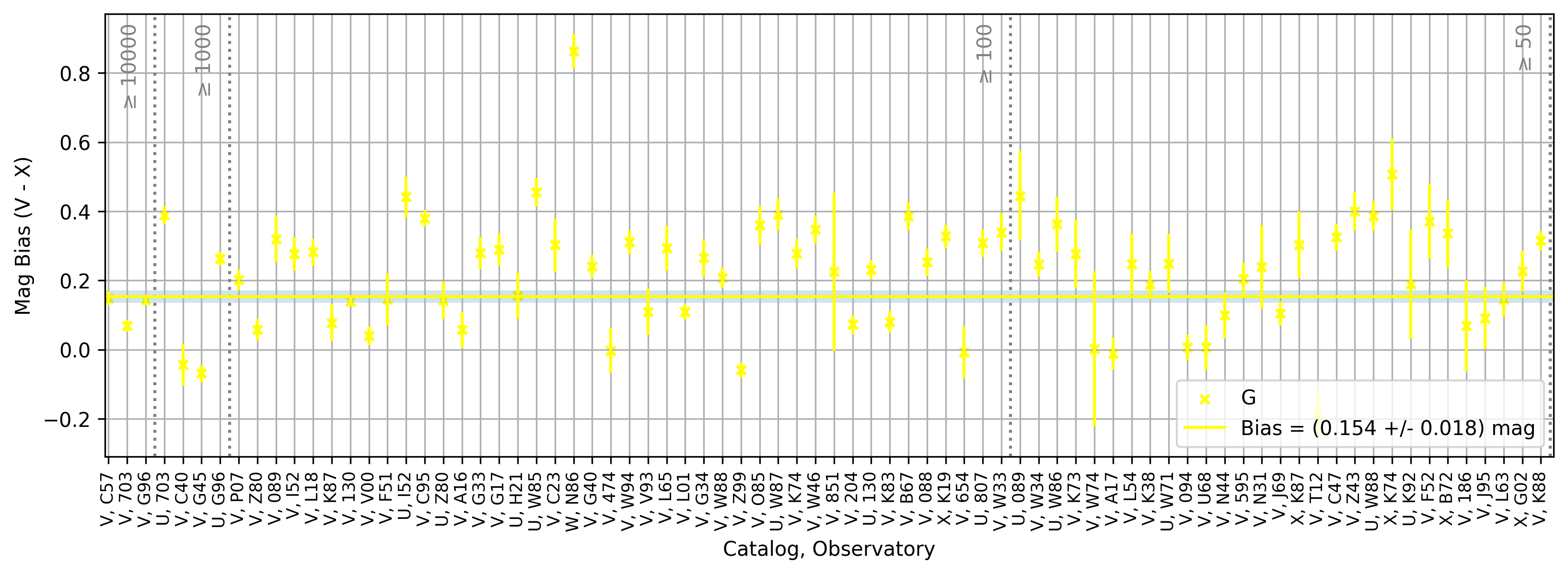}
    \end{subfigure}
    \vspace{2mm}

\raggedleft (Part 1 - continued on next page)
\end{figure*}%
\begin{figure*}[tp]\ContinuedFloat 

    \begin{subfigure}[b]{0.49\textwidth}
        \centering
        \includegraphics[width=\textwidth, trim=0mm 7mm 0mm 2mm,clip]{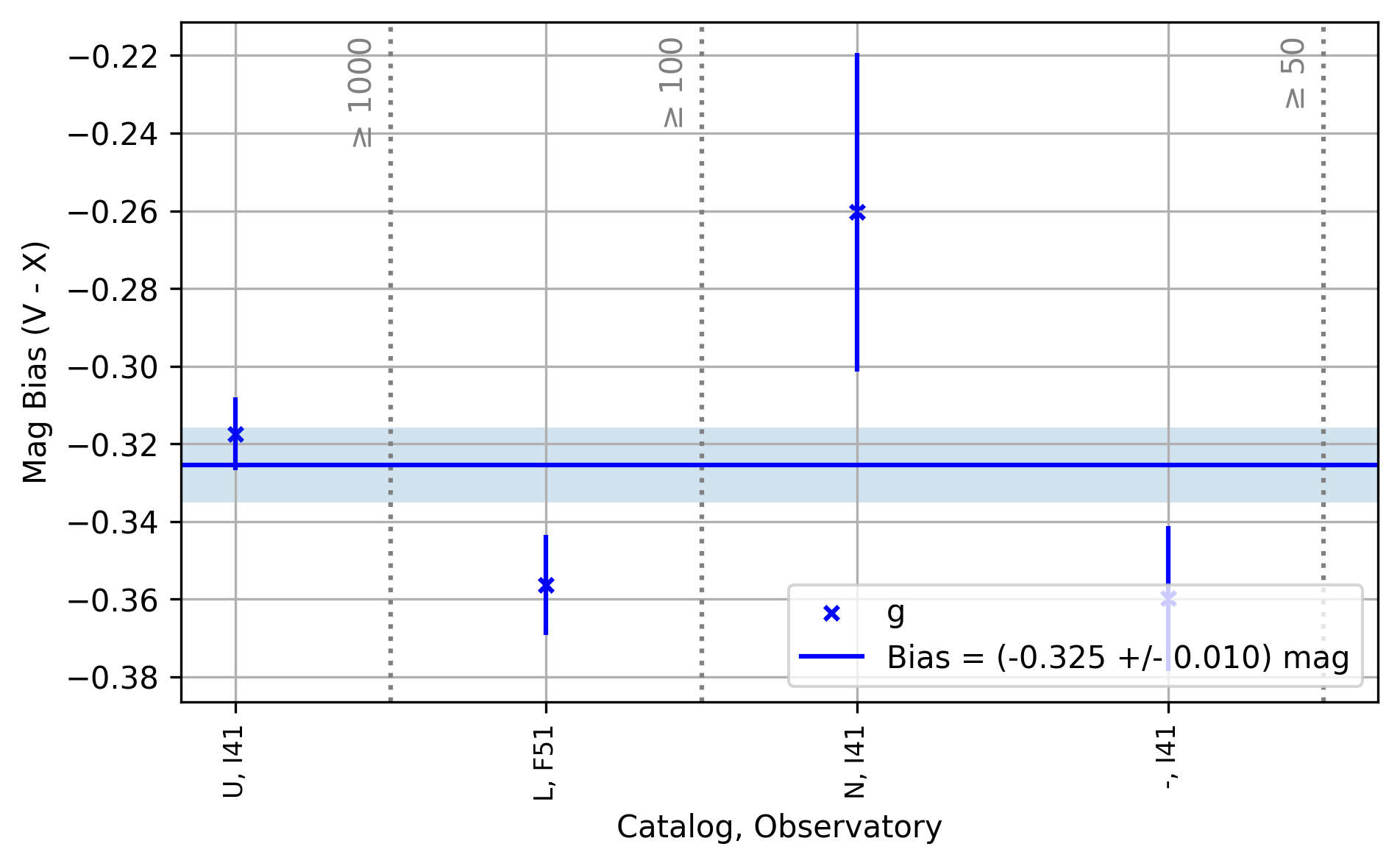}
    \end{subfigure}
    \hfill
    \begin{subfigure}[b]{0.49\textwidth}
        \centering
        \includegraphics[width=\textwidth, trim=0mm 7mm 0mm 2mm,clip]{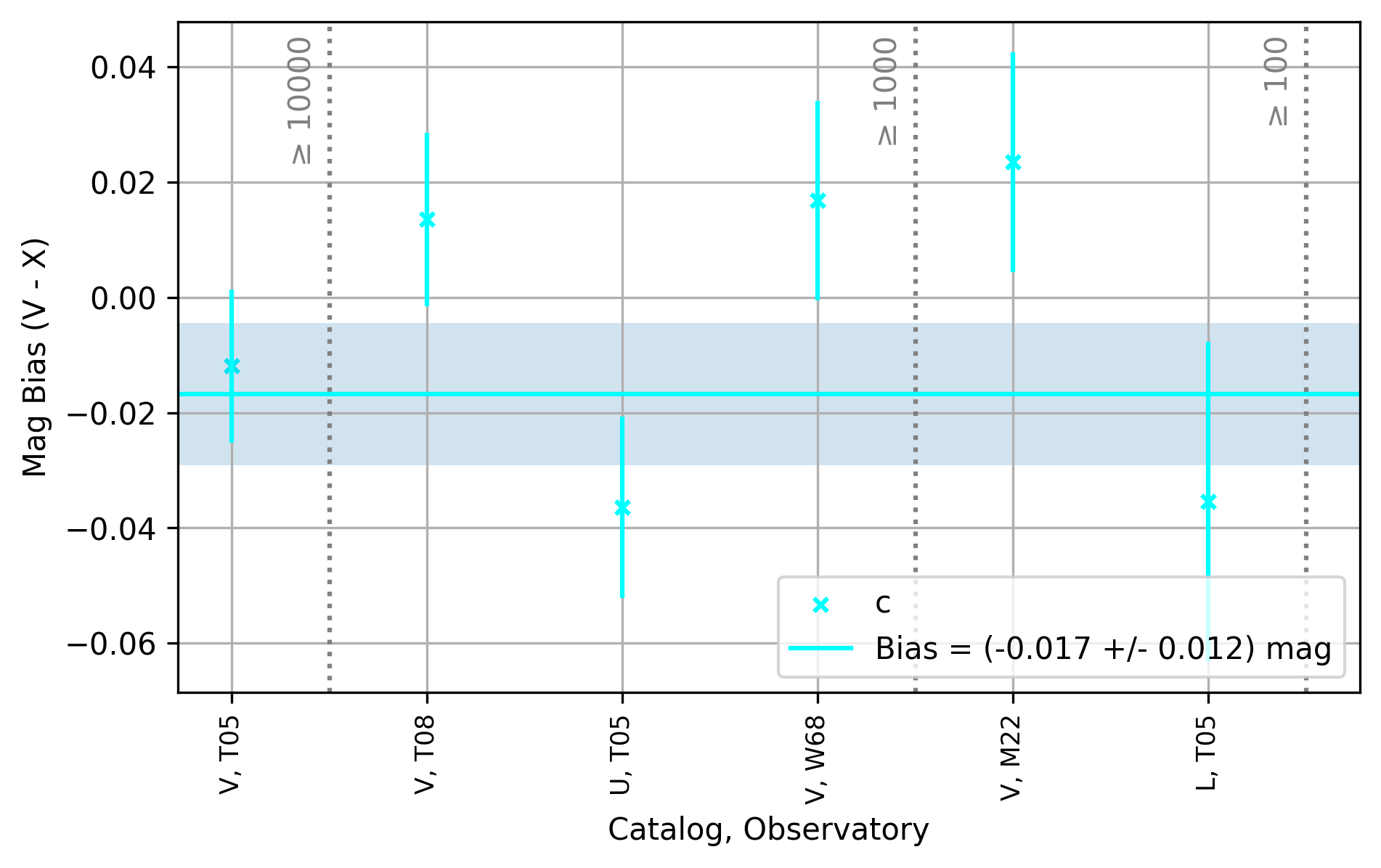}
    \end{subfigure}
    \vspace{2mm}
   
    \begin{subfigure}[b]{0.49\textwidth}
        \centering
        \includegraphics[width=\textwidth, trim=0mm 7mm 0mm 2mm,clip]{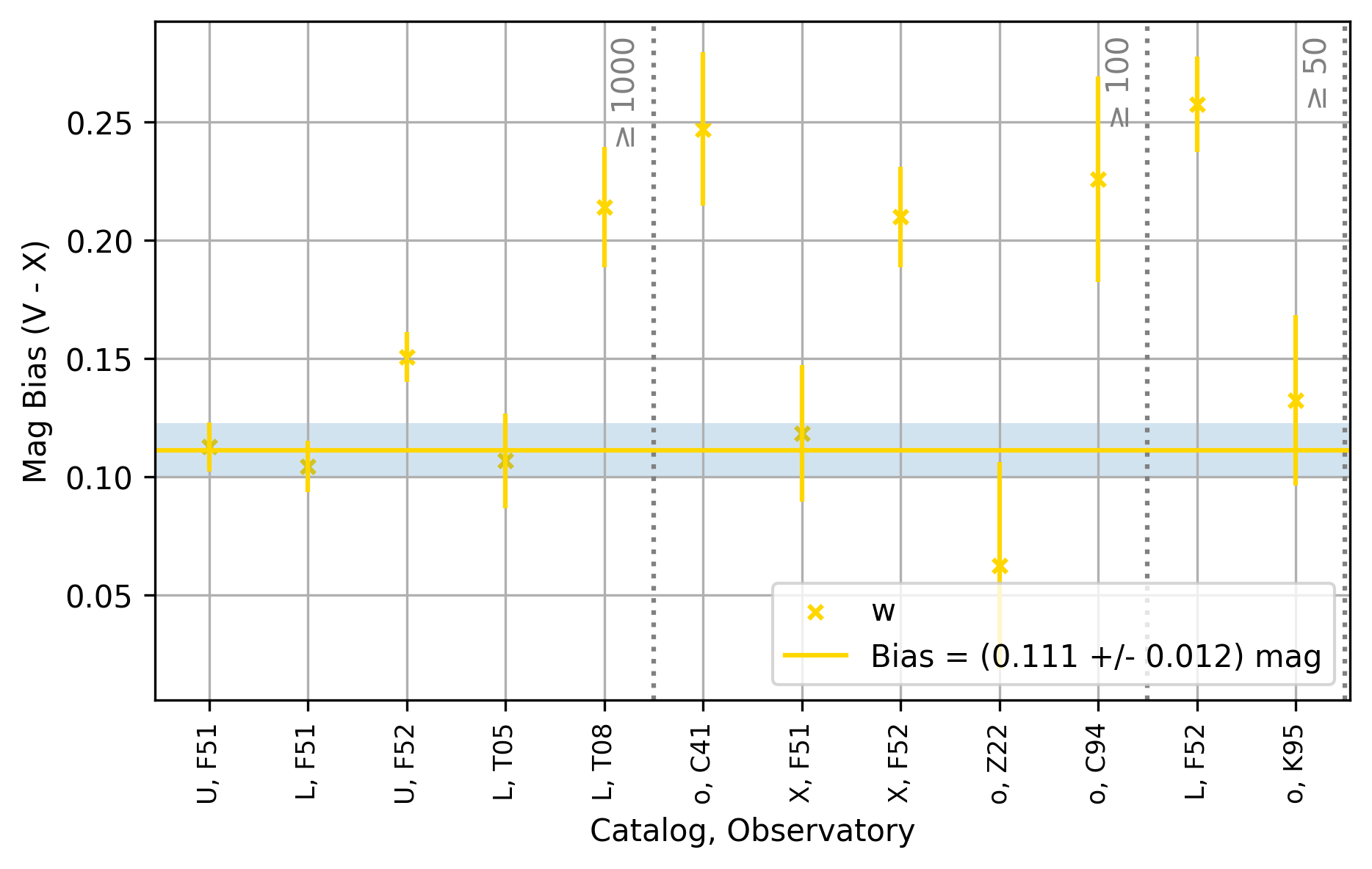}
    \end{subfigure}
    \hfill
    \begin{subfigure}[b]{0.49\textwidth}
        \centering
        \includegraphics[width=\textwidth, trim=0mm 7mm 0mm 2mm,clip]{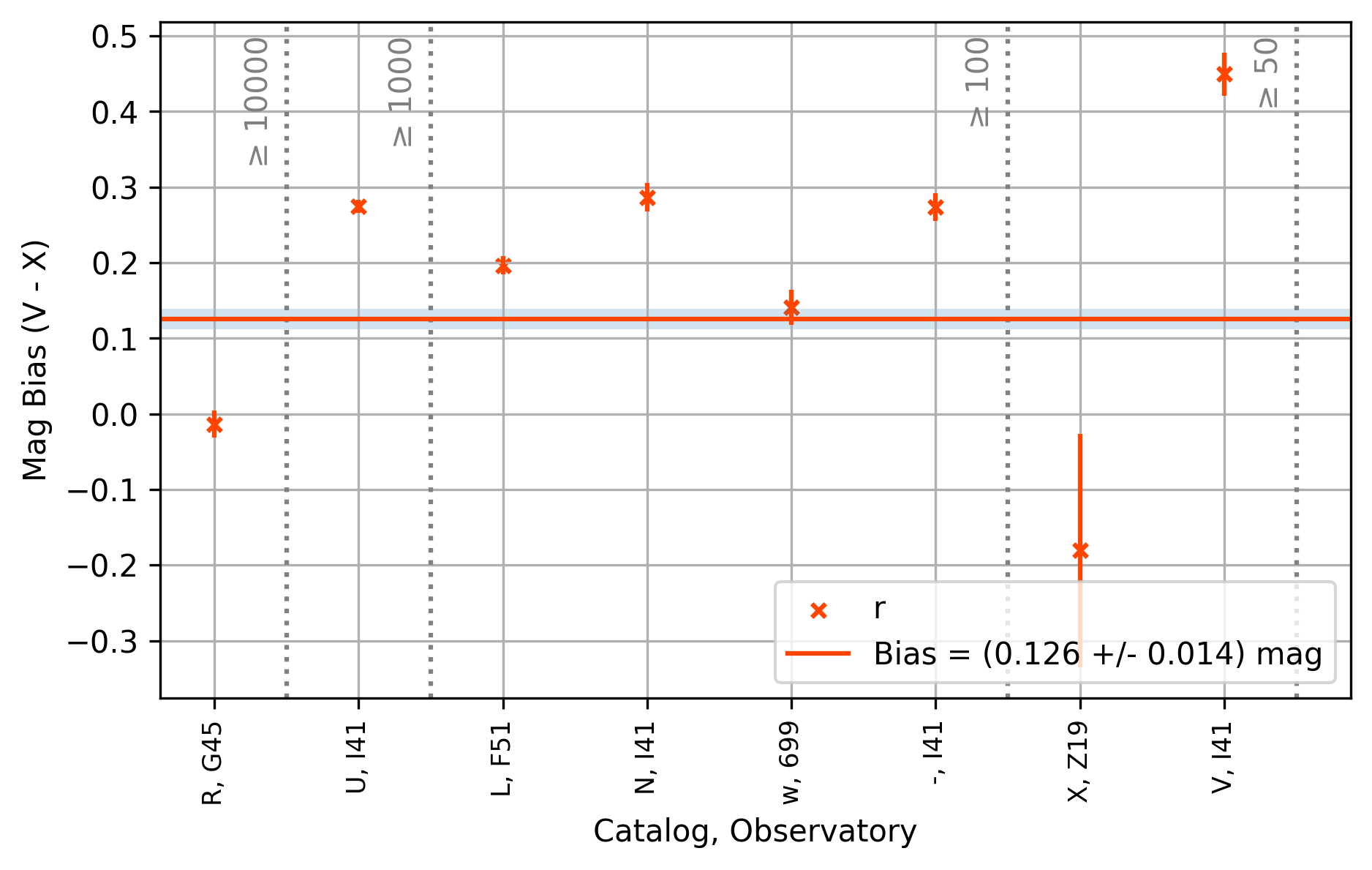}
    \end{subfigure}
    \vspace{2mm}


    \begin{subfigure}[b]{0.49\textwidth}
        \centering
        \includegraphics[width=\textwidth, trim=0mm 7mm 0mm 2mm,clip]{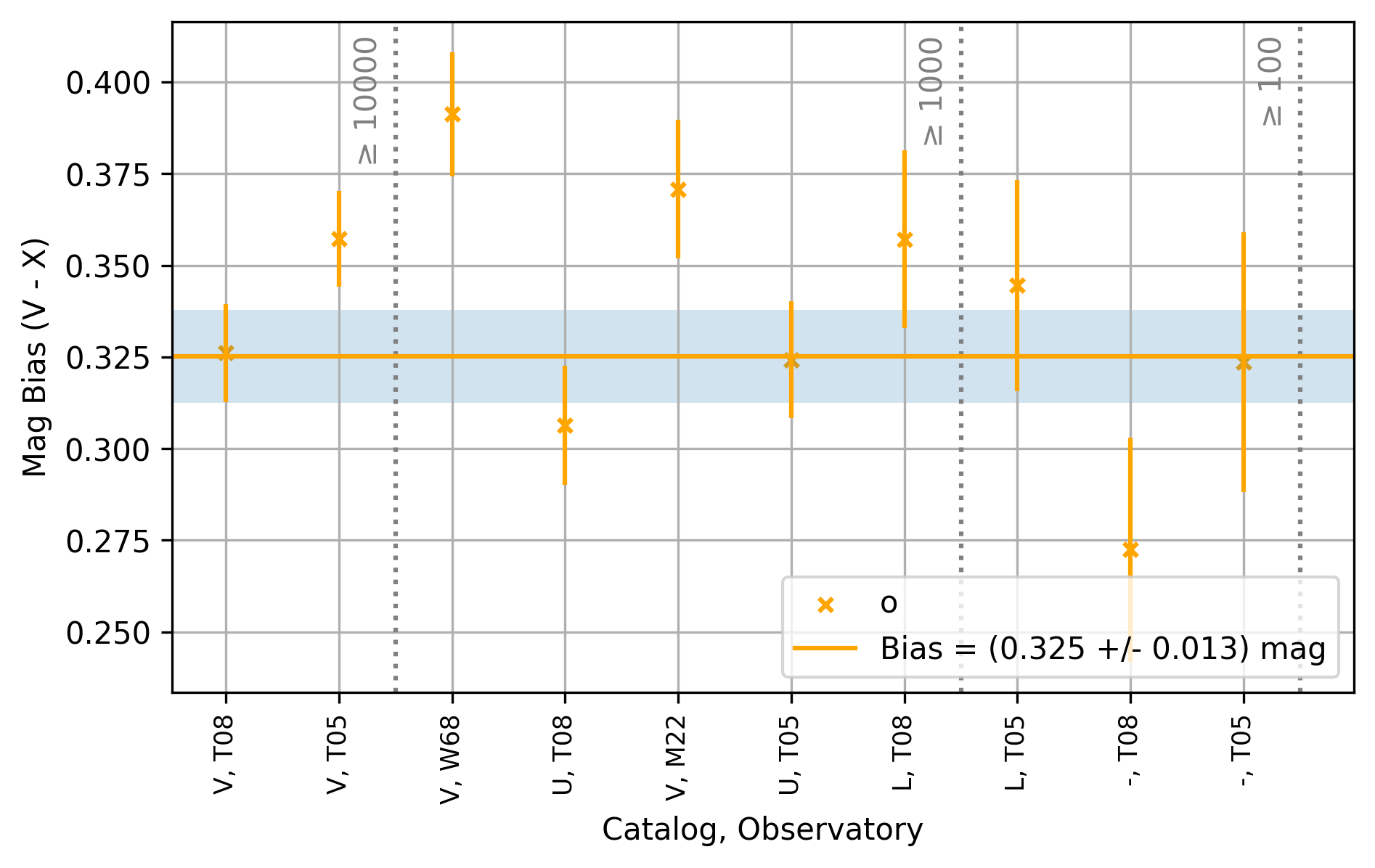}
    \end{subfigure}
    \hfill
    \begin{subfigure}[b]{0.49\textwidth}
        \centering
        \includegraphics[width=\textwidth, trim=0mm 7mm 0mm 2mm,clip]{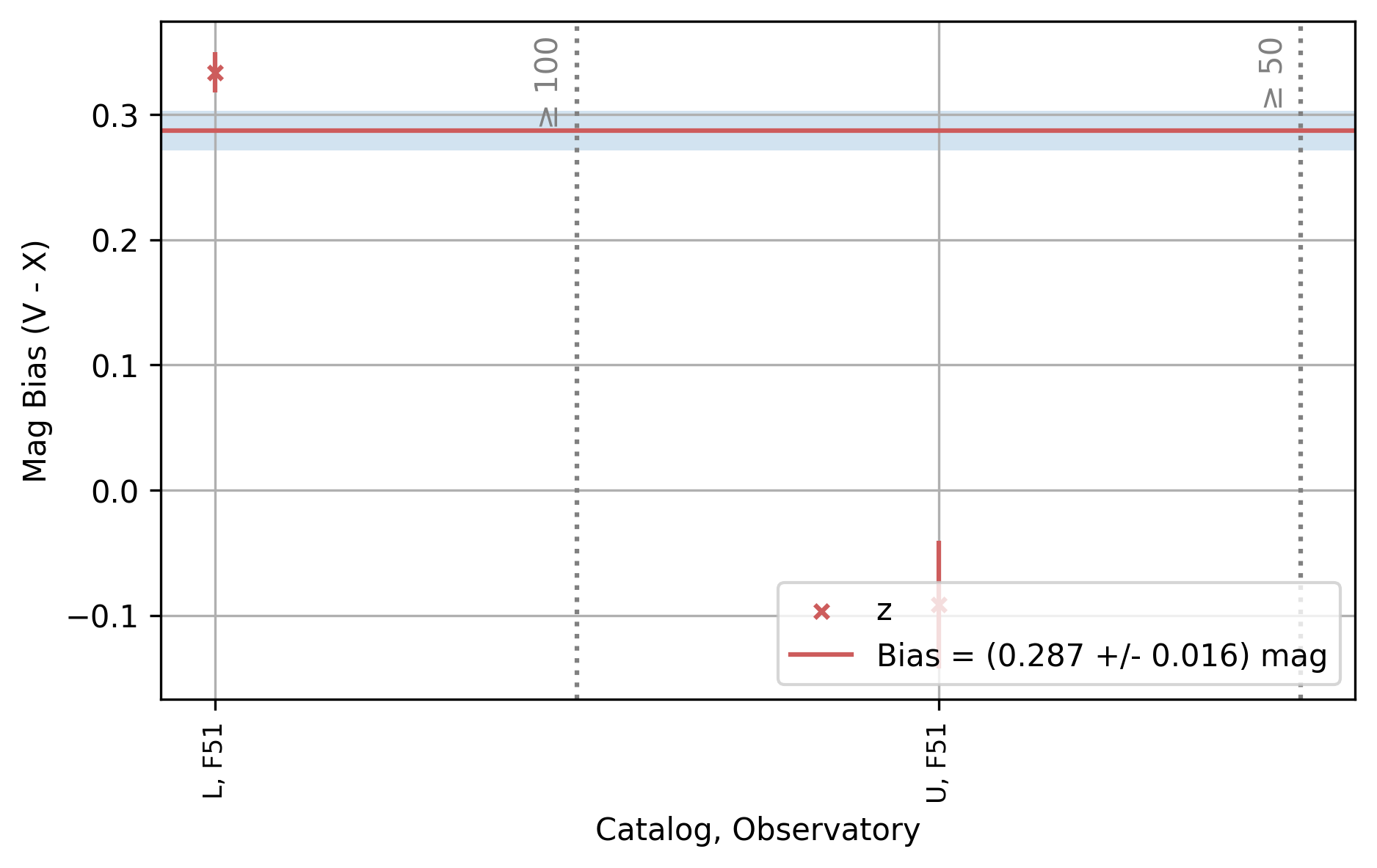}
    \end{subfigure}
    \vspace{2mm}

    \begin{subfigure}[b]{0.49\textwidth}
        \centering
        \includegraphics[width=\textwidth, trim=0mm 7mm 0mm 2mm,clip]{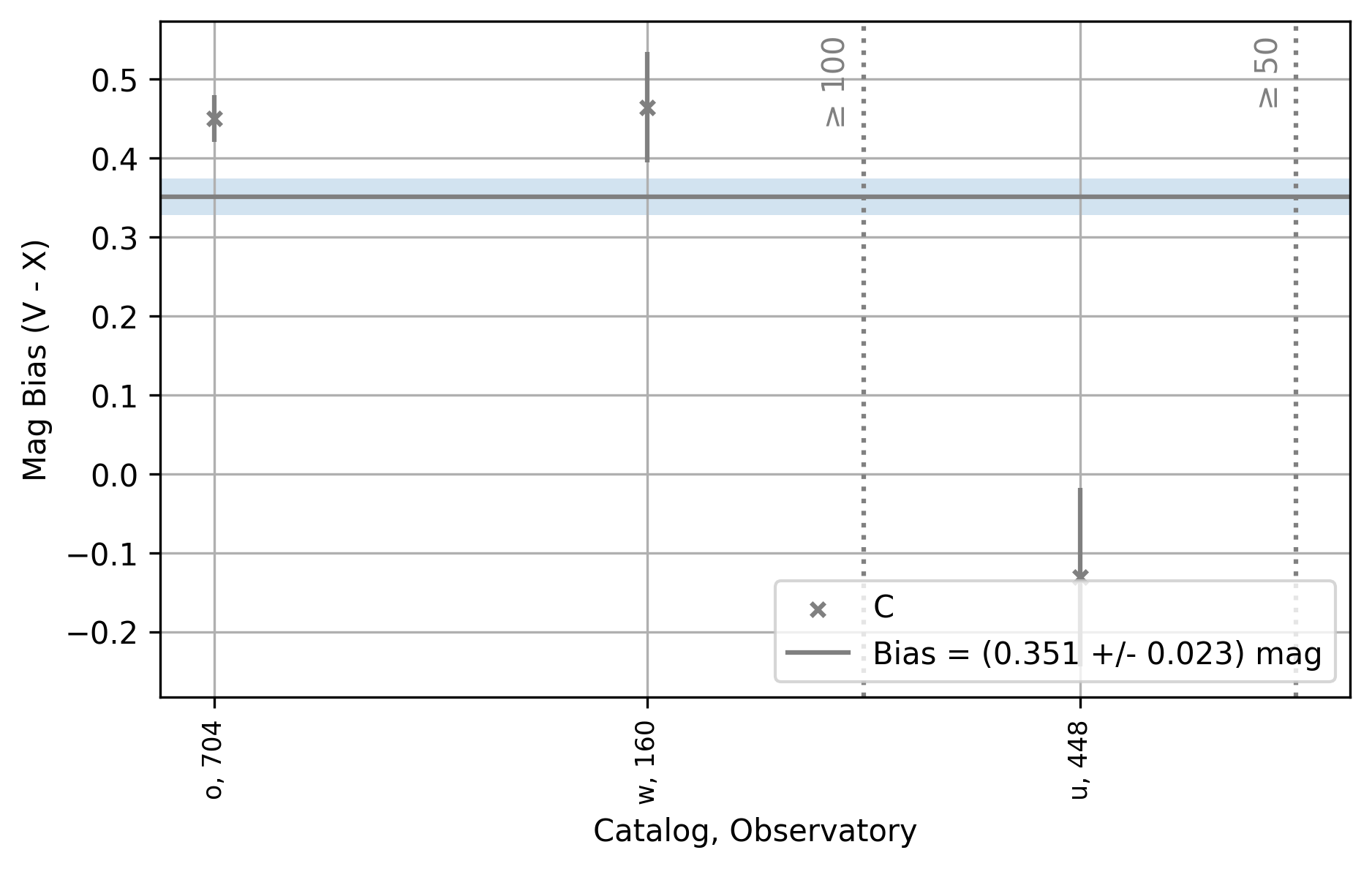}
    \end{subfigure}
    \hfill
    \begin{subfigure}[b]{0.49\textwidth}
        \centering
        \includegraphics[width=\textwidth, trim=0mm 7mm 0mm 2mm,clip]{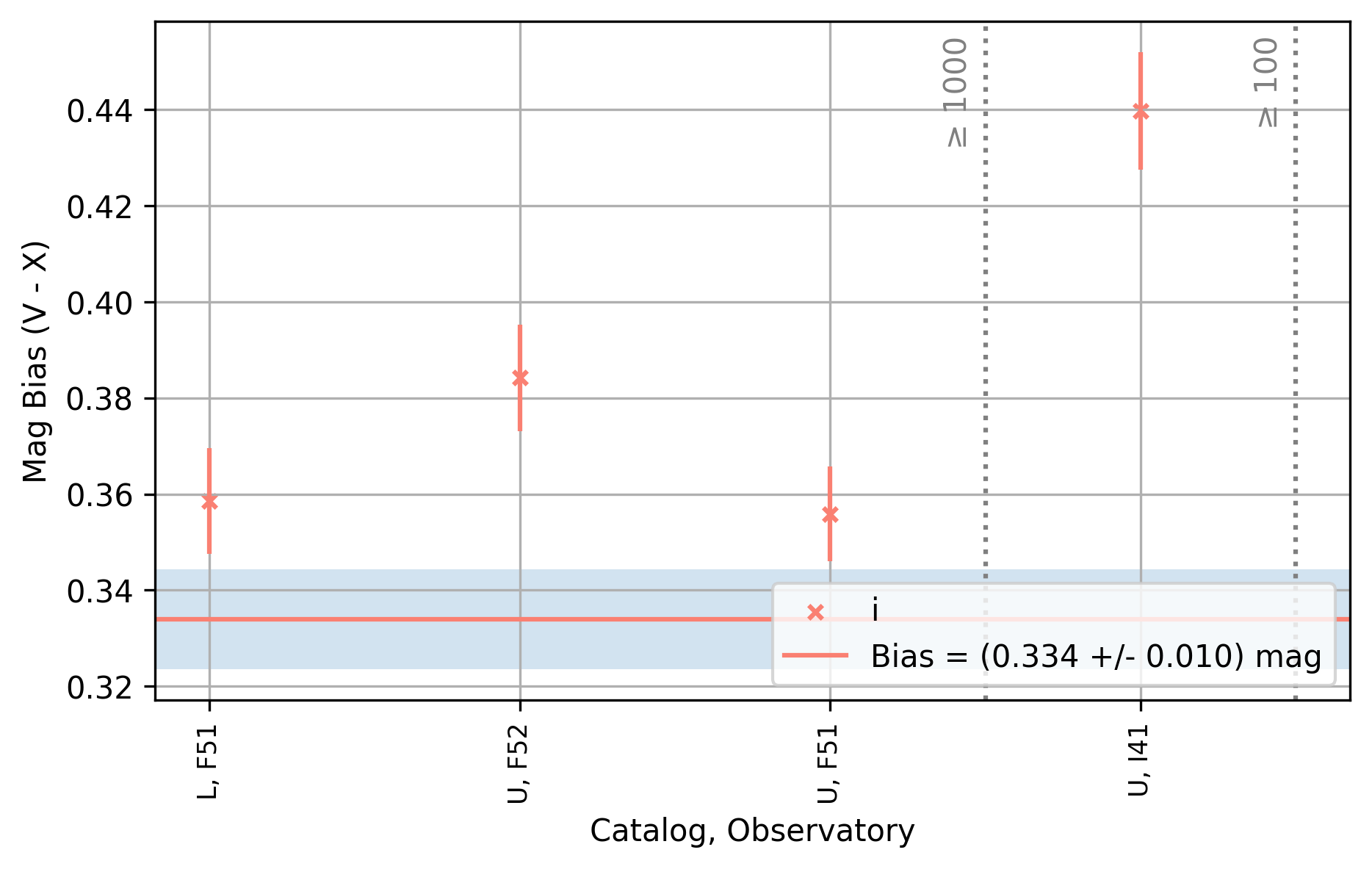}
    \end{subfigure}

    \caption{\footnotesize Results from \textit{aBCO} bias analysis with data from 394 out of 468 asteroids with the average bias and its standard error, sorted in decreasing order of the number of observations (only showing results with more than 50 observations). The horizontal axis lists the combination of the catalog (one letter format, '-' meaning not specified) and the three-letter \ac{MPC} observatory code. The results are separated by color band with according band bias (colored line) including error margins (blue area) from the \textit{aB} analysis in Tab. \ref{tab:res.aB} in comparison, leaving out the following bands with just one observatory: H, J, K, Y (only W91), y (only F51), I (only L10), u (only 309) and B (only 270)}
    \label{fig:res.aBCO}
\end{figure*}

\twocolumn
\begin{figure}[htb]
    \centering
    \vspace{-2cm}
    \begin{subfigure}[b]{0.85\linewidth}
        \centering
        \includegraphics[width=\textwidth, trim=0mm 8mm 0mm 2mm,clip]{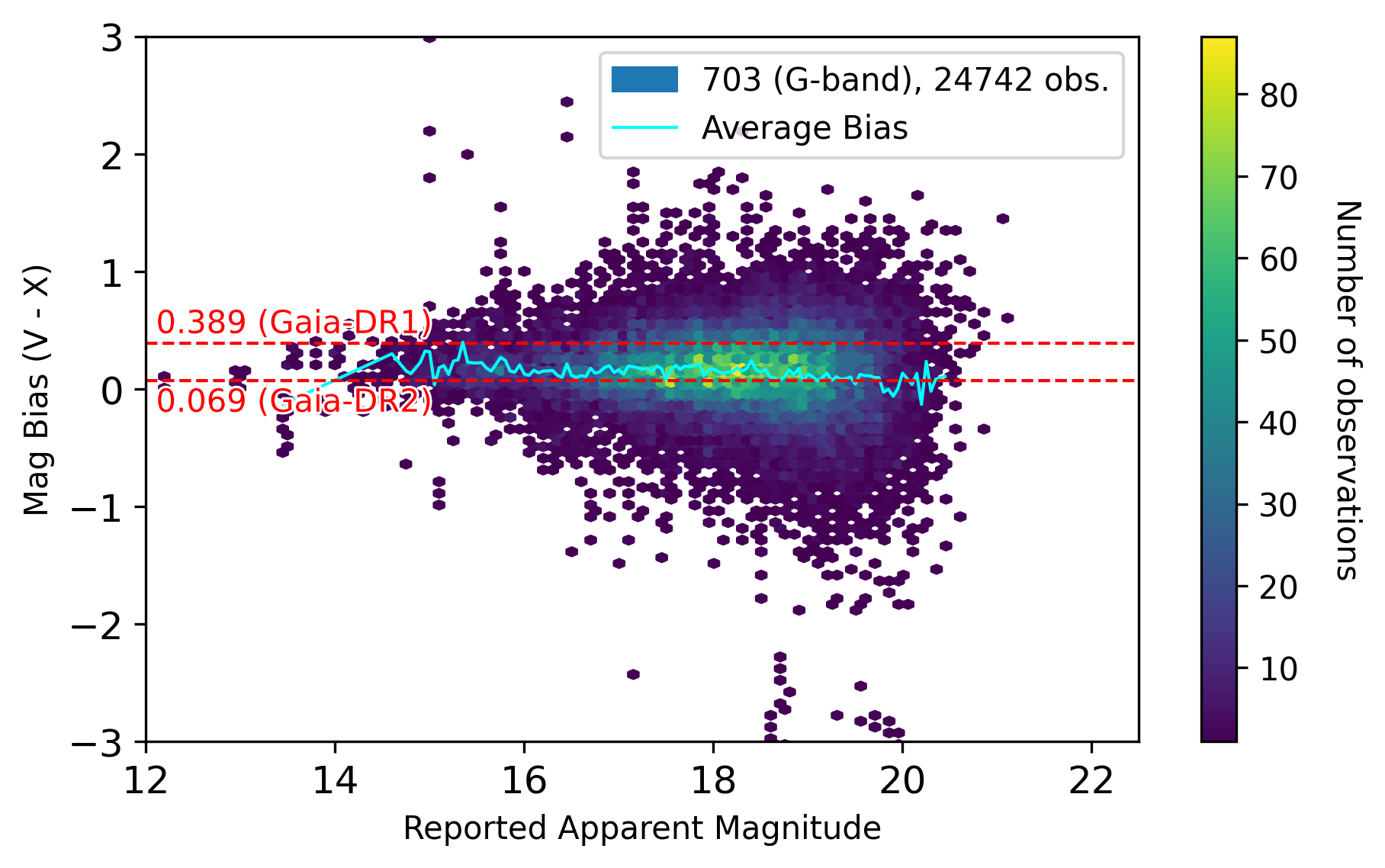}
    \end{subfigure}
    \begin{subfigure}[b]{0.85\linewidth}
        \centering
        \includegraphics[width=\textwidth, trim=0mm 8mm 0mm 2mm,clip]{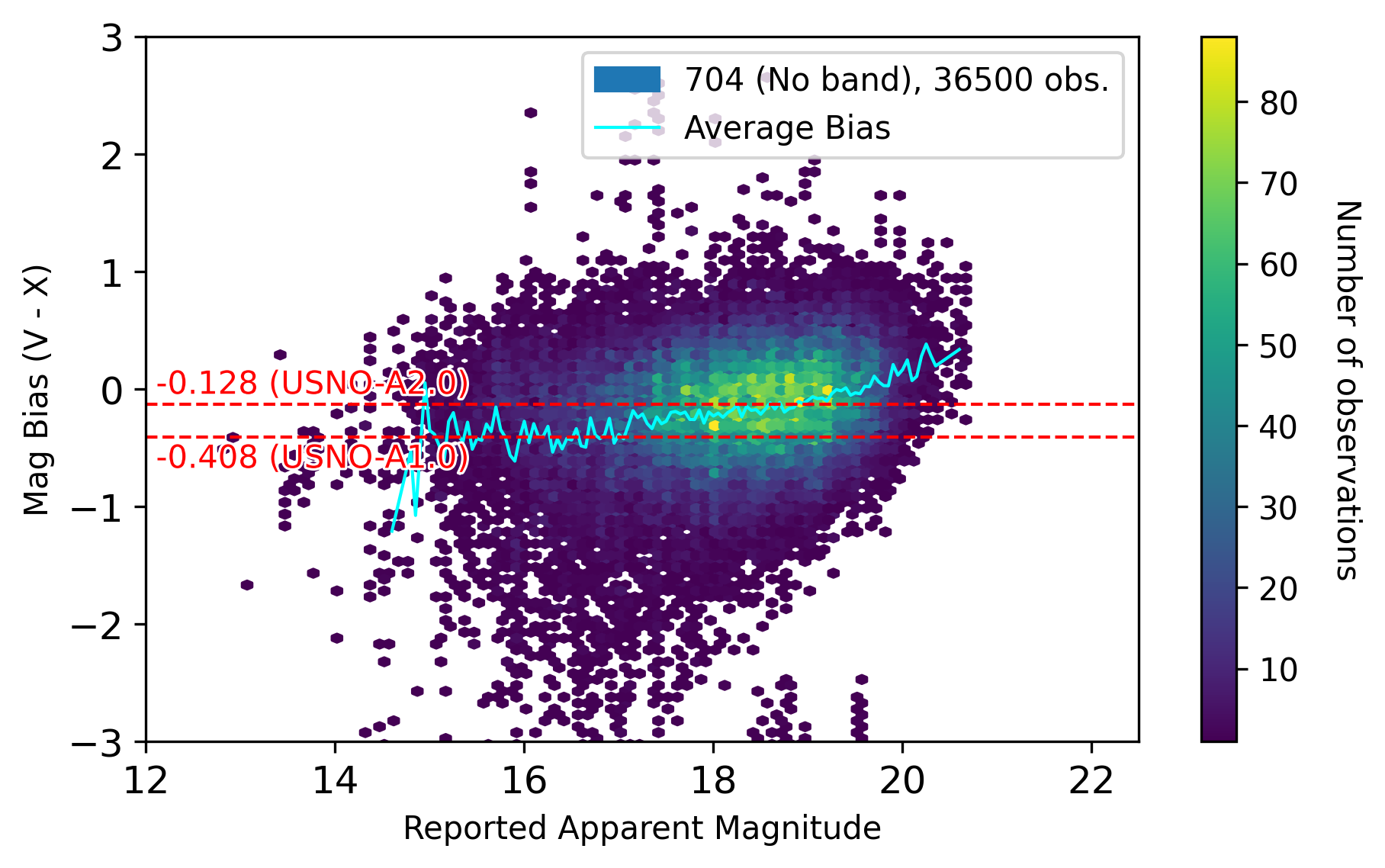}
    \end{subfigure}
    \begin{subfigure}[b]{0.85\linewidth}
        \centering
        \includegraphics[width=\textwidth, trim=0mm 8mm 0mm 2mm,clip]{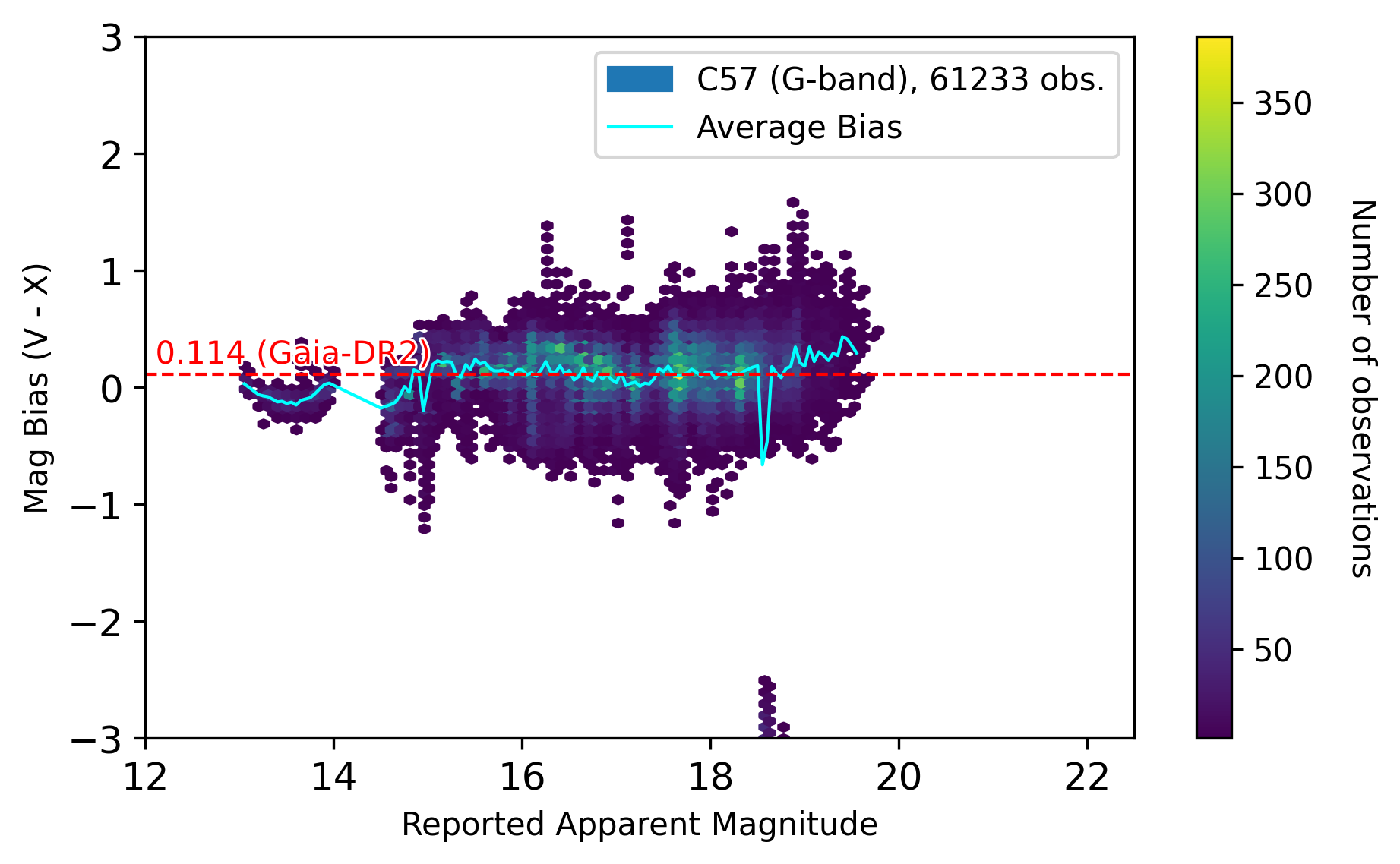}
    \end{subfigure}
    \begin{subfigure}[b]{0.85\linewidth}
        \centering
        \includegraphics[width=\textwidth, trim=0mm 8mm 0mm 2mm,clip]{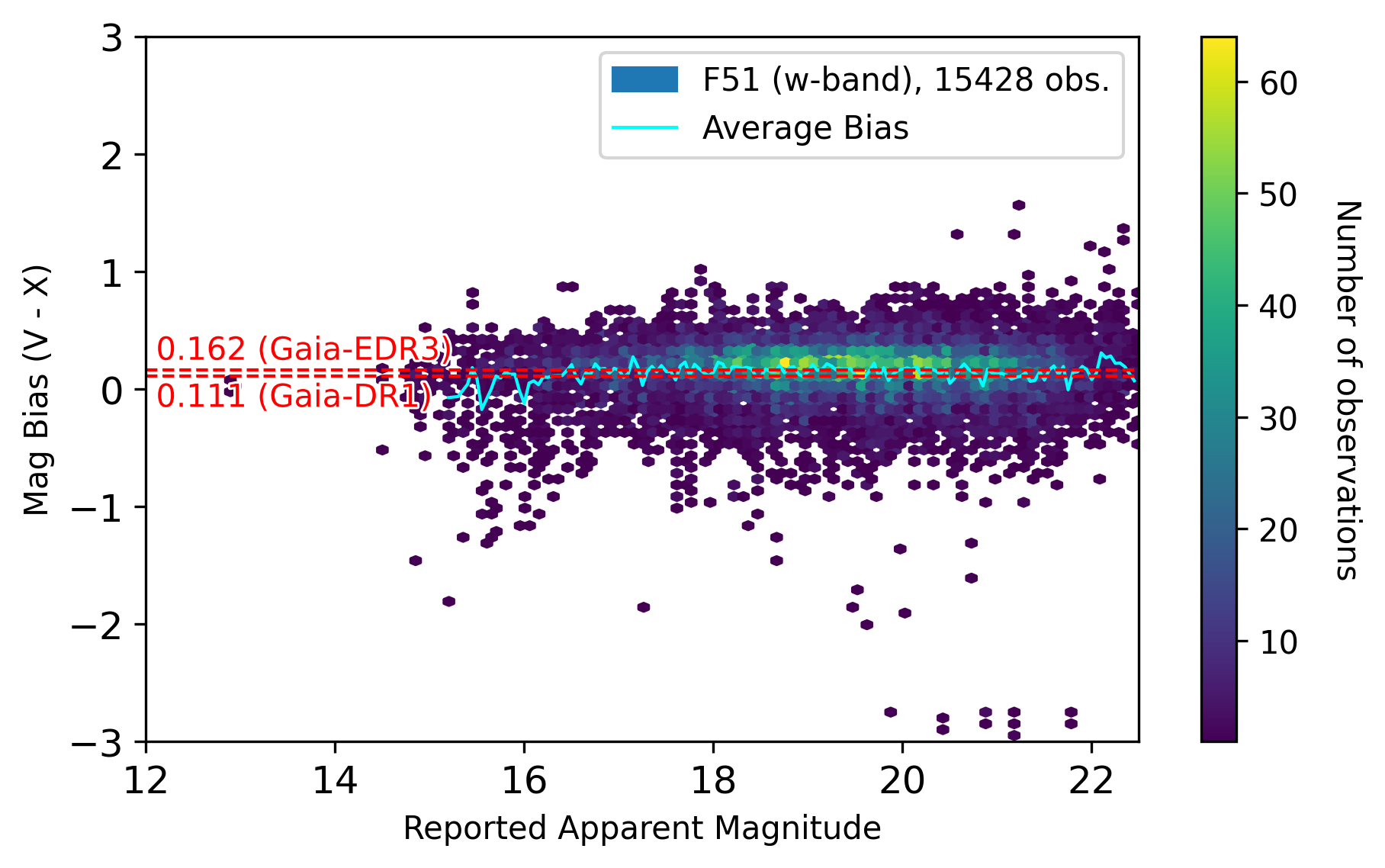}
    \end{subfigure}
    \begin{subfigure}[b]{0.85\linewidth}
        \centering
        \includegraphics[width=\textwidth, trim=0mm 8mm 0mm 2mm,clip]{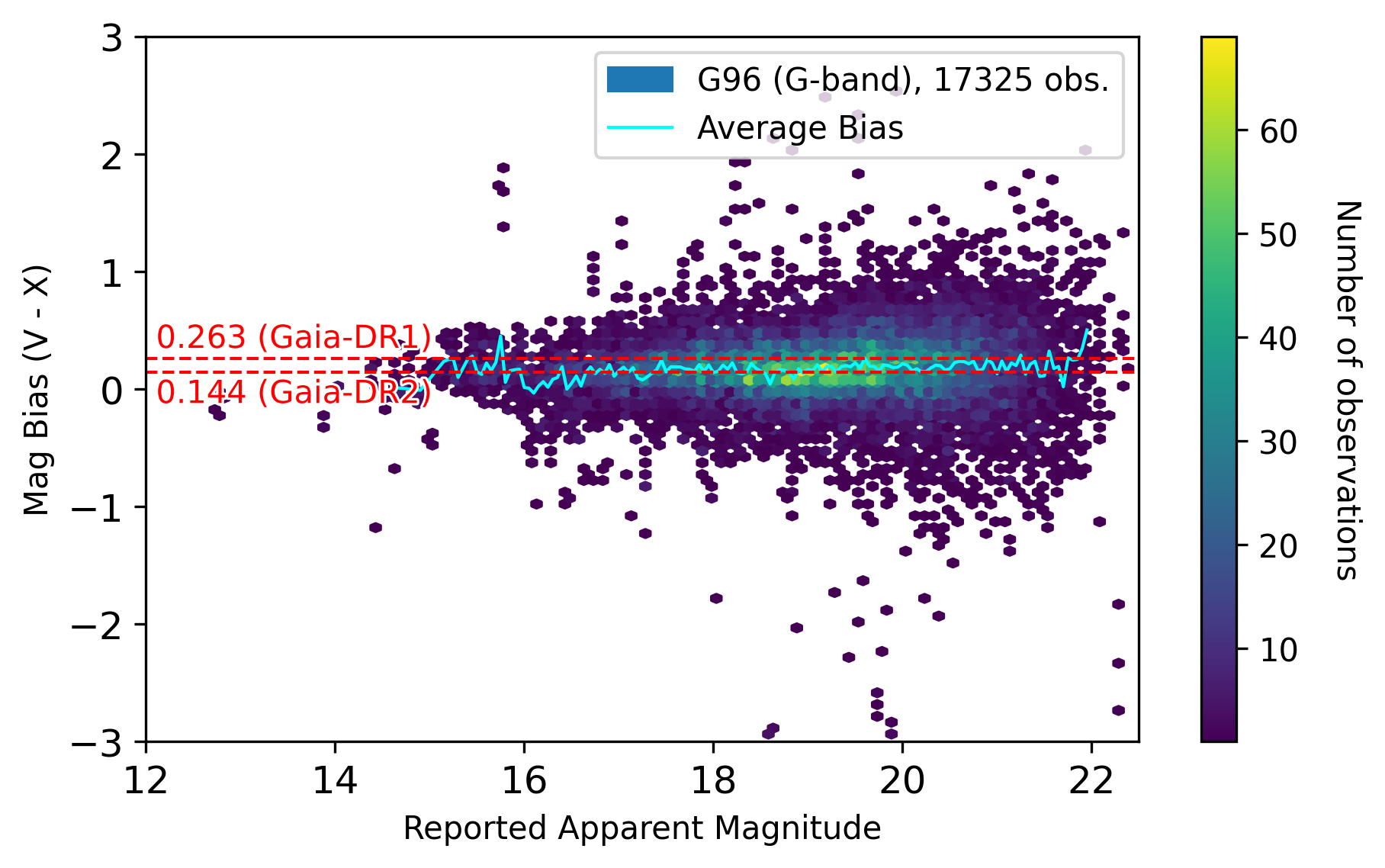}
    \end{subfigure}
    \begin{subfigure}[b]{0.85\linewidth}
        \centering
        \includegraphics[width=\textwidth, trim=0mm 0mm 0mm 0mm,clip]{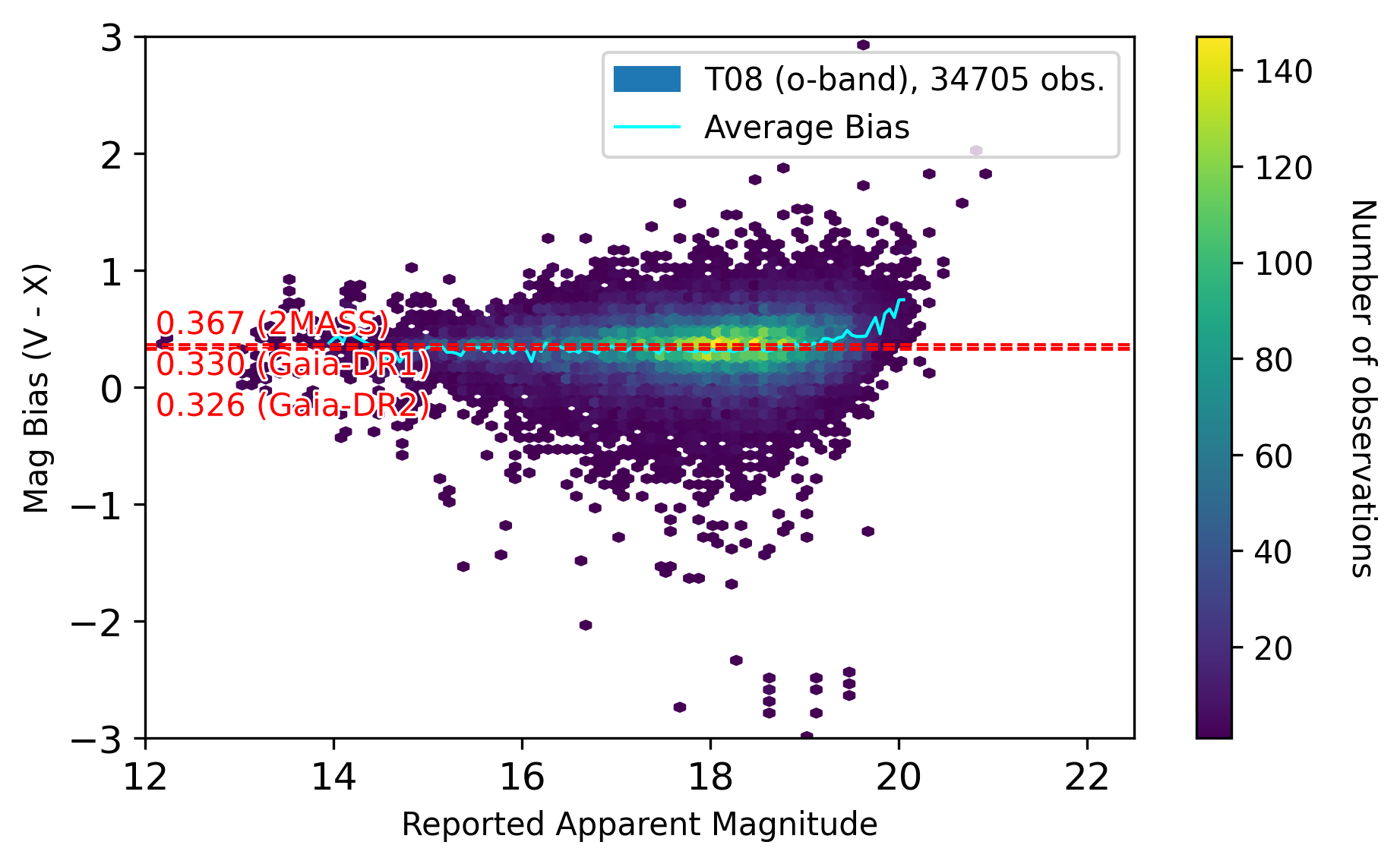}
    \end{subfigure}
    \vspace{-3mm}
    \caption{Dependence of apparent magnitude on photometric bias for large stations (Bin $0.1$~mag/$0.1$~mag).}
    \label{fig:dis.advanced.mag}
\end{figure}

\begin{figure}[htb]
    \centering
    \vspace{-2cm}
    \begin{subfigure}[b]{0.85\linewidth}
        \centering
        \includegraphics[width=\textwidth, trim=0mm 8mm 0mm 2mm,clip]{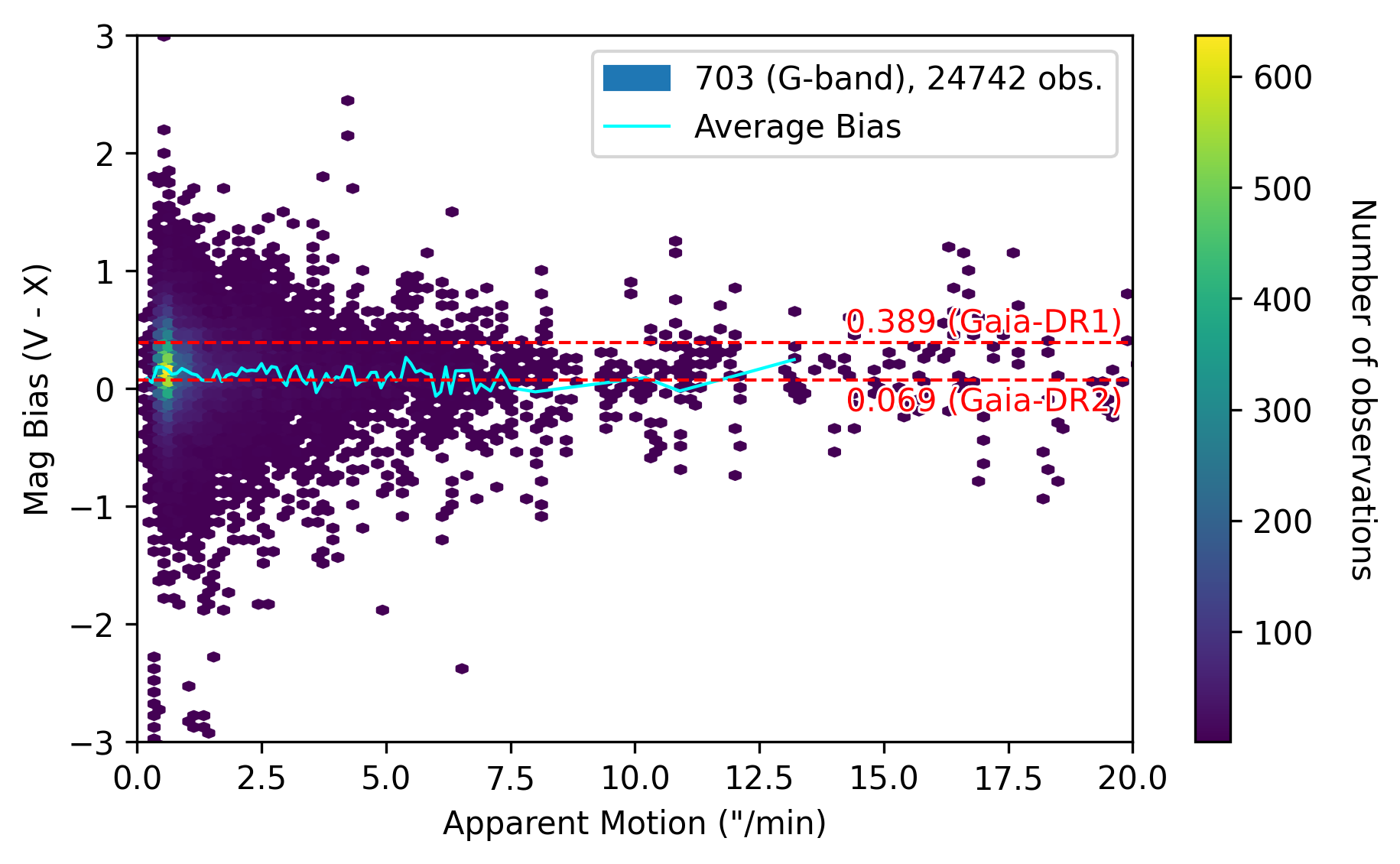}
    \end{subfigure}
    \begin{subfigure}[b]{0.85\linewidth}
        \centering
        \includegraphics[width=\textwidth, trim=0mm 8mm 0mm 2mm,clip]{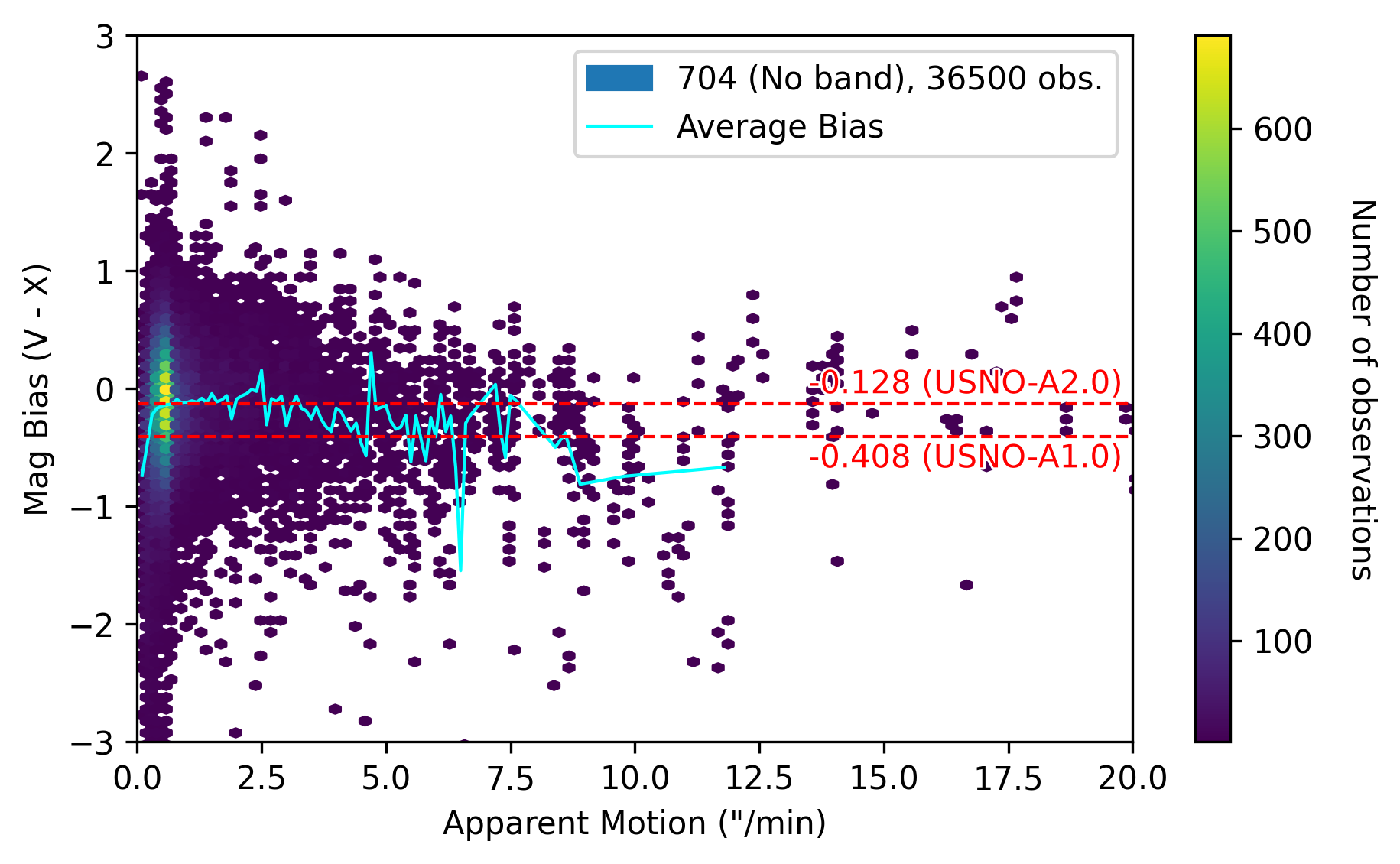}
    \end{subfigure}
    \begin{subfigure}[b]{0.85\linewidth}
        \centering
        \includegraphics[width=\textwidth, trim=0mm 8mm 0mm 2mm,clip]{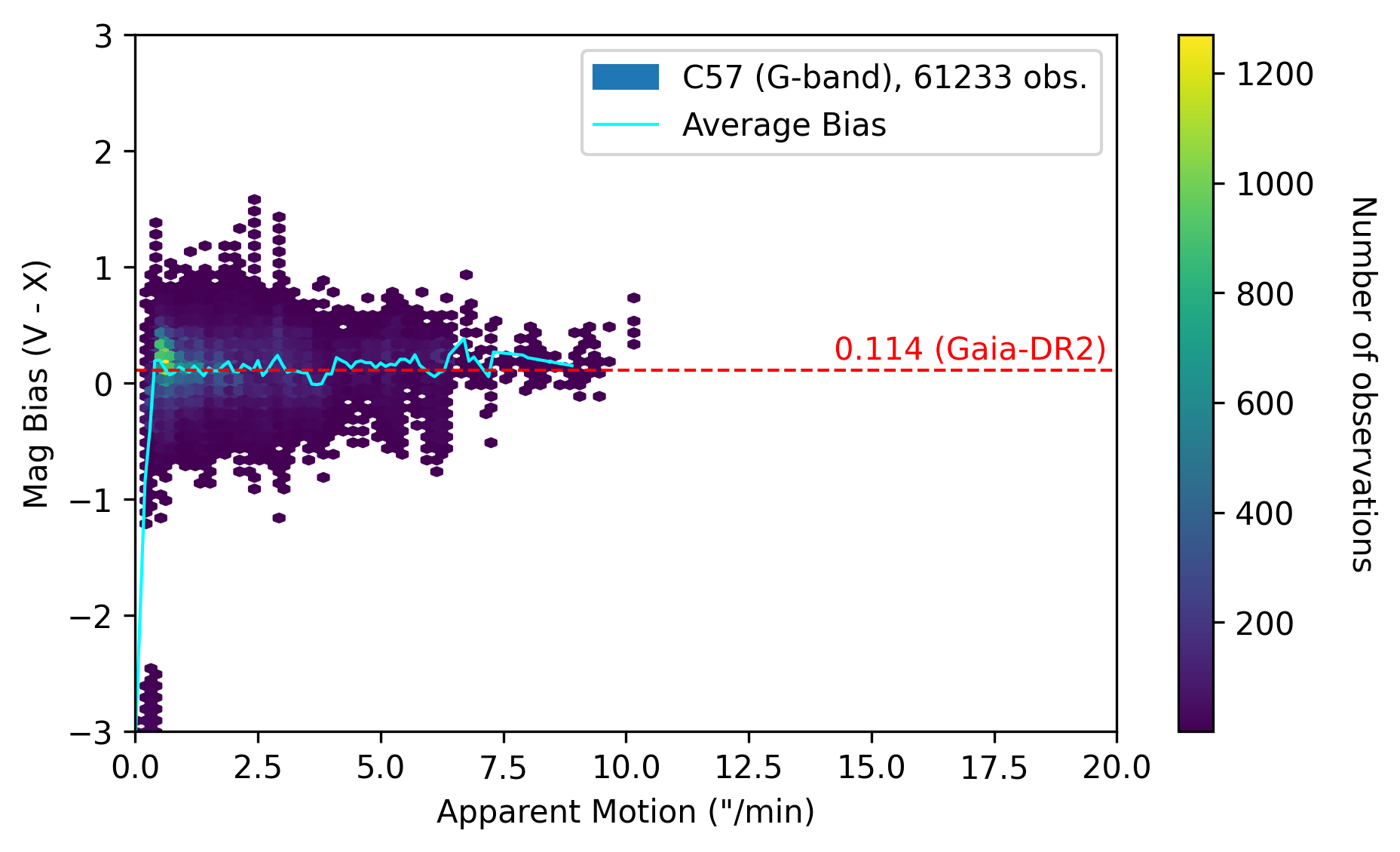}
    \end{subfigure}
    \begin{subfigure}[b]{0.85\linewidth}
        \centering
        \includegraphics[width=\textwidth, trim=0mm 8mm 0mm 2mm,clip]{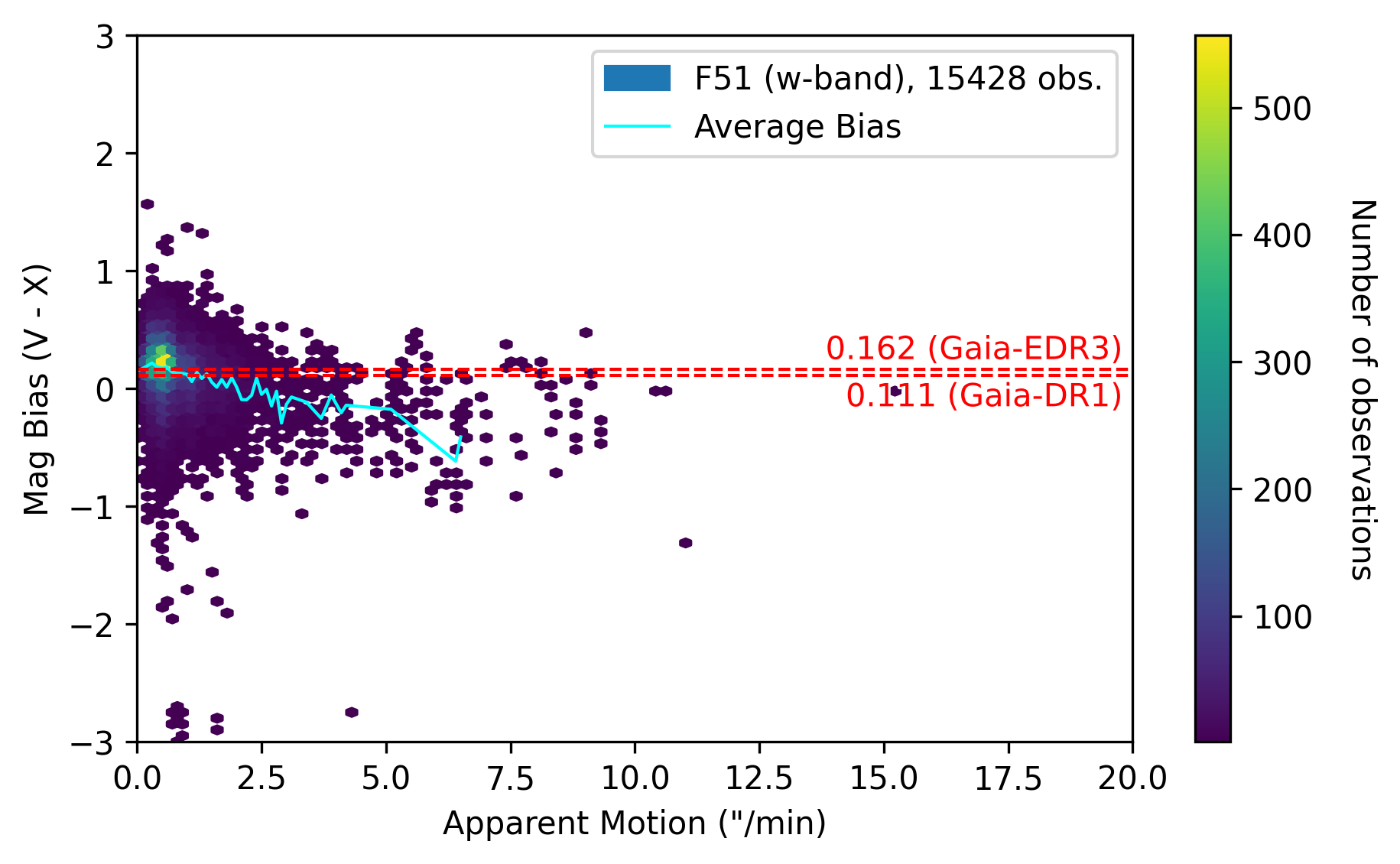}
    \end{subfigure}
    \begin{subfigure}[b]{0.85\linewidth}
        \centering
        \includegraphics[width=\textwidth, trim=0mm 8mm 0mm 2mm,clip]{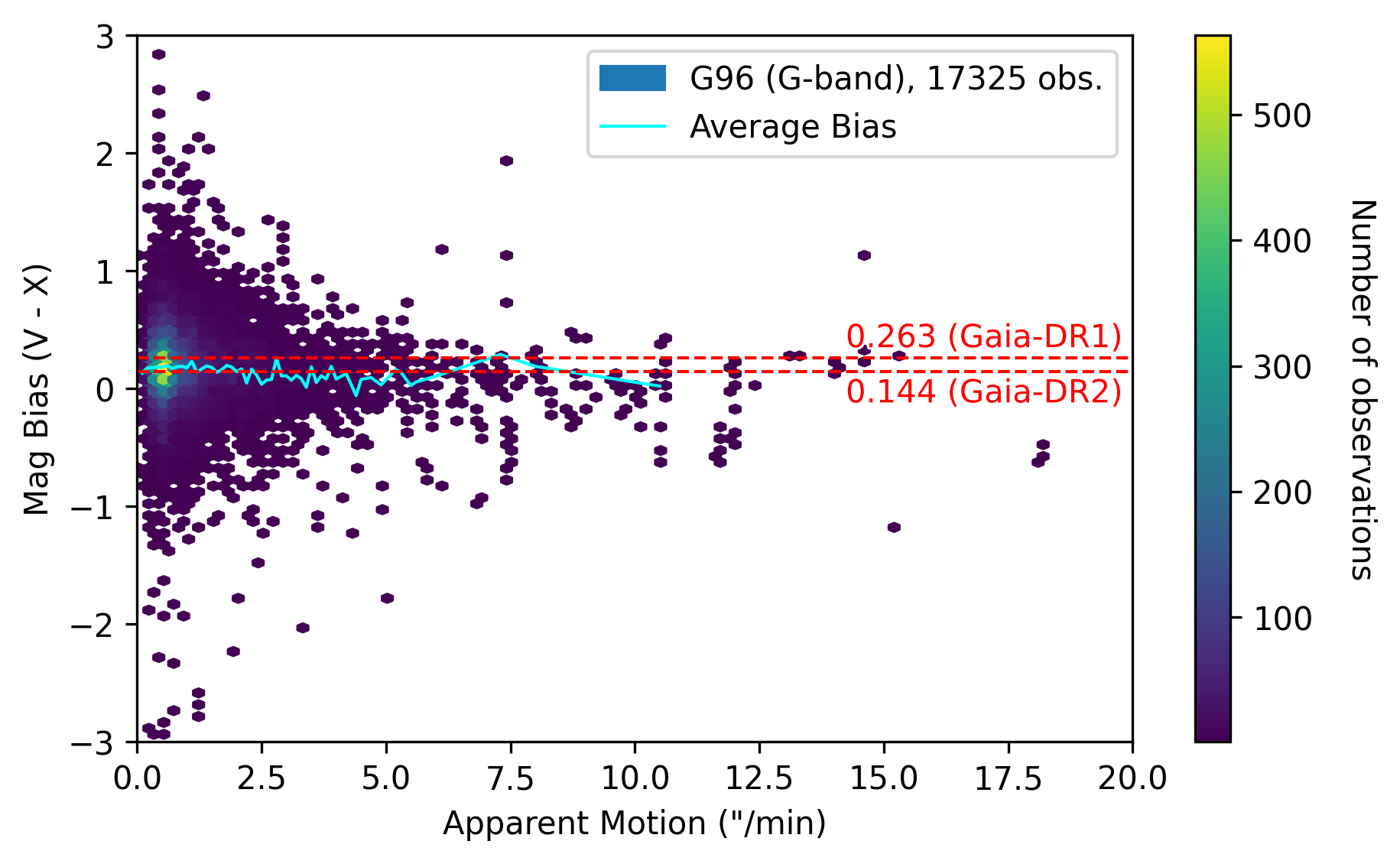}
    \end{subfigure}
    \begin{subfigure}[b]{0.85\linewidth}
        \centering
        \includegraphics[width=\textwidth, trim=0mm 0mm 0mm 0mm,clip]{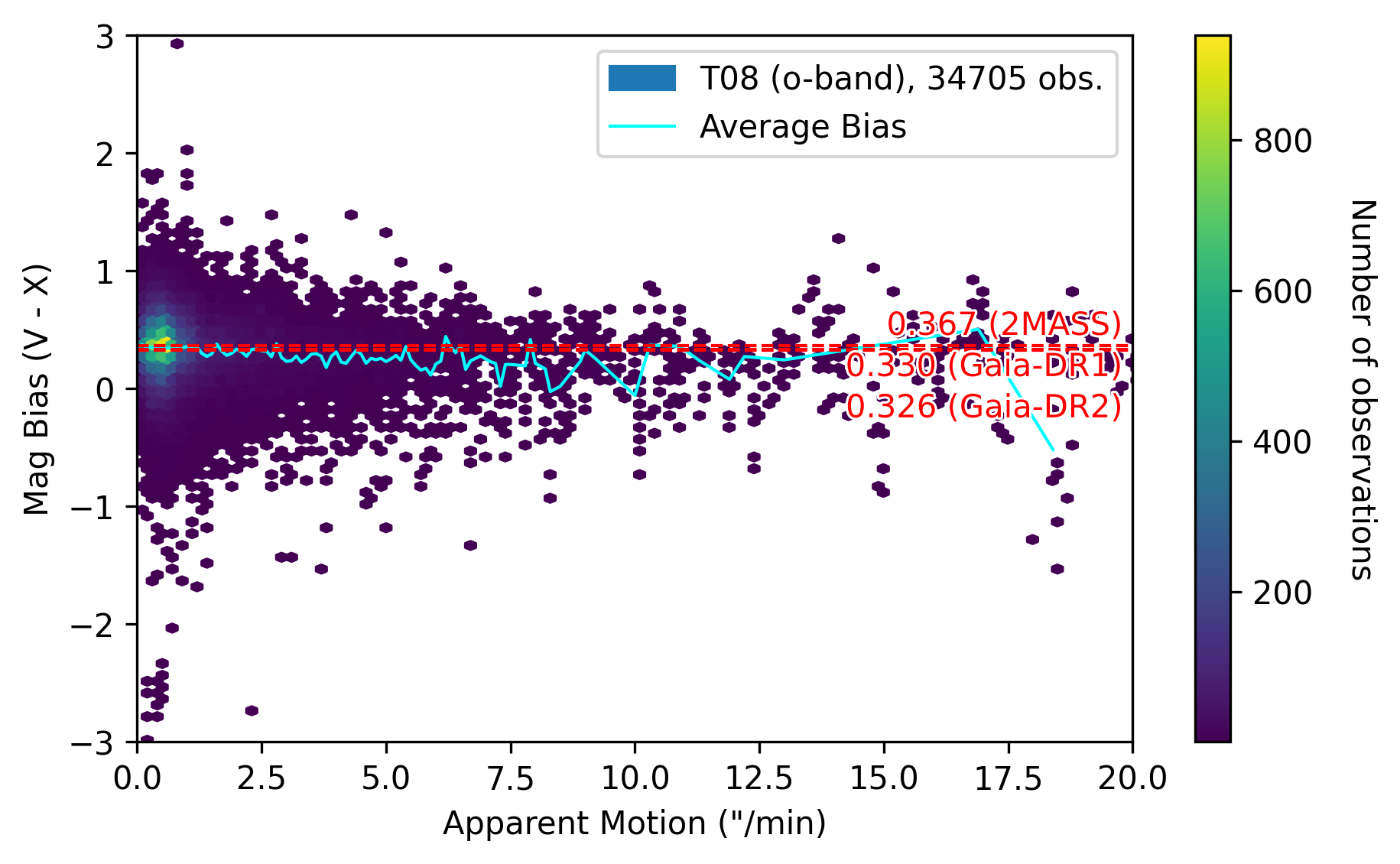}
    \end{subfigure}
    \vspace{-3mm}
    \caption{Dependence of apparent motion on photometric bias for large stations (Bin $0.2"$/min / $0.1$~mag).}
    \label{fig:dis.advanced.motion}
\end{figure}

\begin{figure}[htb]
    \centering
    \vspace{-2cm}
    \begin{subfigure}[b]{0.85\linewidth}
        \centering
        \includegraphics[width=\textwidth, trim=0mm 8mm 0mm 2mm,clip]{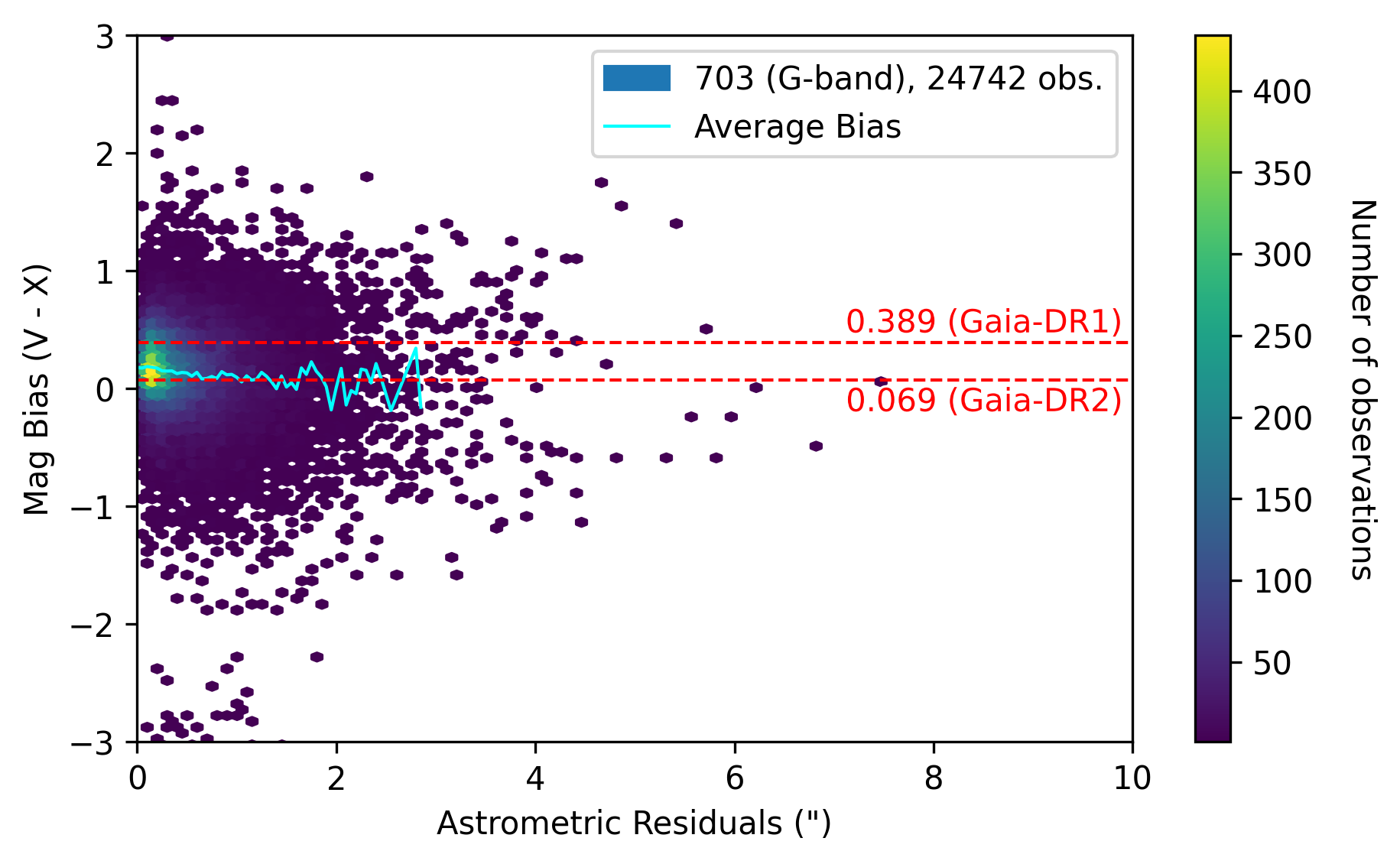}
    \end{subfigure}
    \begin{subfigure}[b]{0.85\linewidth}
        \centering
        \includegraphics[width=\textwidth, trim=0mm 8mm 0mm 2mm,clip]{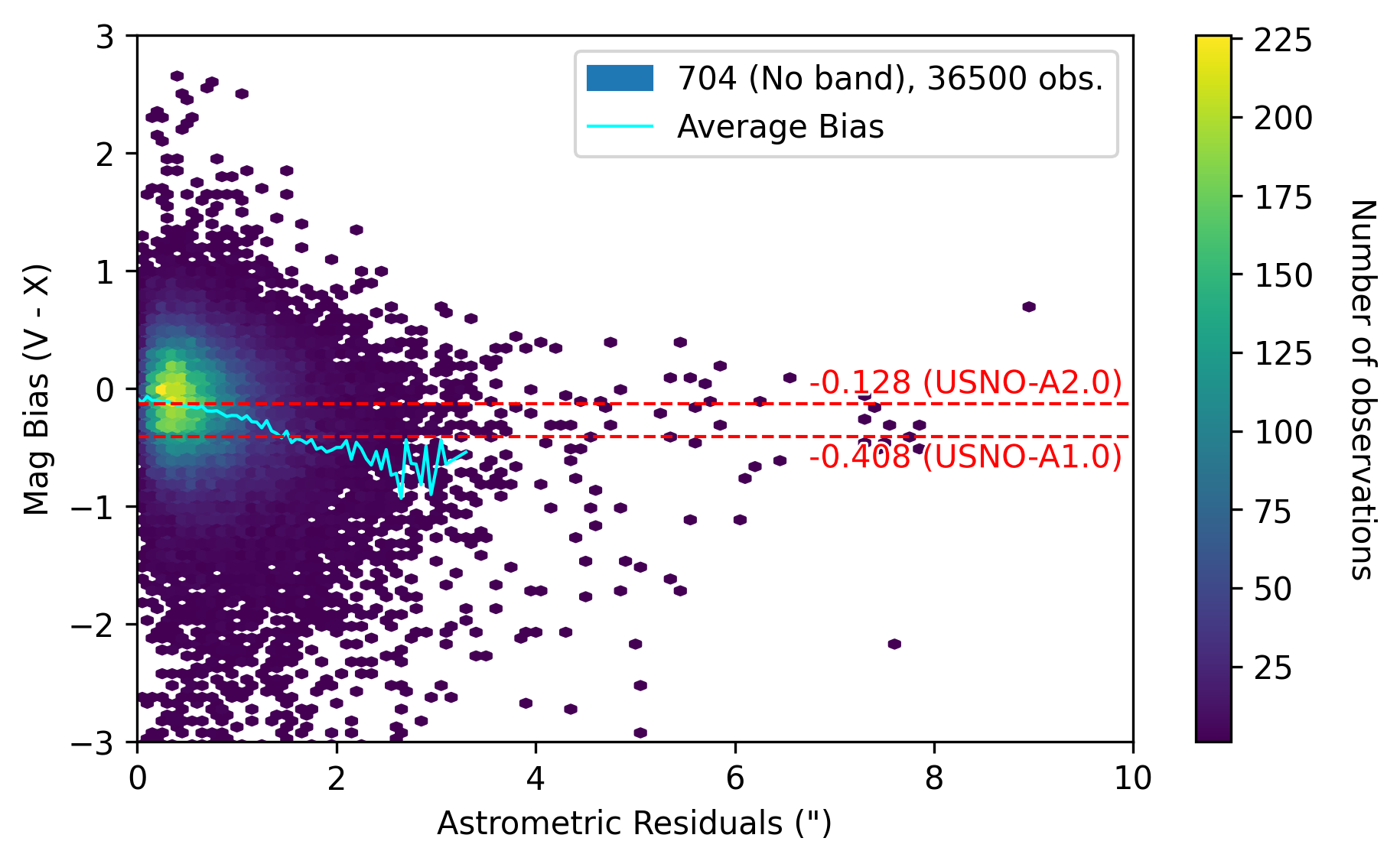}
    \end{subfigure}
    \begin{subfigure}[b]{0.85\linewidth}
        \centering
        \includegraphics[width=\textwidth, trim=0mm 6mm 0mm 2mm,clip]{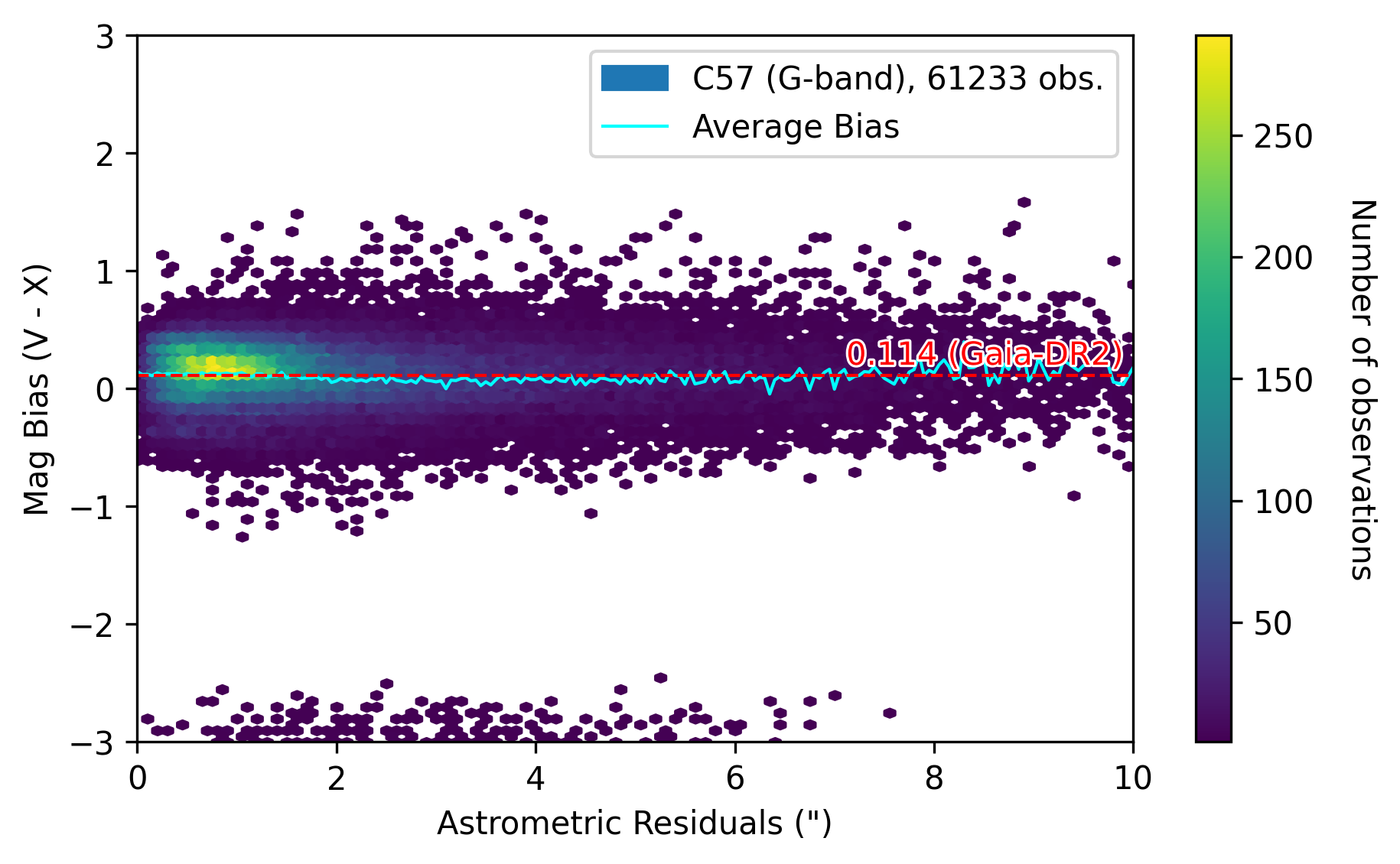}
    \end{subfigure}
    \begin{subfigure}[b]{0.85\linewidth}
        \centering
        \includegraphics[width=\textwidth, trim=0mm 8mm 0mm 2mm,clip]{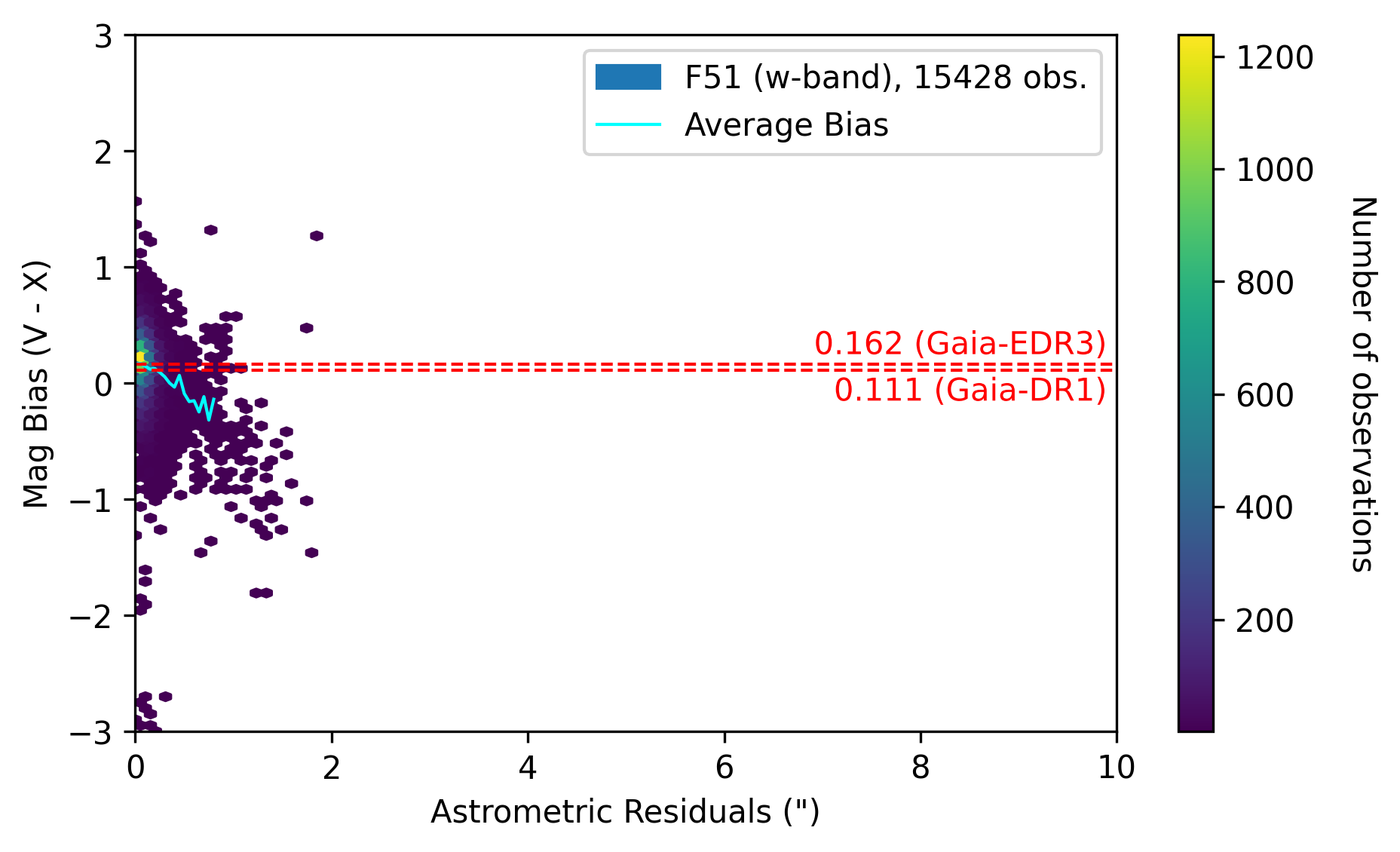}
    \end{subfigure}
    \begin{subfigure}[b]{0.85\linewidth}
        \centering
        \includegraphics[width=\textwidth, trim=0mm 8mm 0mm 2mm,clip]{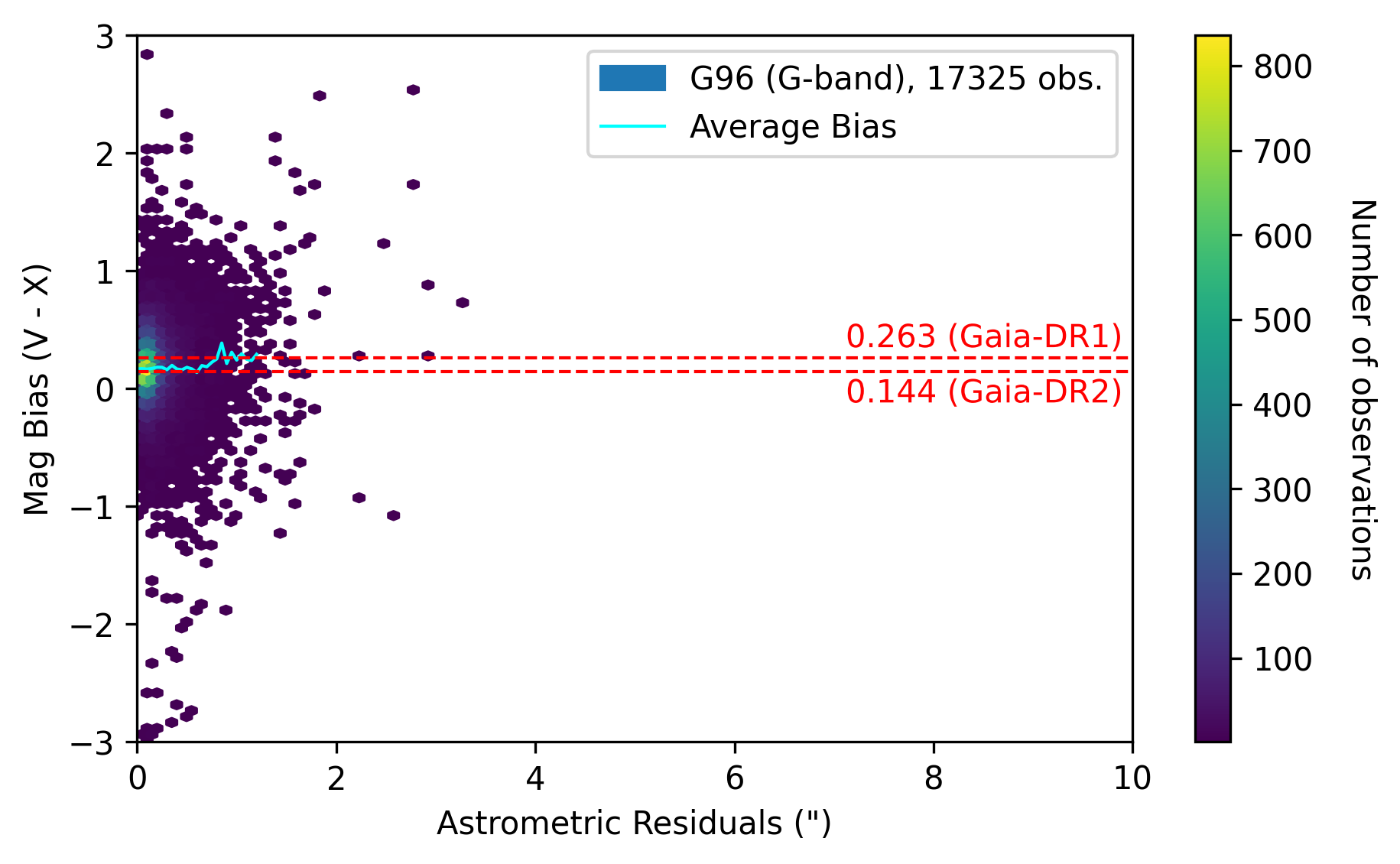}
    \end{subfigure}
    \begin{subfigure}[b]{0.85\linewidth}
        \centering
        \includegraphics[width=\textwidth, trim=0mm 0mm 0mm 0mm,clip]{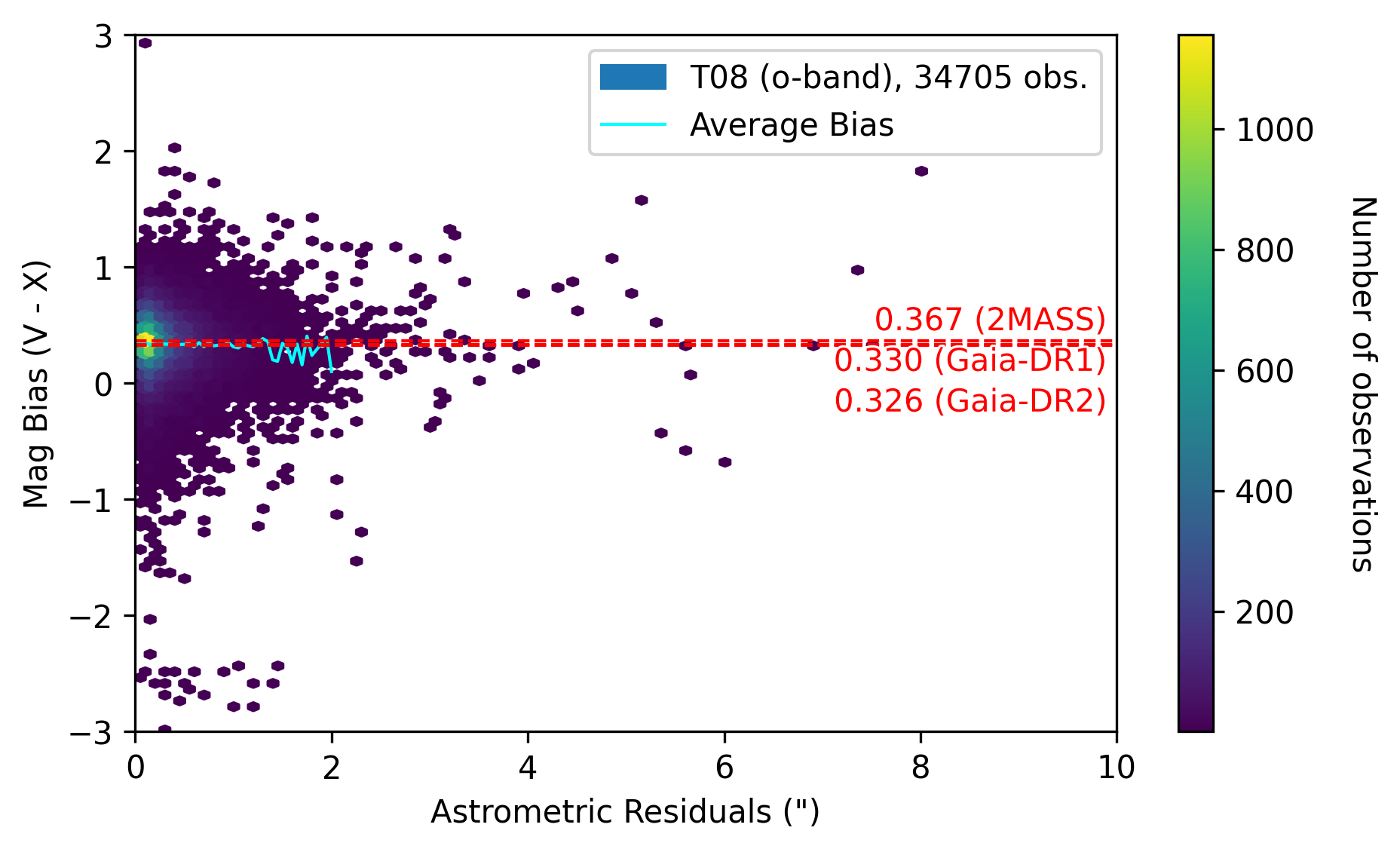}
    \end{subfigure}
    \vspace{-3mm}
    \caption{Dependence of astrometric residuals on photometric bias for large stations (Bin $0.1"$/$0.1$~mag).}
    \label{fig:dis.advanced.resid}
\end{figure}

\begin{figure}[htb]
    \centering
    \vspace{-2cm}
    \begin{subfigure}[b]{0.85\linewidth}
        \centering
        \includegraphics[width=\textwidth, trim=0mm 8mm 0mm 2mm,clip]{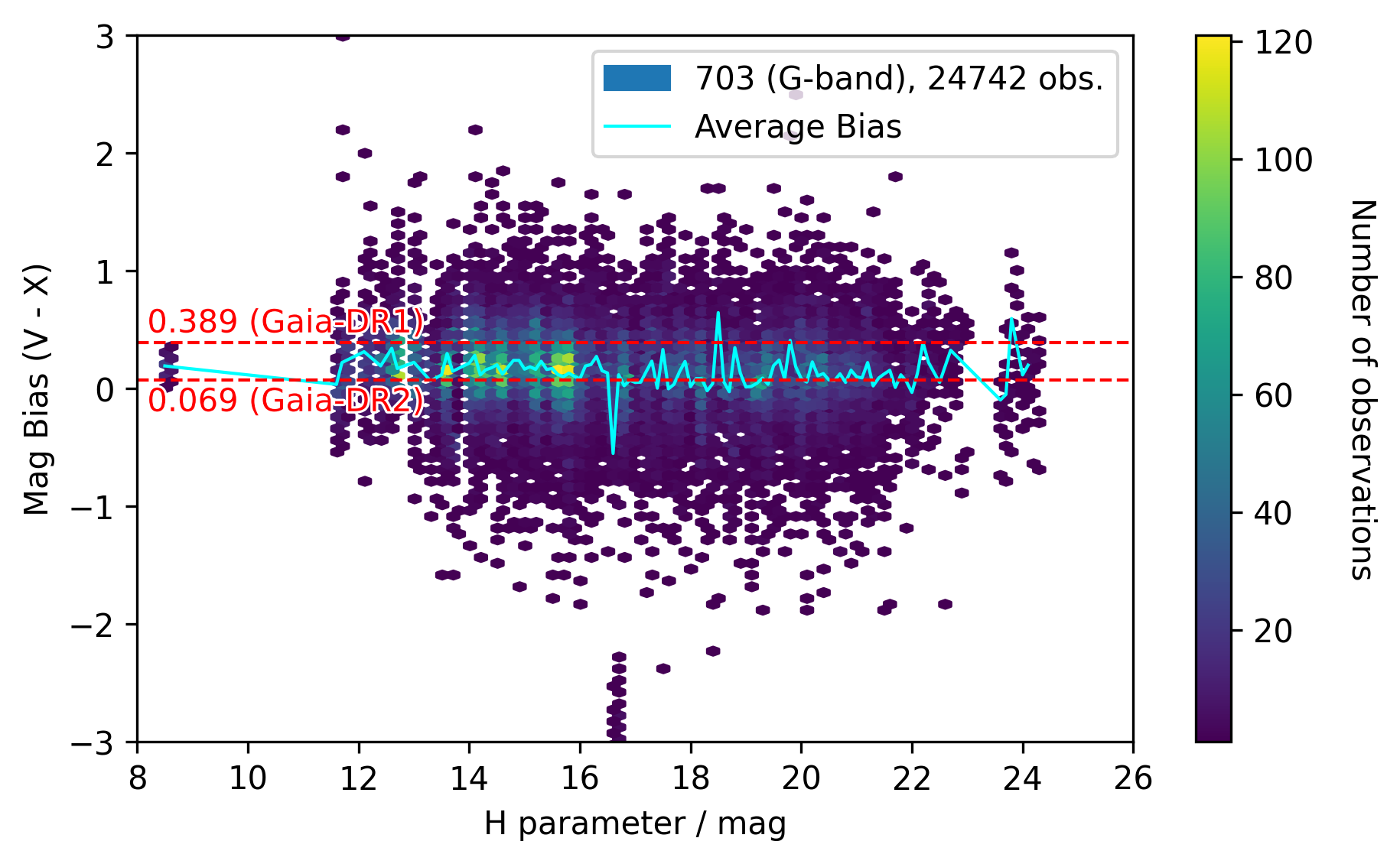}
    \end{subfigure}
    \begin{subfigure}[b]{0.85\linewidth}
        \centering
        \includegraphics[width=\textwidth, trim=0mm 8mm 0mm 2mm,clip]{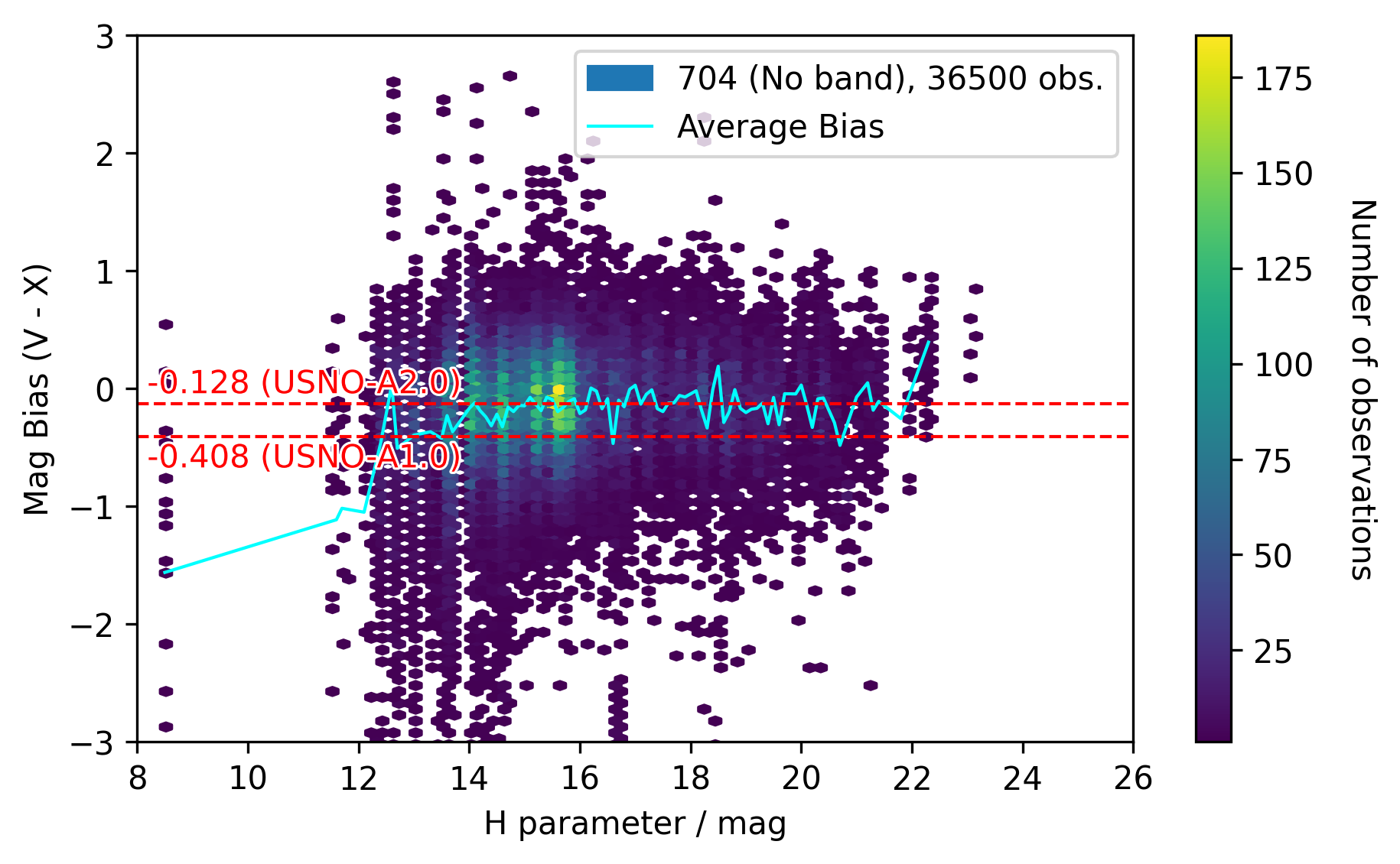}
    \end{subfigure}
    \begin{subfigure}[b]{0.85\linewidth}
        \centering
        \includegraphics[width=\textwidth, trim=0mm 8mm 0mm 2mm,clip]{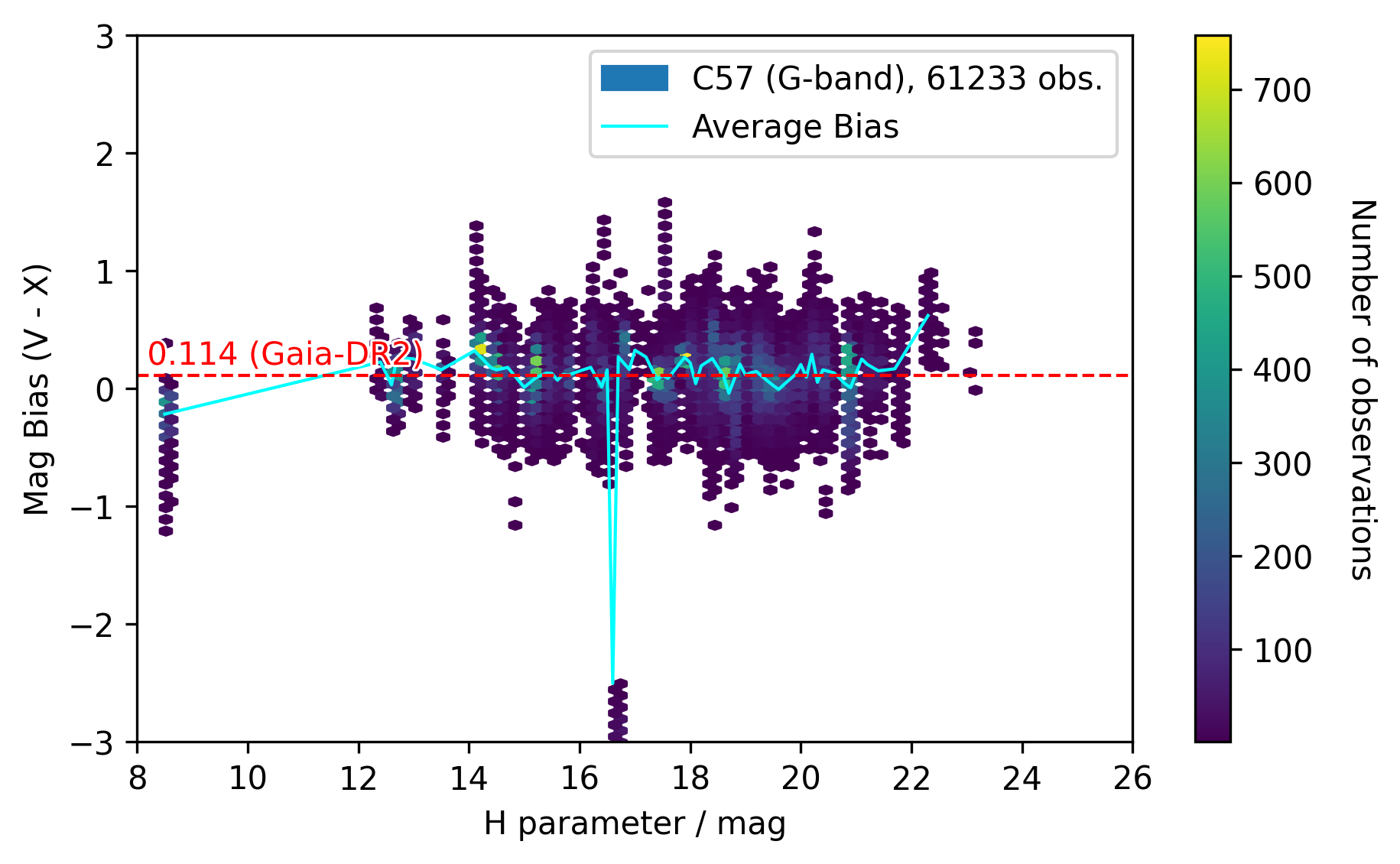}
    \end{subfigure}
    \begin{subfigure}[b]{0.85\linewidth}
        \centering
        \includegraphics[width=\textwidth, trim=0mm 8mm 0mm 2mm,clip]{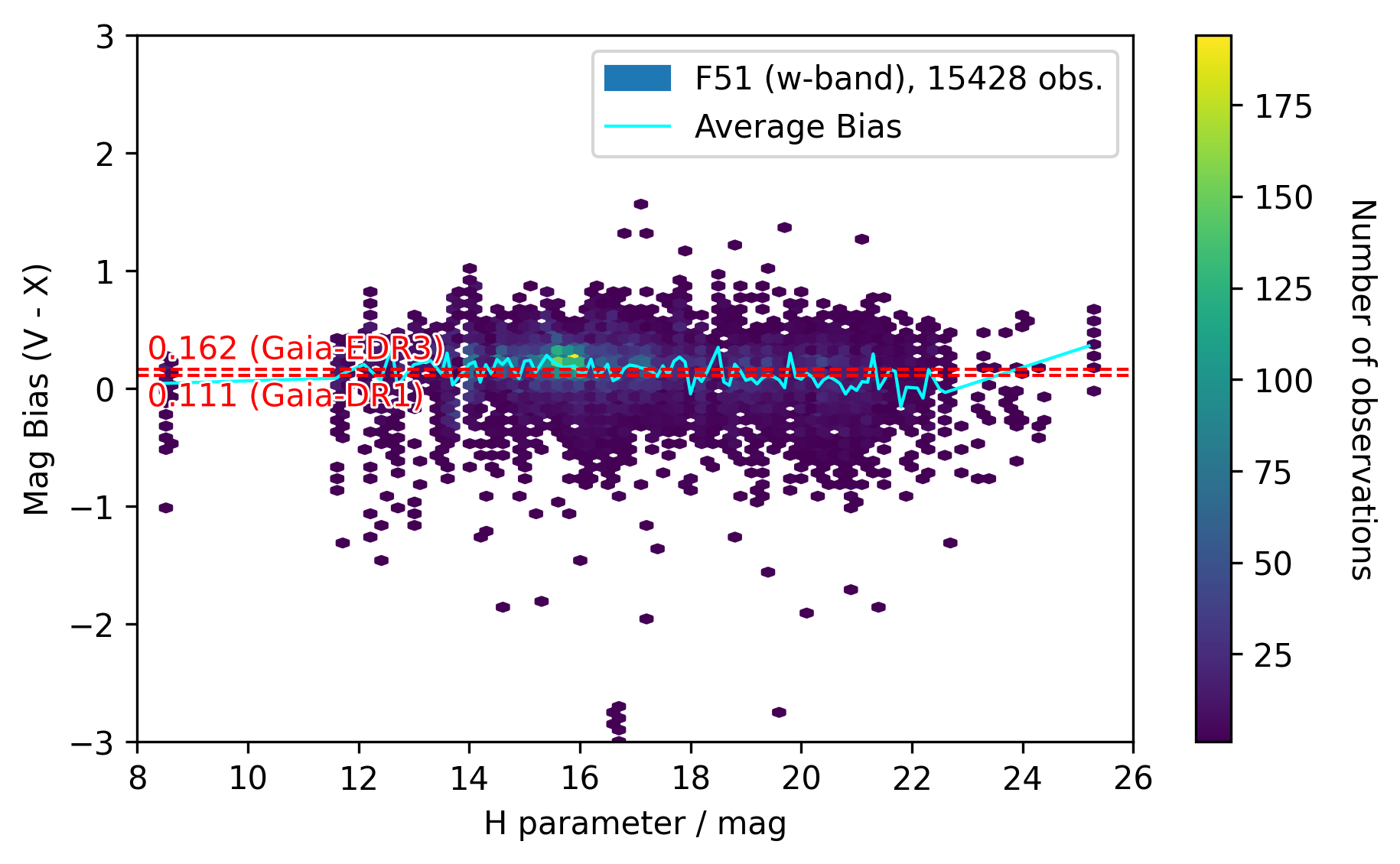}
    \end{subfigure}
    \begin{subfigure}[b]{0.85\linewidth}
        \centering
        \includegraphics[width=\textwidth, trim=0mm 8mm 0mm 2mm,clip]{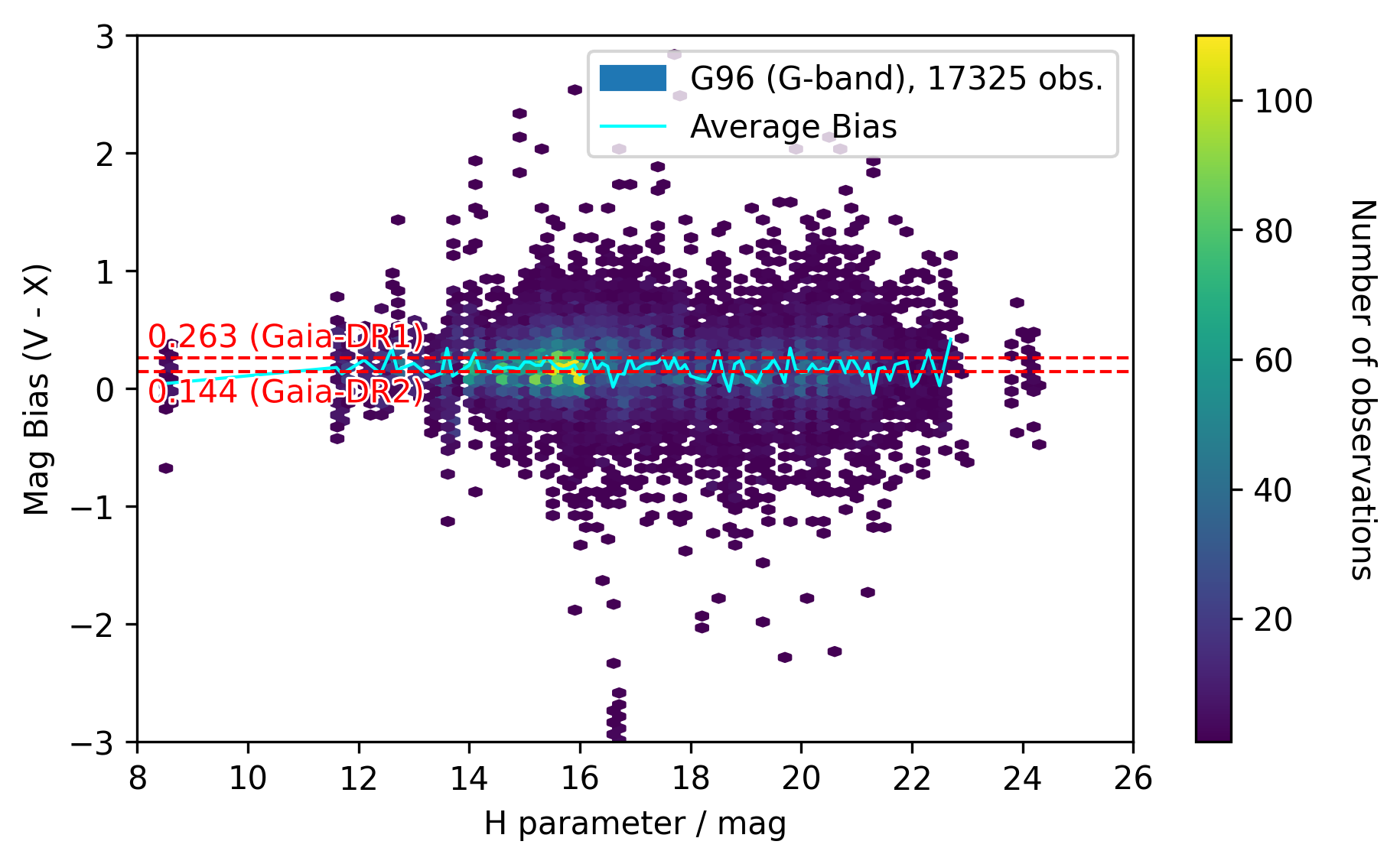}
    \end{subfigure}
    \begin{subfigure}[b]{0.85\linewidth}
        \centering
        \includegraphics[width=\textwidth, trim=0mm 0mm 0mm 0mm,clip]{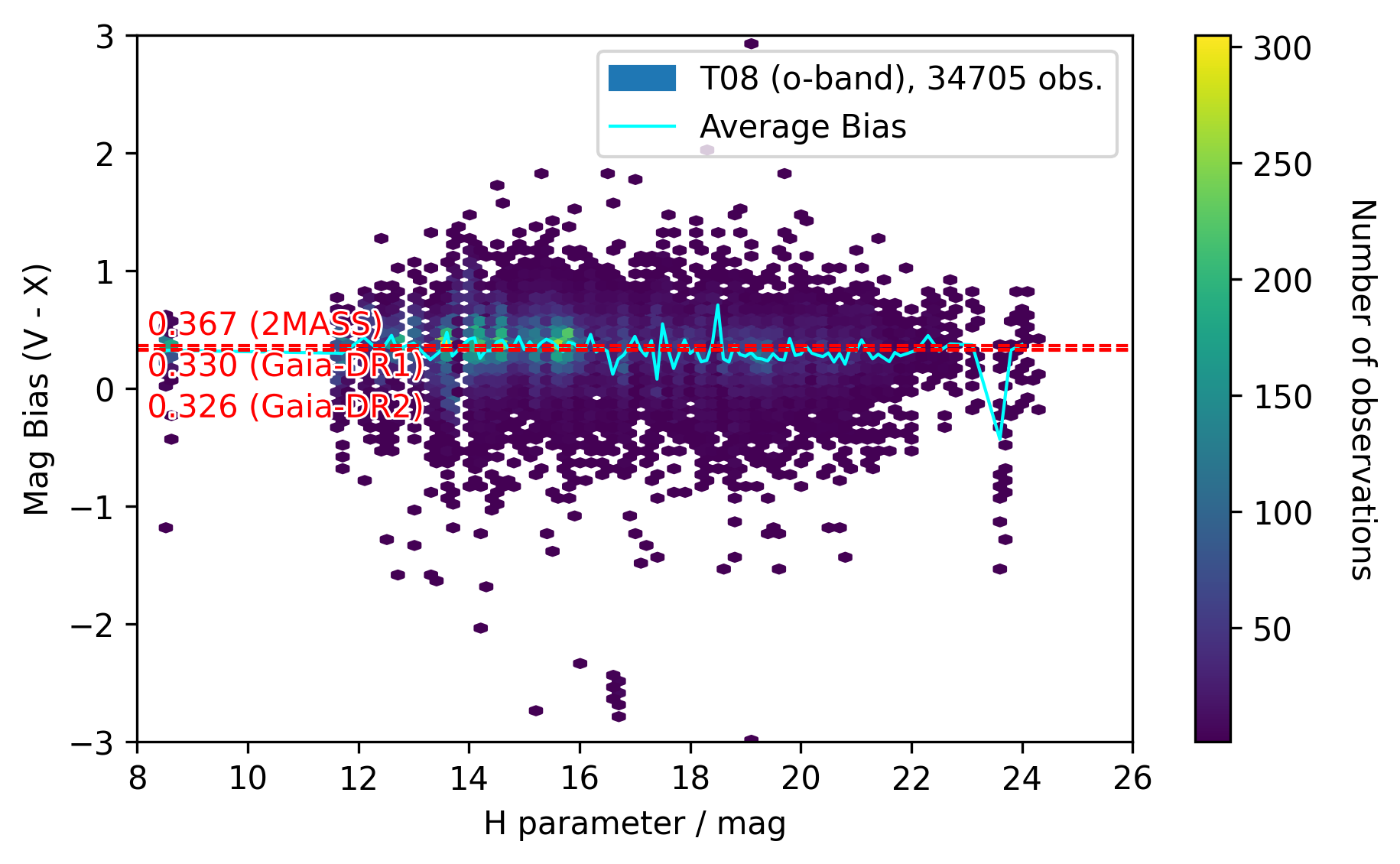}
    \end{subfigure}
    \vspace{-3mm}
    \caption{Dependence of absolute magnitude on photometric bias for large stations (Bin $0.2$~mag/$0.1$~mag).}
    \label{fig:dis.advanced.size}
\end{figure}

\begin{figure}[htb]
    \centering
    \vspace{-2cm}
    \begin{subfigure}[b]{0.85\linewidth}
        \centering
        \includegraphics[width=\textwidth, trim=0mm 8mm 0mm 2mm,clip]{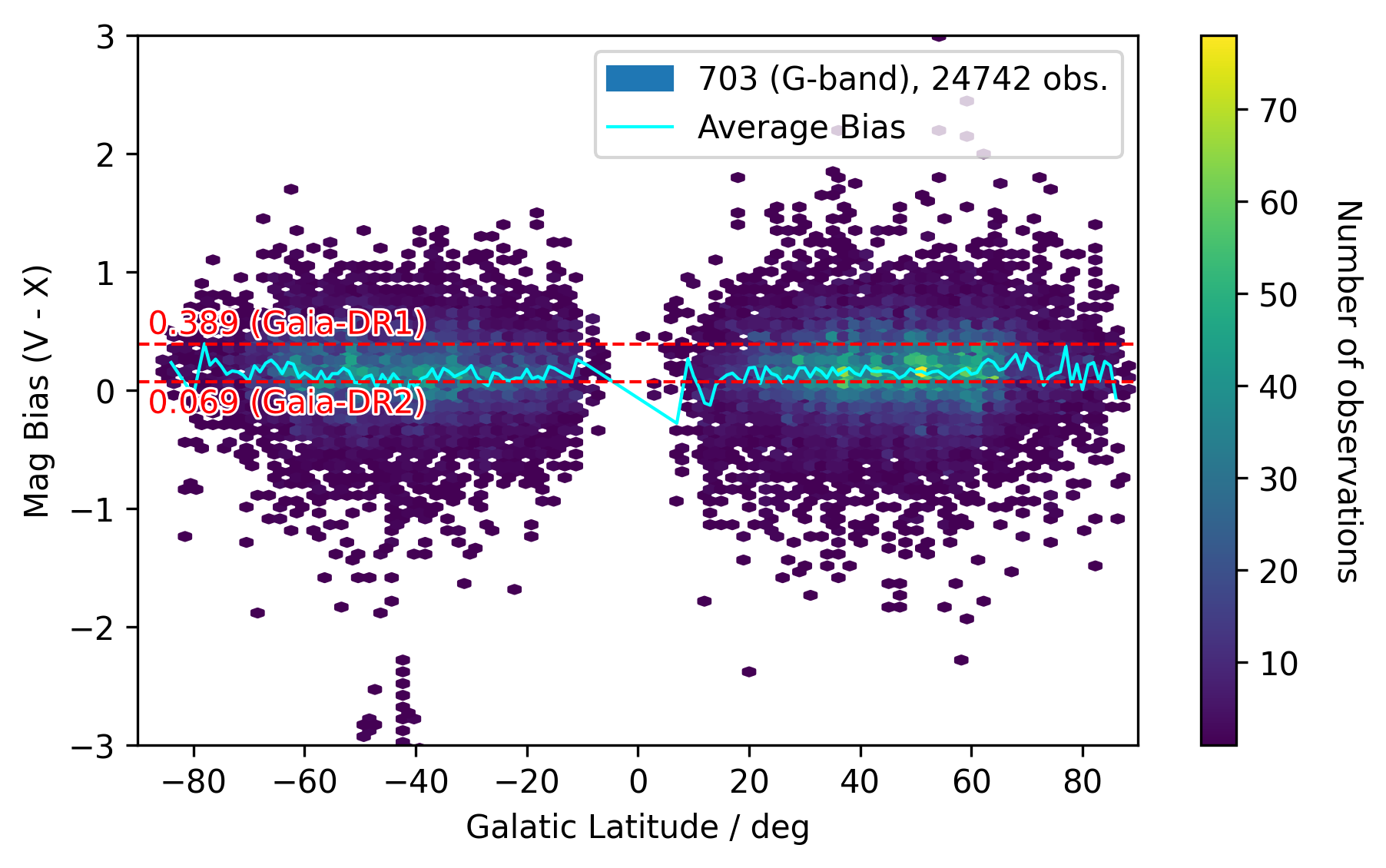}
    \end{subfigure}
    \begin{subfigure}[b]{0.85\linewidth}
        \centering
        \includegraphics[width=\textwidth, trim=0mm 8mm 0mm 2mm,clip]{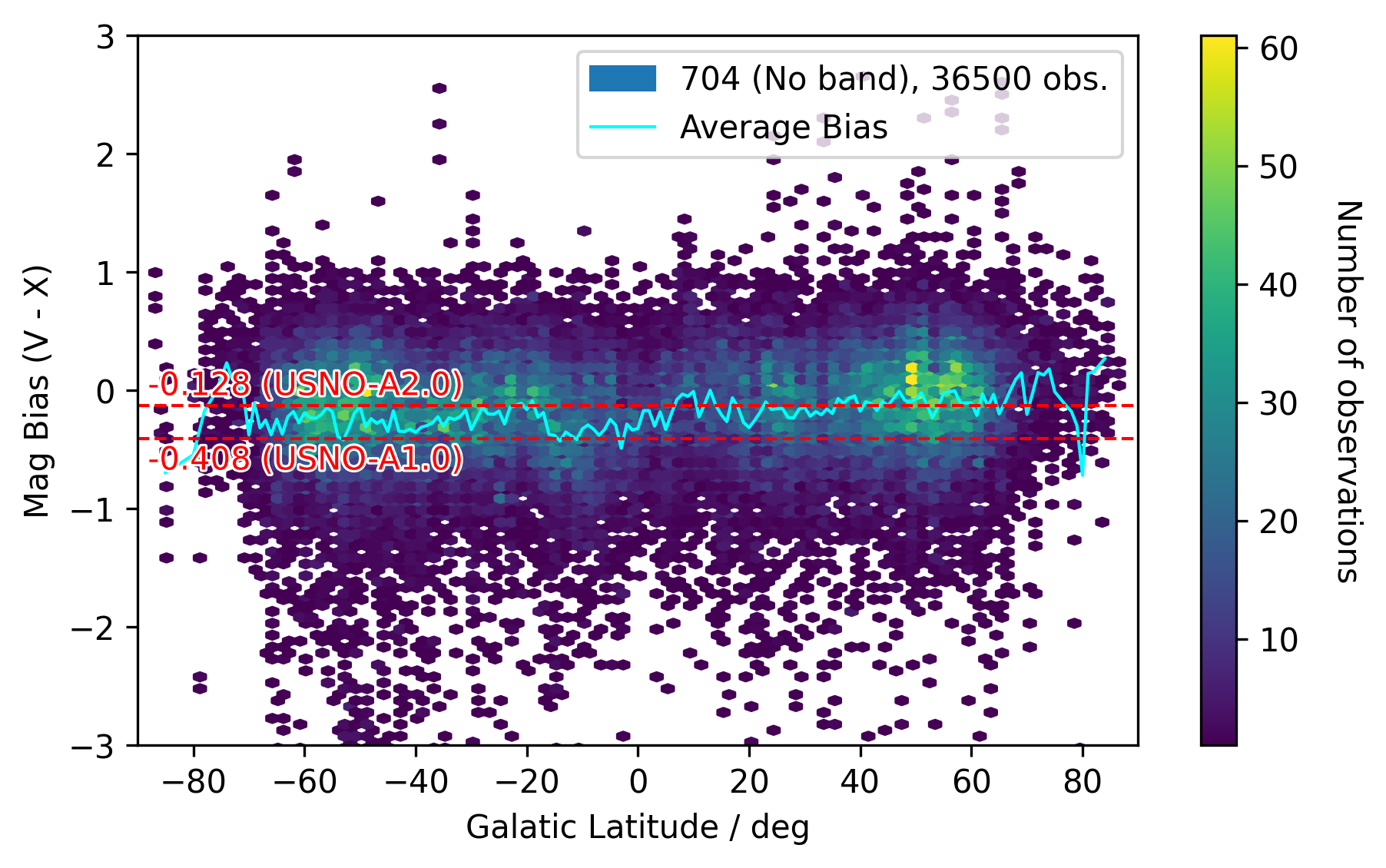}
    \end{subfigure}
    \begin{subfigure}[b]{0.85\linewidth}
        \centering
        \includegraphics[width=\textwidth, trim=0mm 8mm 0mm 2mm,clip]{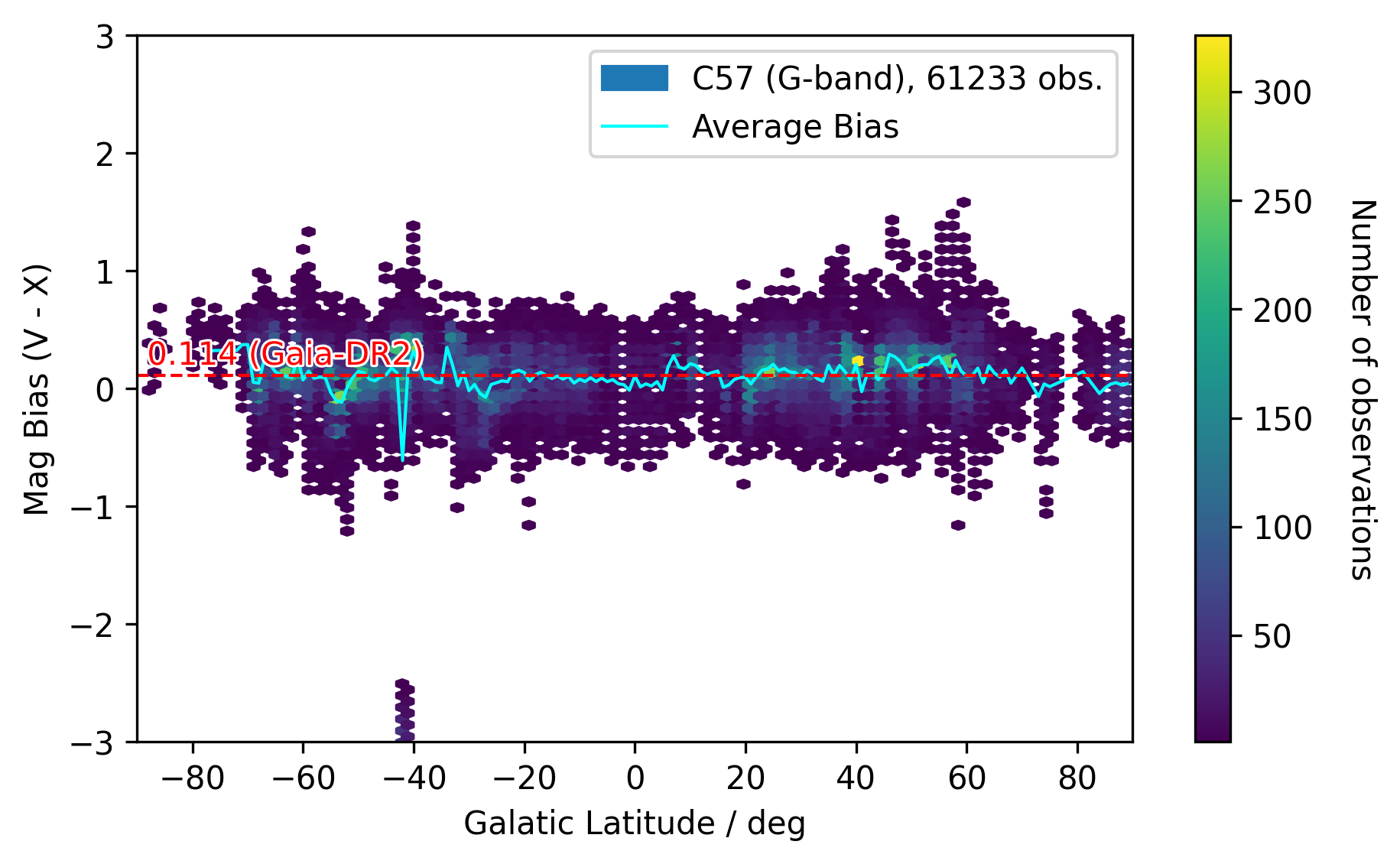}
    \end{subfigure}
    \begin{subfigure}[b]{0.85\linewidth}
        \centering
        \includegraphics[width=\textwidth, trim=0mm 8mm 0mm 2mm,clip]{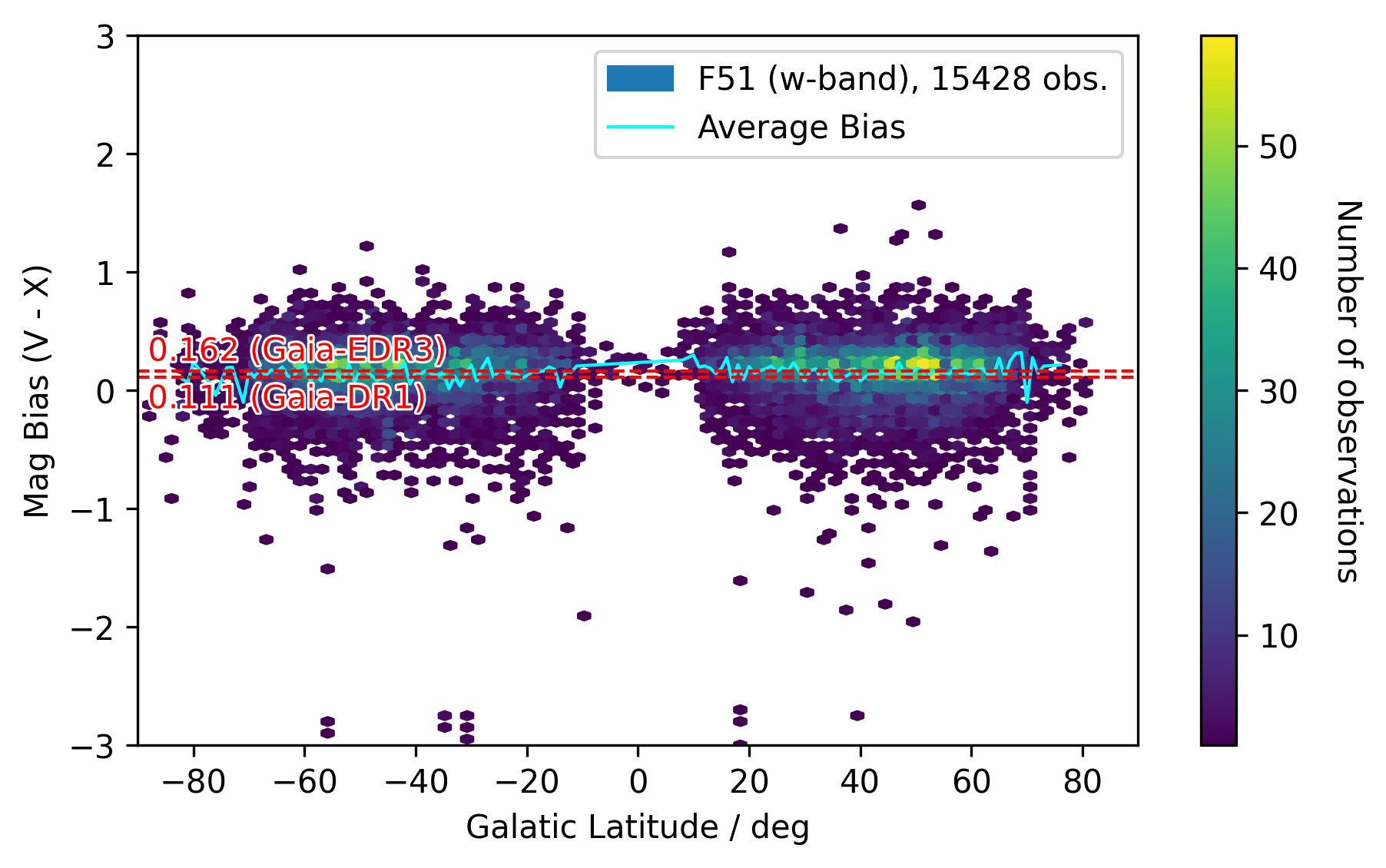}
    \end{subfigure}
    \begin{subfigure}[b]{0.85\linewidth}
        \centering
        \includegraphics[width=\textwidth, trim=0mm 8mm 0mm 2mm,clip]{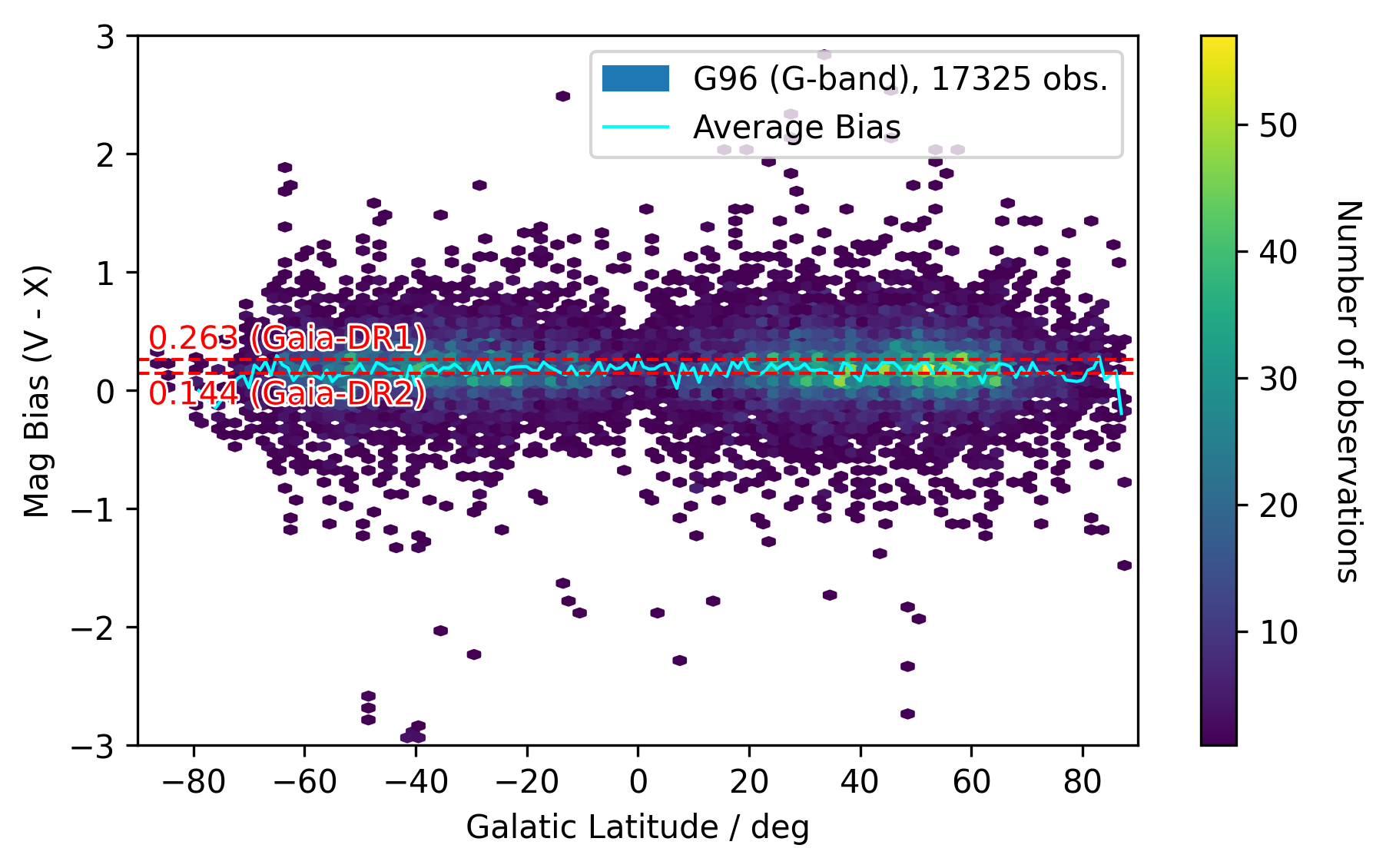}
    \end{subfigure}
    \begin{subfigure}[b]{0.85\linewidth}
        \centering
        \includegraphics[width=\textwidth, trim=0mm 0mm 0mm 0mm,clip]{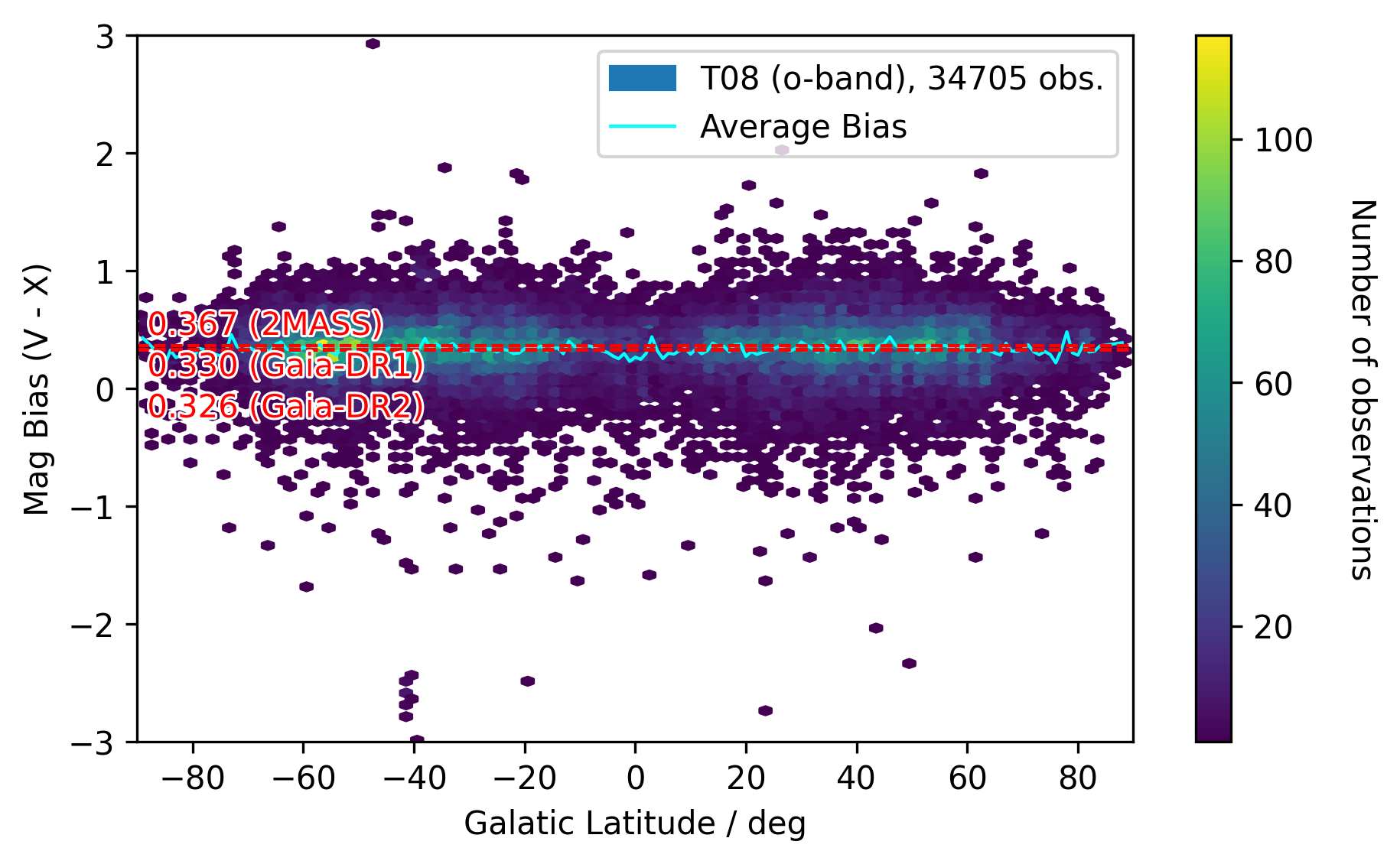}
    \end{subfigure}
    \vspace{-3mm}
    \caption{Dependence of galactic latitude on photometric bias for large stations (Bin $2^\circ$/$0.1$~mag).}
    \label{fig:dis.advanced.glx}
\end{figure}

\begin{figure}[htb]
    \centering
    \vspace{-2cm}
    \begin{subfigure}[b]{0.85\linewidth}
        \centering
        \includegraphics[width=\textwidth, trim=0mm 8mm 0mm 2mm,clip]{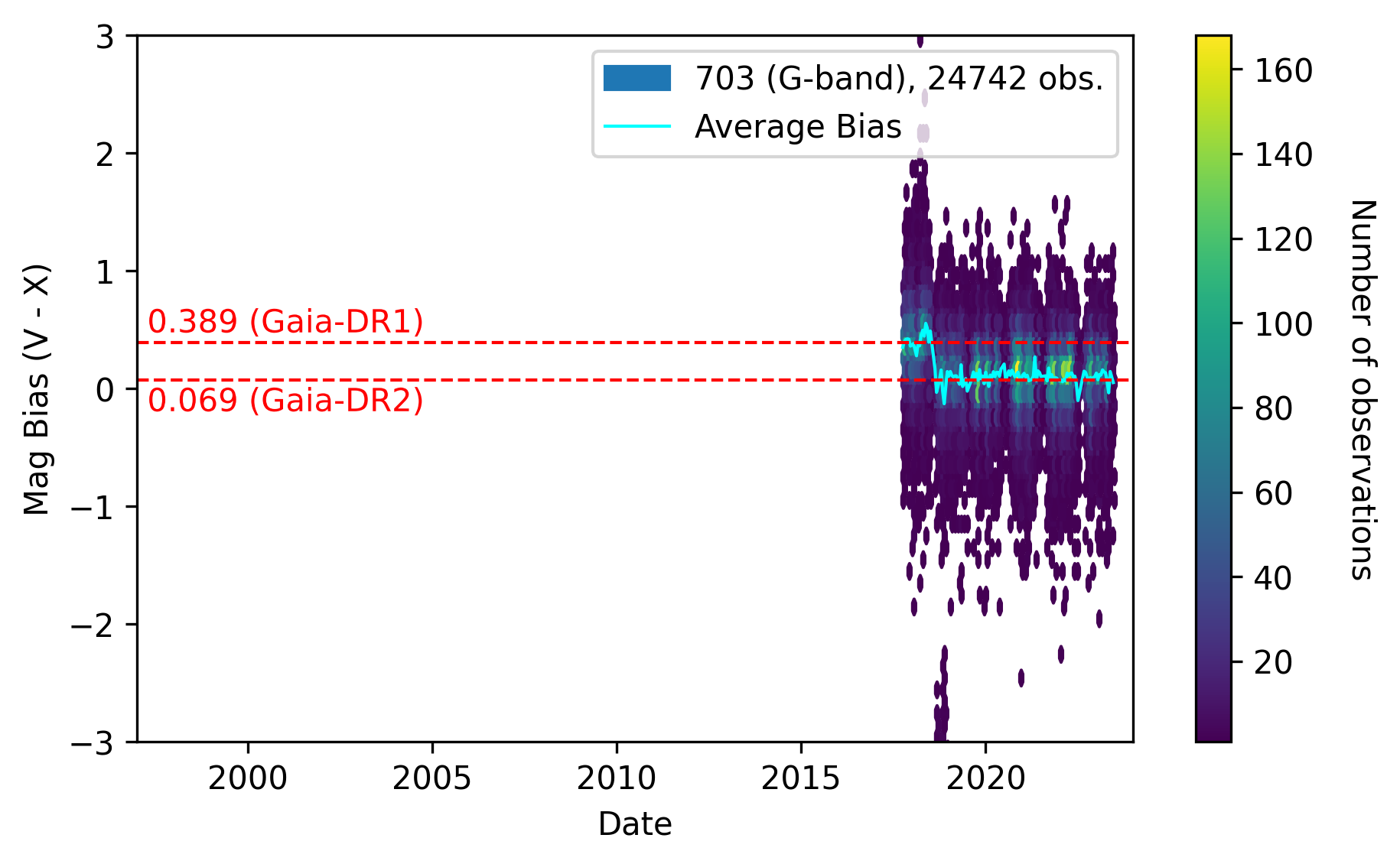}
    \end{subfigure}
    \begin{subfigure}[b]{0.85\linewidth}
        \centering
        \includegraphics[width=\textwidth, trim=0mm 8mm 0mm 2mm,clip]{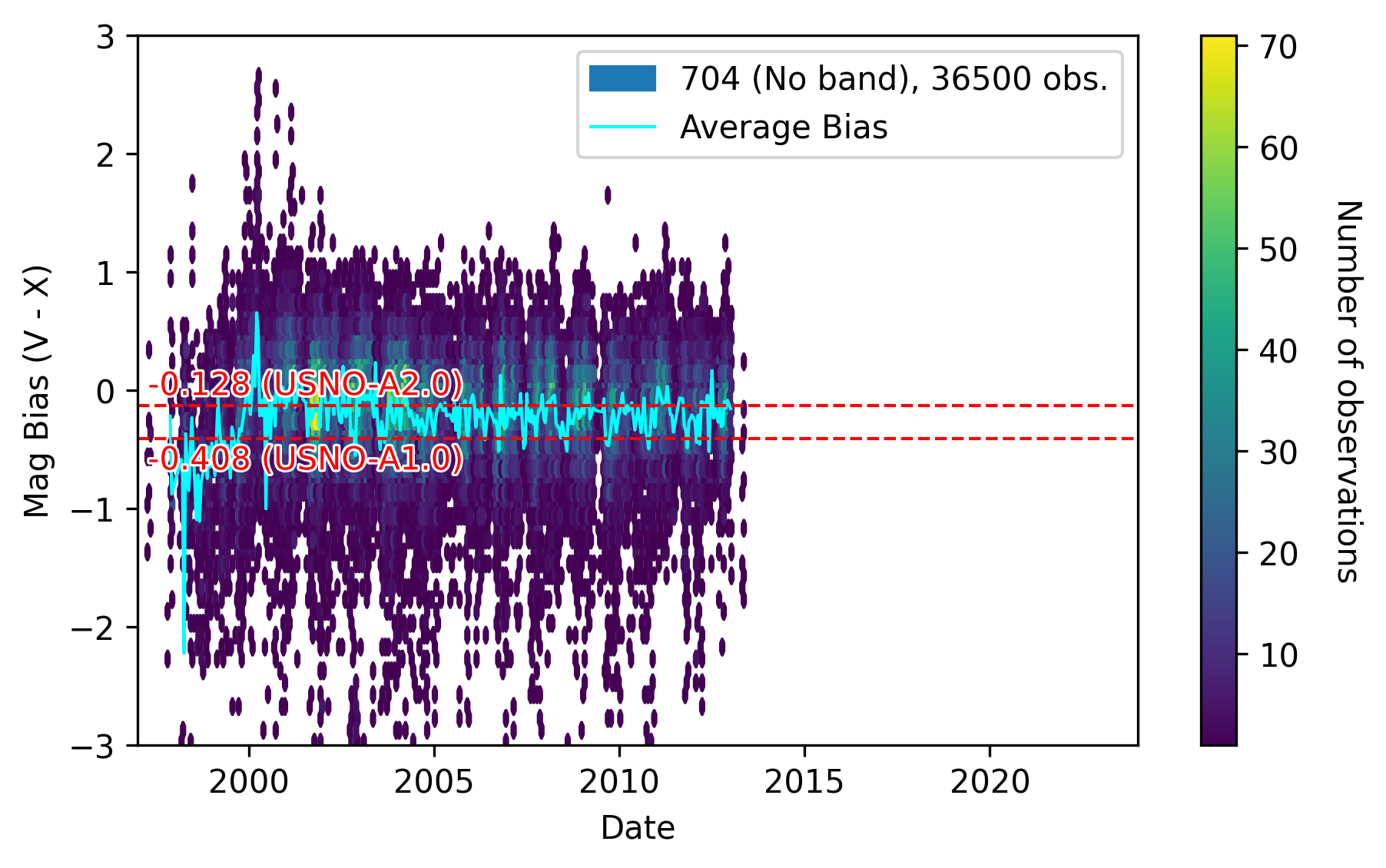}
    \end{subfigure}
    \begin{subfigure}[b]{0.85\linewidth}
        \centering
        \includegraphics[width=\textwidth, trim=0mm 8mm 0mm 2mm,clip]{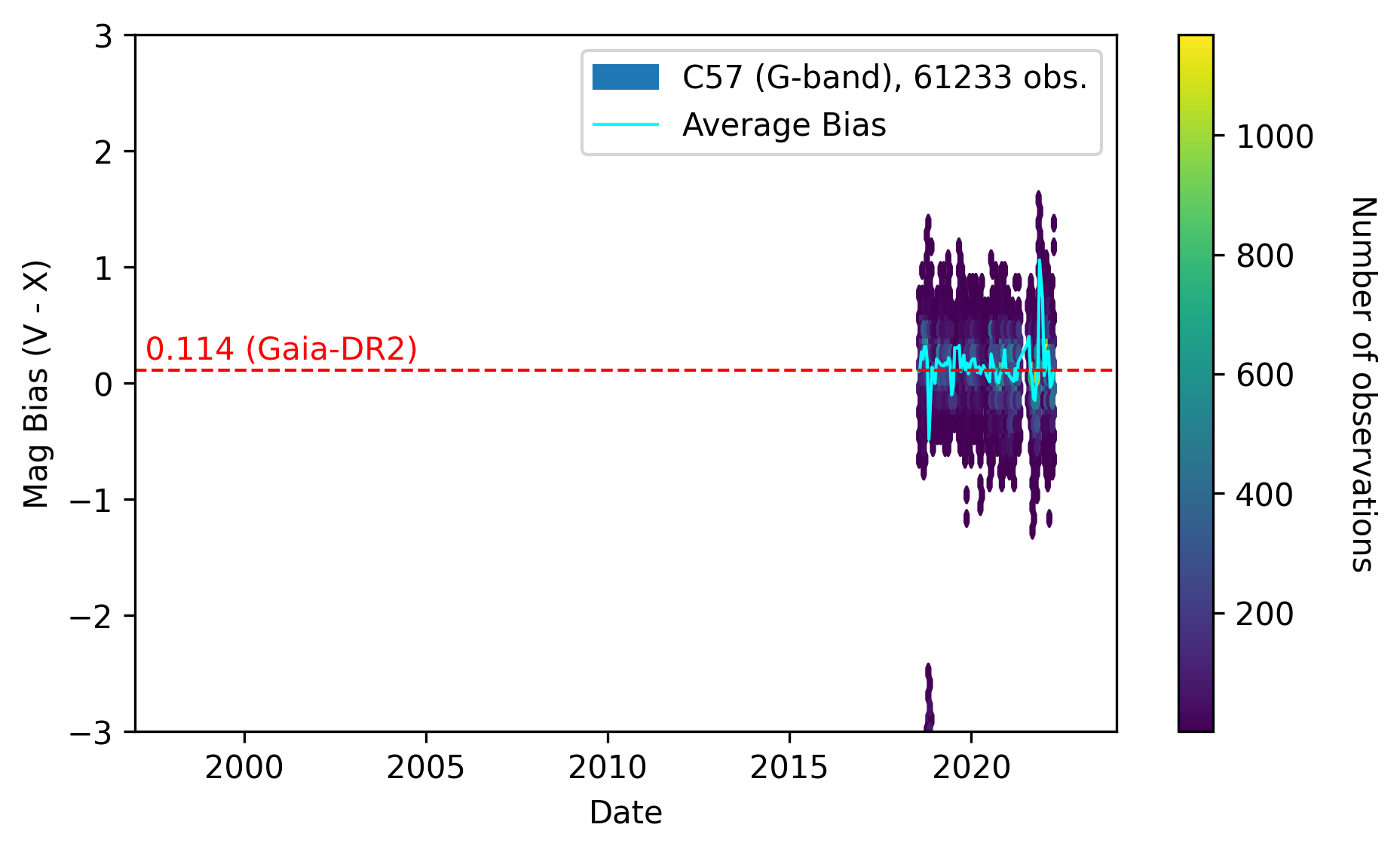}
    \end{subfigure}
    \begin{subfigure}[b]{0.85\linewidth}
        \centering
        \includegraphics[width=\textwidth, trim=0mm 8mm 0mm 2mm,clip]{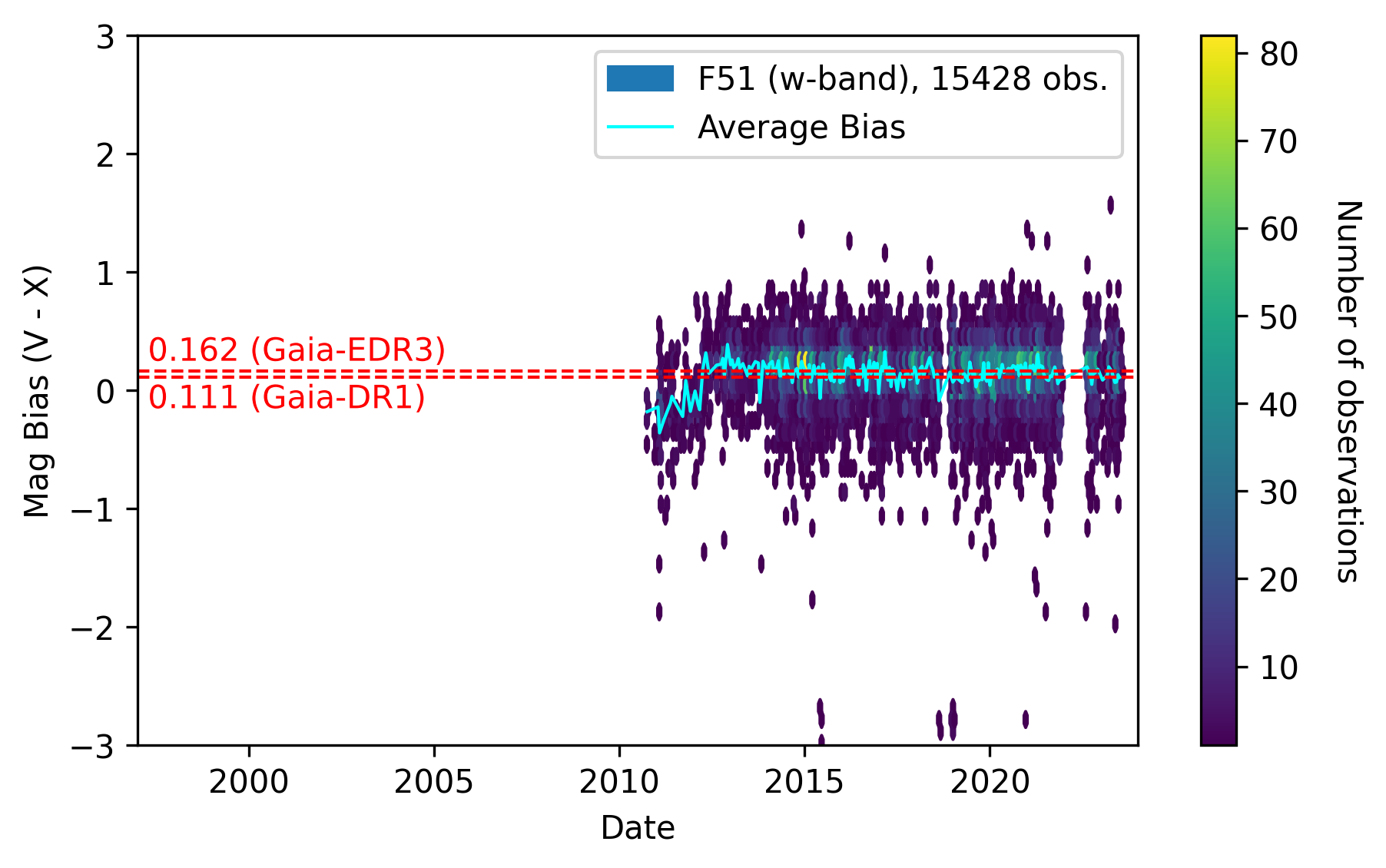}
    \end{subfigure}
    \begin{subfigure}[b]{0.85\linewidth}
        \centering
        \includegraphics[width=\textwidth, trim=0mm 8mm 0mm 2mm,clip]{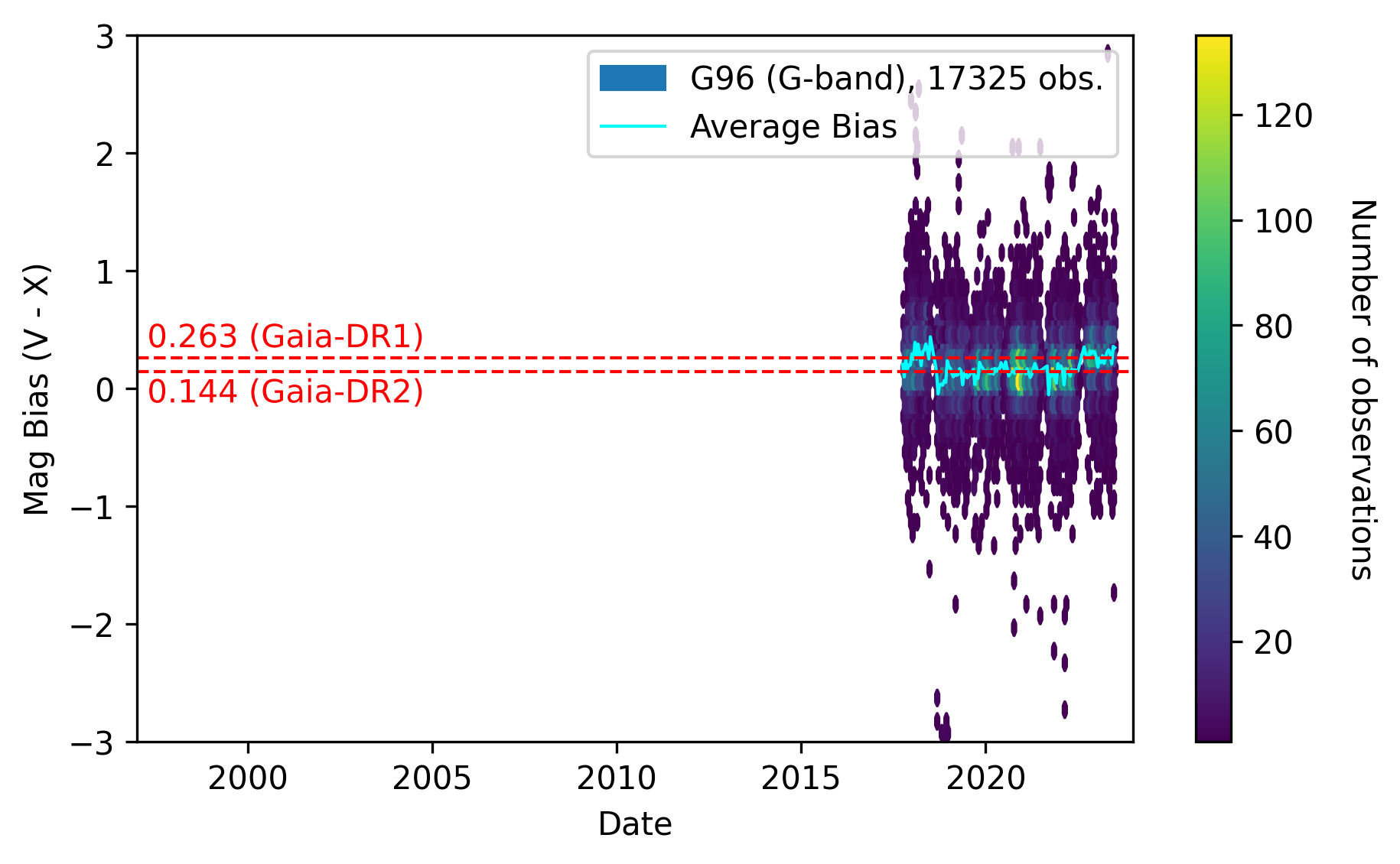}
    \end{subfigure}
    \begin{subfigure}[b]{0.85\linewidth}
        \centering
        \includegraphics[width=\textwidth, trim=0mm 0mm 0mm 2mm,clip]{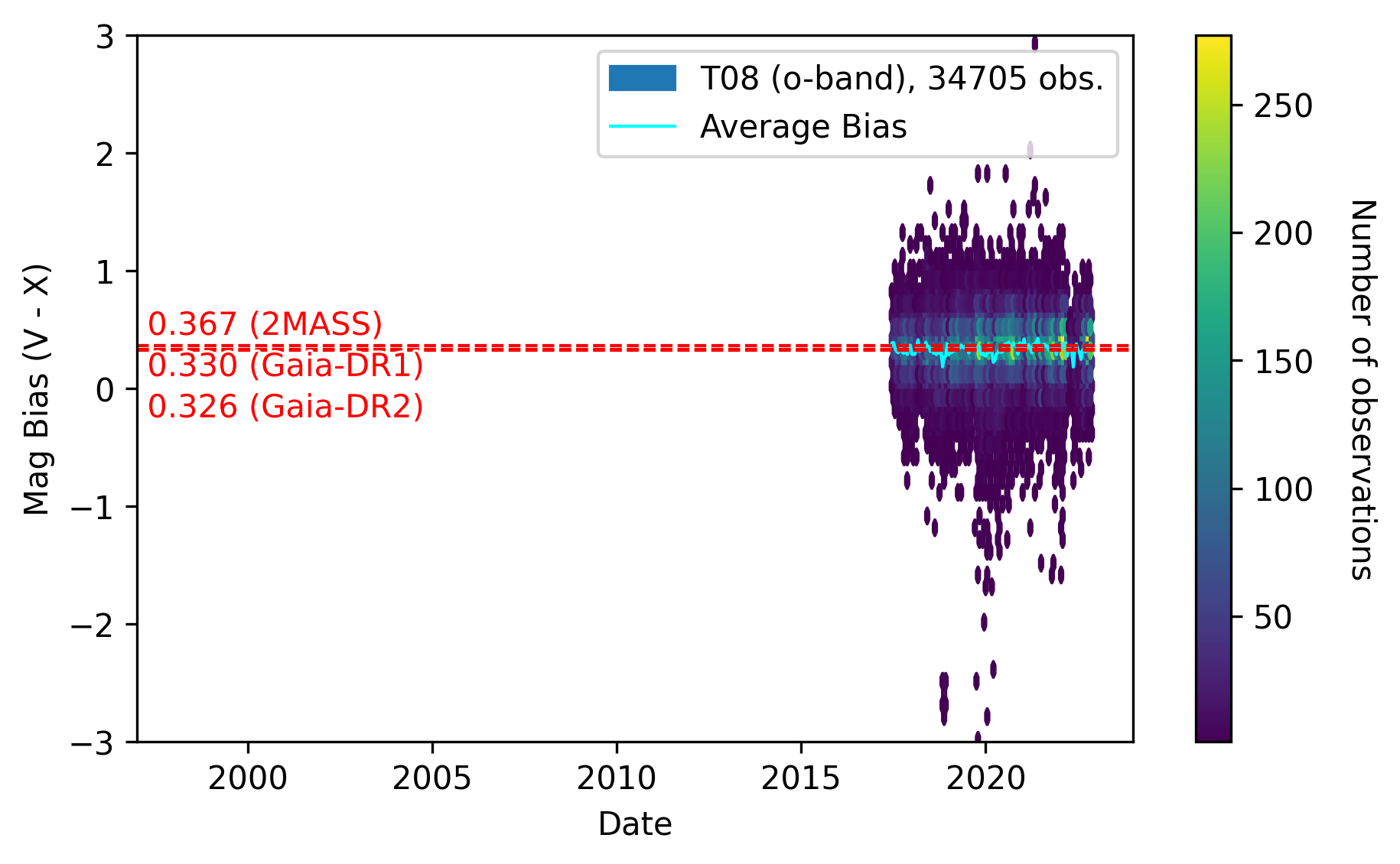}
    \end{subfigure}
    \vspace{-3mm}
    \caption{Dependence of time of observation on photometric bias for large stations (Bin $1$~month/$0.1$~mag).}
    \label{fig:dis.advanced.time}
\end{figure}

\end{document}